\documentclass[12pt,dense]{uvathesis}         %% LaTeX2e document.
\usepackage{amsmath}
\usepackage[final]{graphicx}
\usepackage{xspace}
\graphicspath{{eps/}}

\shorttitle{\GL Systems}
\title{Superheated and Dissipative \GL Systems}
\name{Andrew J.\ Dolgert} \home{Springfield, Virginia}
\degrees{B.A.\ Williams College, 1992} \department{Physics}
\monthyear{May, 1999} \numberofmembers{4}

\newcommand{\GL}{Landau-Ginzburg\xspace}
\newcommand{\del}{\boldsymbol{\nabla}}
\newcommand{\pad}[2]{\ensuremath{\frac{\partial #1}{\partial #2}}}
\newcommand{\fud}[2]{\ensuremath{\frac{d #1}{d #2}}}
\newcommand{\re}{\text{Re}}
\newcommand{\ncap}[2][{}]
  {\begin{center}
   \begin{minipage}{.8\hsize}
   \caption[#1]{#2}
   \end{minipage}\end{center}}

\newlength{\figwidth}

\begin{document}

\maketitle \copyrightpage
\begin{frontmatter}

\begin{acknowledgements}
An enormous number of people helped me complete this dissertation.
When Alan Dorsey accepted me as an advisee, he said he would be
happy to have a collaborator.  He, himself, has always been an
excellent collaborator, choosing projects thoughtfully, shaping
them into coherence, and deciding when they were done.  John
Di~Bartolo worked on the first half of them with me.  Tom Blum
worked on the second half as did Dr.~Michael Fowler, who has been
my patient advisor since Dr.~Dorsey's departure. When I needed to
talk with someone about a sticky point, there were always willing
professors around the department including Dr.~Vittorio Celli,
Dr.~Bascom Deaver, Dr.~Paul Fendley, Dr.~Hank Thacker, and
especially Dr.~Eugene Kolomeisky.

\medskip

Thanks Mom and Dad.  This would not have happened without Aunt
Tish's love and support.  Lastly, dear Warren has been the best
friend anyone could imagine, and that surely must count as a
helpful gift towards finishing this dissertation.

\medskip

NSF Grant DMR~96-28926 funded one Summer's research, and Dr.~Lou
Bloomfield lent me $64\:\text{MB}$ of memory and a Barracuda.

\end{acknowledgements}

\begin{abstractpage}
We study two superconducting systems using the \GL equations.  The
first is a superconducting half-space with an applied magnetic
field parallel to the surface.  We calculate the maximum applied
field that still supports superconductivity in the material by
solving the \GL equations analytically in asymptotic regimes of
the \GL parameter, $\kappa$.  These results are then checked
against numerical studies.

\medskip

The other system is a thin, current-carrying strip.  Here we do
the first systematic study of nucleation of superconductivity in a
normal strip and of the motion of a normal-superconducting
interface in the presence of a current.  This dissipative system
requires solving the time-dependent \GL equations analytically and
numerically.
\end{abstractpage}

\tableofcontents \listoffigures\listoftables
\end{frontmatter}

\chapter{Introduction---\GL Equations}
We are going to examine superconductors in large applied fields
that push them to the verge of nonequilibrium or out of
equilibrium.  We are interested in the nonlinear dynamics at a
mesoscopic level.  The mechanisms of superconductivity are not
intrinsic to this study.  Of more interest is how we can describe
balances among populations of superconducting and normal electrons
or the magnetic field and the Meissner effect.

We study two systems with phase boundaries.  One is the critical
field of a first-order phase transition, the other the progression
of a phase transition in time.  One first analyzes such systems in
the bulk by looking at free energies and possibilities of
metastable states, but studying spatial variations in the applied
fields and responses of the metallic states yields valuable
insight into how systems either maintain equilibrium or temper
their tendency to collapse.

The \GL (GL) equations will be our model for these systems.  It is
a venerable mean-field theory which continues to provide insight
into the actual behavior of metals.  These equations describe
general phase-transitions in terms of spatially-dependent
variables.  The \GL equations describe very nonlinear, and hence
complex, systems with little price in terms of calculating
specific material parameters.\footnote{The recent tendency in
literature to use the name Landau-Ginzburg equations rather than
the traditional Ginzburg-Landau equations seems to indicate an
interest in ascribing primacy to Landau.  In either case, it has
been nearly impossible to avoid abbreviating the name with either
GL or TDGL (for time-dependent \GL equations.)}

The \GL equations, as we work with them, are coupled nonlinear
differential equations.  All of the new research in this
dissertation deals directly with those differential equations. The
focus of the work, therefore, is essentially mathematical and of
interest in its own right.  All of it, however, is introduced,
executed, and summarized in light of the physical content.  In
fact, we begin this introduction to the \GL equations with a
description of the free energy.

\section{Thermodynamic Free Energy}
It can be shown from thermodynamic arguments that the transition
of a metal from normal to superconducting is first order if the
system is in an applied magnetic field.  When the metal is normal,
the magnetic field penetrates the sample.  The Gibbs free energy
is the free energy of the metal plus that of the magnetic field
\begin{equation}
    G_n = F_n+\frac{H^2}{8\pi}.
\end{equation}
When the system becomes superconducting, it expels the magnetic
field.  The additional work required to expel the field is equal
to the original field energy~\cite{degennes}
\begin{equation}
    G_s = F_s + \frac{H^2}{4\pi}.
\end{equation}
Here, $F_s$ is the free energy of the metal in the superconducting
state.  The latent heat of transition from normal to
superconducting is thus
\begin{equation}
    L = T(S_n-S_s) = -\frac{T}{4\pi}H\frac{dH}{dT}.
\end{equation}
Type-i metals abruptly expel applied magnetic fields at a critical
field or temperature so that $dH/dT$ is nonzero for all applied
fields. This is a first-order phase transition that will exhibit
hysteresis.  When there is no applied field, the transition is
second-order.

We want to look at the mechanics of the first-order phase
transition on a mesoscopic scale with the \GL equations.  The \GL
free energy permits spatial variations in the free energy
difference between the normal and superconducting states,
$G_n-G_s$, so that we can find what spatial distributions minimize
the free energy.

\section{\GL Free Energy}

In this section, we write down the free energy of a superconductor
and derive \GL differential equations describing its phases.  This
derivation is well documented in several
works~\cite{degennes,landau50}. Our goal here is to demonstrate
physical meanings of the nondimensionalized forms used in the rest
of the work. The \GL equations are derived from a free energy
postulated to be valid near phase transitions from a less ordered
phase to a more ordered one.  While the form of the free energy
near a phase transition is general, we will examine these
equations solely in the context of the transition from
superconducting to normal.

\begin{figure}
    \centerline{\includegraphics[width=\figwidth]{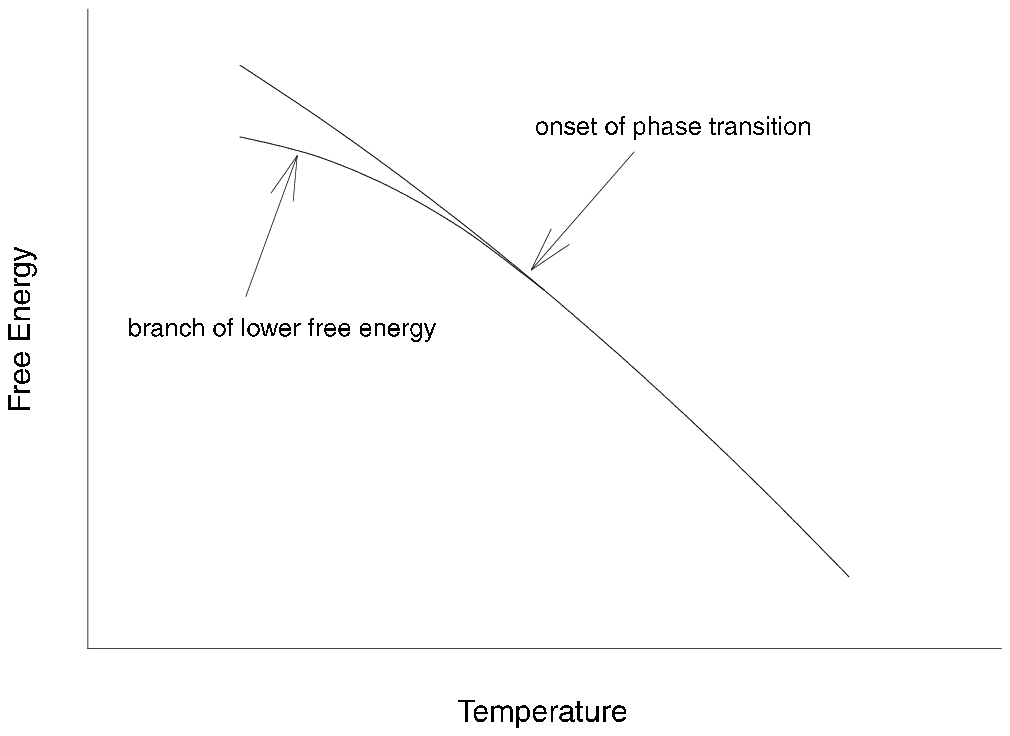}}
    \ncap[Free Energy Diagram of Second-order Phase Transition]{This figure
    shows the free energy of the normal and superconducting phases
    as a function of temperature. Below the critical temperature, $T_c$,
    the system undergoes a second-order phase change to the
    superconducting phase.\label{easyFreeEnergy}}
    \end{figure}

We are going to look at a phase transition from a higher
temperature, disordered state to a lower temperature state with
more order and less symmetry.  Above an onset temperature called
the critical temperature, $T_c$, the system has one available
state.  Below that temperature, the superconducting state with
lower free energy becomes accessible as shown in
Figure~\ref{easyFreeEnergy}.\footnote{The Free Energy decreases
with increasing temperature as seen from the thermodynamic
relation $S=-\left(\pad{A}{T}\right)_V$, where $A$ is the
Helmholtz Free Energy.}
If we can find how the Free Energy of the system depends on its
superconducting state, then standard thermodynamic arguments will
tell us when and how the system is superconducting. We measure how
superconducting the system is with an order parameter, $\psi =
|\psi|e^{i\theta}$. When the order parameter is zero, the system
is normal.  When the order parameter is nonzero, the system is
superconducting.  The density of superconducting electrons is
associated with the norm of the order parameter, $n_s \propto
|\psi|^2$.

The most significant independent variables are the temperature and
order parameter. Because the order parameter is small at the onset
of the phase transition, we want to write the free energy density
as a series in the order parameter
    \begin{equation}
    f = f_n + a\psi+b\psi^2+c\psi^3+d\psi^4+\dots
    \end{equation}
where the total free energy is the free energy of the normal state
plus an excess due to superconductivity.  The system, thus the
order parameter, will minimize the free energy.  Because the order
parameter, $\psi$, is complex (isomorphic to the SO(2) symmetry
group), the lowest order invariants are $|\psi|^2$ and $|\psi|^4$.
No first or third-order terms can appear in the free energy.  The
correct form, to fourth order, is
    \begin{equation}
    f = f_n + \alpha(T)|\psi|^2+
    \frac{\beta(T)}{2}|\psi|^4.\label{TheGLEquation}
    \end{equation}
The coefficients $\alpha$ and $\beta$ are constants at a given
temperature and depend on the system.
    \begin{figure}
    \centerline{\includegraphics[width=\figwidth]{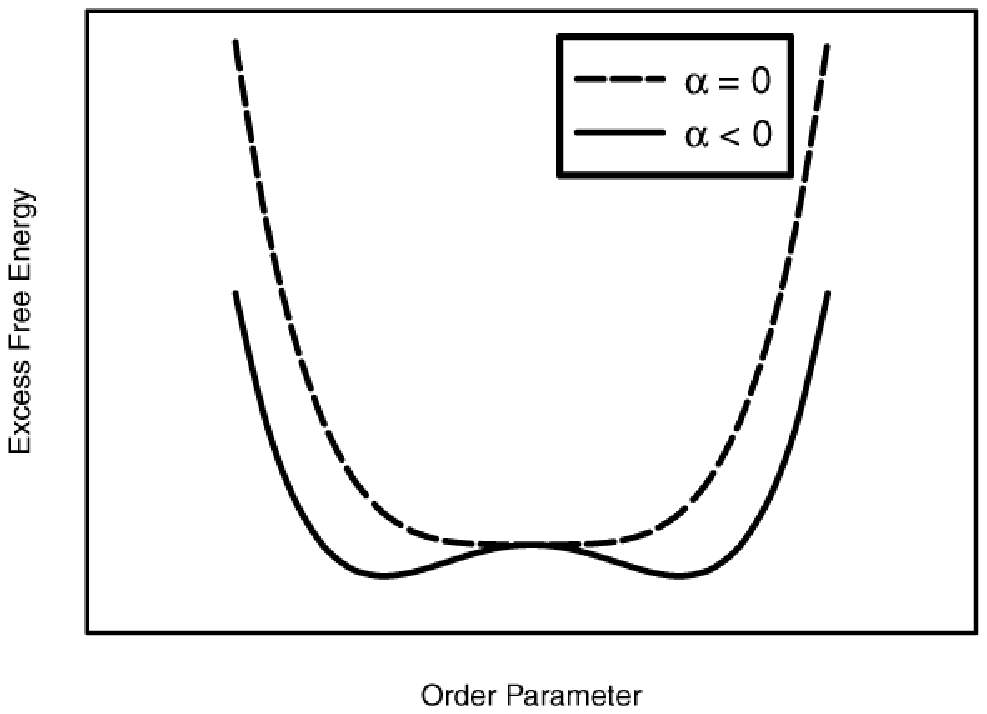}}
    \ncap[The \GL Energy Mexican Hat]{For lower temperatures where $\alpha<0$, the free energy
    finds its minimum at a nonzero order parameter.  At higher
    temperatures, $\alpha=0$, and the normal state is most stable.\label{easyMexicanHat}}
    \end{figure}%
There can be a free energy minimum for a nonzero order parameter
only if $\alpha<0$ and $\beta>0$.  In this case, the free energy
as a function of order parameter is as shown in
Figure~\ref{easyMexicanHat}, and the equilibrium state is $\psi =
\sqrt{-\smash[b]{\alpha/\beta}}$.  The parameters, $\alpha$ and
$\beta$, depend on temperature, and they define a maximum value of
the order parameter so that it varies between $\psi=0$ and
$\psi=\sqrt{-\smash[b]{\alpha/\beta}}$.
Equation~\ref{TheGLEquation} is \emph{the} \GL equation because it
contains the the basic physics of a bulk system.

Describing a system with any spatial variation requires
modifications to the \GL free energy above to account for the
energy cost of bending the order parameter.  The lowest order
derivative of the order parameter gives us
    \begin{equation}
    f = f_n + \alpha|\psi|^2+\frac{\beta}{2}|\psi|^4+
    \frac{1}{2m}\left|\hbar\del\psi\right|^2.
    \end{equation}
Lastly, if the metal is in an applied magnetic field,
$\mathbf{H}$, we can include the vector potential, $\mathbf{A}$,
of the magnetic induction, $\mathbf{B}=\del\times\mathbf{A}$,
gauge-invariantly to find Ginzburg and Landau's expansion of the
free energy density
    \begin{equation}
    f = f_n +
    \alpha|\psi|^2+ \frac{\beta}{2}|\psi|^4+
    \frac{|\mathbf{H}|^2}{8\pi}+
    \frac{1}{2m}\left|\left(-i\hbar\del-
    \frac{e\mathbf{A}}{c}\right)\psi\right|^2.
    \end{equation}
$f_n$ is the free energy of the normal phase.  The second to last
term is just the energy of the magnetic field and last term a
familiar representation of kinetic energy, here the kinetic energy
of superconducting charge carriers.

\section{Time-dependent \GL Equations}
Because we want to look at a metal at constant temperature, we
switch to the Gibbs free energy density.  The Gibbs free energy is
generally used for chemical systems and includes the work required
to expel the magnetic field from the sample. That energy is of the
form $U = \frac{1}{8\pi}\int \mathbf{B}\cdot\mathbf{H}$
    \begin{equation}
    g = f-\frac{\mathbf{B}\cdot\mathbf{H}}{4\pi}.
    \end{equation}
The total Gibbs free energy for the sample, integrated over all space, is then
    \begin{equation}
    G=\int\:d^3x\left\{ f_n + \alpha|\psi|^2+
    \frac{\beta}{2}|\psi|^4+\frac{|\mathbf{H}|^2}{8\pi}+
    \frac{1}{2m}\left|\left(-i\hbar\del-
    \frac{e\mathbf{A}}{c}\right)\psi\right|^2
    -\frac{\mathbf{B}\cdot\mathbf{H}}{4\pi}\right\}.
    \end{equation}
The state variables, $(\psi,\mathbf{A})$, will minimize the free
energy in steady state.  The minimization conditions can be
written as functional derivatives
    \begin{equation}
    \frac{\delta G}{\delta \psi} = 0\qquad\text{and}\qquad
    \frac{\delta G}{\delta\mathbf{A}} = 0.\label{disseqn1}
    \end{equation}
Equation~\ref{disseqn1} defines the time-independent
Ginzburg-Landau equations if $G$ is the GL free energy. If we want
to describe nonequilibrium systems, we can construct a
relaxational model using the same free energy
\begin{gather}
    \pad{\psi}{t} = -\Gamma\frac{\delta G}{\delta \psi^*}\label{RELAX}
    \qquad\text{and}\qquad\pad{\mathbf{A}}{t} = -\Gamma\frac{\delta
    G}{\delta\mathbf{A}}.
\end{gather}
This form is called the Glauber model or Type A dynamics.  The
rate of relaxation of the system is controlled by $\Gamma$, called
an Onsager coefficient.  While the order parameter and vector
potential are here shown with the same relaxation rate, we will
later choose separate constants, $\Gamma$, for each.

The Glauber model does not generate entirely general dynamical
equations. For instance, important work on thin superconducting
strips uses a model including higher order time
derivatives~\cite{schmid74}
\begin{equation}
    \frac{1}{\Gamma_1}\pad{\psi}{t} +\frac{1}{\Gamma_2}\pad{^2\psi}{t^2}=
    -\frac{\delta G}{\delta \psi^*}.
\end{equation}
The second-order derivative in time permits oscillatory,
self-driving solutions. Modern field theories typically include a
Langevin term describing noise
\begin{equation}
    \pad{\psi}{t} = -\Gamma\frac{\delta G}{\delta \psi^*}+\xi.
\end{equation}
The noise term is necessary in order to include thermally-driven
transitions explicitly in the theory.  Without it, the system
would always remain in any metastable state.  The pure
relaxational dynamics of the Glauber model assume the system will
always proceed towards a lower free energy without any
self-driving terms or noise.

The plan for the rest of the chapter is first to follow the
prescription above to derive a preliminary version of the
time-dependent \GL equation then write down the improved and
preferred equations derived by Gorkov and
Eliashberg~\cite{gorkov68}. The point is that, while Gorkov and
Eliashberg derived time-dependent \GL equations from microscopics,
the results are only a minor modification of those derived from
the simple relaxational dynamics above.

If we compute the two functional derivatives, $\frac{\delta
G}{\delta\psi^*}$ and $\frac{\delta G}{\delta \mathbf{A}}$
(Appendix~\ref{sec:derivetdgl}), we find
\begin{gather}
    \frac{1}{\Gamma}\pad{\psi}{t}+\alpha\psi+\beta|\psi|^2\psi+
    \frac{1}{2m}(-i\hbar\del-\frac{e}{c}\mathbf{A})^2\psi = 0 \label{eqn:td1}\\
    \frac{1}{\Gamma}\pad{\mathbf{A}}{t}+
            \frac{1}{4\pi}\del\times(\del\times\mathbf{A}-\mathbf{H})
    +\frac{e}{2mc}\psi(-i\hbar\del+\frac{e}{c}\mathbf{A})\psi^*
    +\frac{e}{2mc}\psi^*(i\hbar\del+\frac{e}{c}\mathbf{A})\psi=0.\label{eqn:td2}
\end{gather}
The first equation, describing the motion of the order parameter,
is largely described above.  We can see in this equation that in
the bulk where $\psi$ does not vary, it will have the value
$\psi_0=\sqrt{-\alpha/\beta}$.  In the second equation, we can
identify first the term $\del\times\del\times\mathbf{A}$ as the
total current generating the local magnetic field.  If we notice
that a stationary system has $\pad{\mathbf{A}}{t}=0$, we see the
last two terms must constitute the supercurrent.  For instance,
one of the boundary conditions on this system is that the total
current out of the sample be zero, or
\begin{align}
    (i\hbar\del+\frac{e}{c}\mathbf{A})\psi & = 0 \\
    \del\times\mathbf{A} &= \mathbf{H}.
\end{align}
A more common way to see the second equation written is
\begin{gather}
    \frac{1}{\Gamma}\pad{\mathbf{A}}{t}+\frac{1}{4\pi}\del\times(\del\times\mathbf{A}-\mathbf{H})
    -\frac{i\hbar{}e}{2mc}(\psi\del\psi^*-\psi^*\del\psi)
    +\frac{e^2}{mc^2}|\psi|^2\mathbf{A}=0.\label{tdgl22}
\end{gather}
The first term, $\pad{\mathbf{A}}{t}$ is related to the total
current for a system out of equilibrium.  Because it expresses the
inertia of the superconducting electrons accelerating in the
presence of a current, it is called the kinetic inductance.

The equation as it stands describes only supercurrent in a sample.  Since we
are also interested in samples with normal current, we should add an electric
field somewhere.  If we start from simple E\&M, we know that the total current
is related to the magnetic field by
\begin{equation}
    \del\times \mathbf{H} = \frac{4\pi}{c}\mathbf{J} = \frac{4\pi}{c}
    (\mathbf{J}_n+\mathbf{J}_s).\label{eqn:twofluid}
\end{equation}
Distinguishing the current as either normal or superconducting is
called the two-fluid model.  It implies there are distinct
populations of normal and superconducting electrons.  A more
current understanding is that the norm of the order parameter
represents the proportion of electrons \emph{participating in
Cooper pairing.} Comparing Eqn.~\ref{eqn:twofluid} with
Eqn.~\ref{tdgl22}, we see the supercurrent is
\begin{equation}
    \mathbf{J}_s = \frac{ie\hbar}{2m}(\psi\del\psi^*-\psi^*\del\psi)
        -\frac{e^2}{mc}|\psi|^2\mathbf{A}.
\end{equation}
Because the supercurrent does not dissipate energy, the normal
current alone is responsible for any electric field.  The
two-fluid model seems to identify
$\Gamma^{-1}\partial\mathbf{A}/\partial t$ with the normal
current.  That term is not normal current but the kinetic
inductance of the superconducting electrons.  Accelerating Cooper
pairs to an equilibrium superfluid velocity generates an ephemeral
voltage in the sample even though there is no dissipation.

If we want to include a normal current in our model, we can look
at replacing the kinetic inductance with a traditional term of the
form $\mathbf{J}_n = \sigma\mathbf{E}$.  Maxwell's Equations tell
us we can always express the electric field with a scalar
potential $\phi$ of the form
\begin{equation}
    \mathbf{E}+\frac{1}{c}\pad{\mathbf{A}}{t} = -\del\phi.
\end{equation}
If we solve that equation for $\mathbf{J}_n$ and substitute it
into Eqn.~\ref{tdgl22}, we find the kinetic inductance appears
naturally as part of the electric field. Assembling the pieces in
the form $\mathbf{J}_t = \mathbf{J}_n+\mathbf{J}_s$, we find
\begin{gather}
    \frac{c}{4\pi}\del\times\del\times\mathbf{A} =
    -\frac{\sigma}{c}\pad{\mathbf{A}}{t}-\sigma\del\phi
    +\frac{i\hbar{}e}{2m}(\psi\del\psi^*-\psi^*\del\psi)
    -\frac{e^2}{mc}|\psi|^2\mathbf{A}\label{tdgl23}
    +\frac{c}{4\pi}\del\times\mathbf{H}
\end{gather}
In addition, we have found that Maxwell's Equations determine the
Onsager Coefficient, $\Gamma$ for the relaxation equation for the
vector potential.  While there was a single relaxation constant,
$\Gamma$, for the time dependence of both the order parameter and
magnetic field, we will now set one of them to agree with
Maxwell's equations.  This is just a likely choice, and those who
look to \GL equations for more precise correspondence with
physical values will often reserve the right to vary independently
relaxation coefficients for the order parameter and magnetic
field.

One can find in the above equation a gauge symmetry whereby the
equation is unchanged under transformations of the type
\begin{equation}
    \psi \rightarrow \psi e^{\frac{ie}{\hbar c}\chi}\qquad
    \mathbf{A} \rightarrow \mathbf{A}+\del\chi\qquad
    \phi \rightarrow \phi-\frac{1}{c}\pad{\chi}{t}.
\end{equation}
In order that both the equation for the magnetic potential and the
order parameter be gauge invariant, we should add a
gauge-invariant term for $\phi$ to Eqn.~\ref{eqn:td1} to get
\begin{gather}
    \frac{1}{\Gamma}\pad{\psi}{t}+\frac{ie}{\hbar\Gamma}\phi
    +\alpha\psi+\beta|\psi|^2\psi+
    \frac{1}{2m}(-i\hbar\del-\frac{e}{c}\mathbf{A})^2\psi = 0.\label{tdgl11}
\end{gather}
Our insertion of an electric field by hand has effectively
switched to proper canonical variables.  We have been seeking a
sensible form similar to that derived from microscopics by Gorkov
and Eliashberg~\cite{gorkov68} and quoted in Du et
al.~\cite{du94}\footnote{When quoting this equation, I have
switched the gauge field, $\mathbf{A}$, to match the sign of my
definition, $\mathbf{B} = \del\times\mathbf{A}$.  Du et al.\
explained the sign changes by switching the sign of the electric
field, a choice not generally welcome to physicists.}
\begin{gather}
    \hbar\pad{\psi}{t}+ie\Phi\psi-D(\hbar\del+ie\mathbf{A})^2\psi+
    (\beta|\psi|^2+\alpha)\psi = 0 \\
    \nu \del\times\del\times\mathbf{A} = \mathbf{E}+2\tau
    \left[|\psi|^2\mathbf{A}+\frac{i\hbar}{2e}(\psi^*\del\psi-\psi\del\psi^*)
    +\del\times\mathbf{H}\right].
\end{gather}
Without defining the constants, we can say that these equations
are, term for term, proportional to those we have derived.  For
reasonable choices of constants, they behave like those derived
straight from the \GL free energy.

The derivations of Eqns.~\ref{tdgl11} and \ref{tdgl23} may have
seemed correct enough with respect to Maxwell's Equations that the
version of the time-dependent \GL equations derived by Gorkov and
Eliashberg have an unnecessary number of parameters.  A quick way
to see how we have been duped by treading well-worn paths is to
look for a continuity equation for the charge associated with the
superconducting current
\begin{equation}
    \pad{|\psi|^2}{t}+\del\cdot\mathbf{J}_s = 0.
\end{equation}
A note by Neu~\cite{neu} points out that such a continuity
equation exists, but, in our variables, we would need to choose
$\Gamma=-ie/\hbar$.  That choice of the Onsager coefficient would
violate our relaxational dynamics, but it gives the nonlinear
Schr\"odinger equation. In practice, choosing a complex
coefficient allows one to study the Hall Effect in
superconductors, but we will generally choose $\Gamma$ such that
the time derivative of $\psi$ has no factor in front.  The
implication is that $|\psi|^2$ is only loosely associated with the
density of superconducting electrons. What we have instead is a
non-conserved order parameter.  Our version of a continuity
equation is
\begin{equation}
    \frac{1}{\Gamma}\pad{|\psi|^2}{t}+\frac{\hbar^2}{2m}\left(\del^2|\psi|^2
    -2\del\psi\cdot\del\psi^*\right)-\frac{2}{c}\mathbf{J}\cdot\mathbf{A} = 0
\end{equation}
which is more of a balance between energy in the system and
dissipation in the current.

\section{Dimensionless Form of \GL Equations}

It now falls to us to rewrite the Landau-Ginzburg equations in
dimensionless form.  In the process, we will expose the important
length scales in the problem.  The two length scales in the system
are the magnetic field penetration depth and the superconducting
coherence length. The illustrations of length scales from the
time-dependent \GL equations are very simple and are shown here
much as they appear in Du, Gunzburger, and Peterson~\cite{dgp92}.

The magnetic penetration depth, $\lambda$, is a measure of how far
an applied magnetic field penetrates a sample.  If we look at a
perfectly conducting sample occupying a halfspace and apply a
constant magnetic field, then the time-dependent \GL equations,
Eqns.~\ref{eqn:td1} and~\ref{eqn:td2}, simplify to
\begin{equation}
    \frac{c}{4\pi}\del\times\del\times\mathbf{A} =
    -\frac{e^2}{m c}\psi_0^2\mathbf{A}.
\end{equation}
Taking the curl of this equation, we arrive at the London equation,
\begin{equation}
    \del\times\del\times\mathbf{H}+\frac{1}{\lambda^2}\mathbf{H} = 0
\end{equation}
where the magnetic penetration depth is
\begin{equation}
    \lambda = \sqrt{\frac{mc^2}{4\pi e^2\psi_0^2}}.
\end{equation}
The London equation predates the \GL equations.  It says that a
magnetic field applied to the surface will have a decay length of
$\lambda$.

We can make a similar calculation for the coherence length.  If we
imagine a metal normal to the left, superconducting to the right,
then the order parameter has to rise from $\psi=0$ to
$\psi=\sqrt{-\smash[b]{\alpha/\beta}}$.  The \GL equations in the
absence of magnetic field reduce to
\begin{equation}
    \alpha\psi-\beta|\psi^2|\psi-\frac{\hbar^2}{2m}\del^2\psi=0.
\end{equation}
The solution to this equation with the desired boundary conditions is
\begin{equation}
    \psi = \sqrt{\frac{-\alpha}{\beta}}\tanh\left(\frac{x-x_0}{\sqrt{2}\xi}\right)
\end{equation}
where
\begin{equation}
    \xi = \left(\frac{\hbar^2}{-2m\alpha}\right)^{1/2}.
\end{equation}
We call $\xi$ the coherence length.  It is the order parameter's recovery
length to perturbations.

Using the length scales shown above, we can (see
Appendix~\ref{sec:dimensionless}) rescale the variables so that only two remain
\begin{gather}
    \gamma\left(\pad{\psi}{t}+i\kappa\phi\psi\right)
        +\psi-|\psi|^2\psi
        -(\frac{i}{\kappa}\del+\mathbf{A})^2\psi = 0 \label{eqn:dimless1}\\
    \pad{\mathbf{A}}{t}
    +\del\phi
    +\del\times\del\times\mathbf{A}
        -\frac{i}{2\kappa}(\psi\del\psi^*-\psi^*\del\psi)
        +\mathbf{A}|\psi|^2 = 0.\label{eqn:dimless2}
\end{gather}
The switch to dimensionless variables is nicely summarized in Du,
Gunzburger, and Peterson~\cite{dgp92}.  A more complete summary is
shown in Table~\ref{tab:transform}.
\begin{table}
\begin{alignat*}{3}
  \lambda & = \left(\frac{mc^2}{4\pi e^2|\psi_0|^2}\right)^{1/2}\qquad
 & \xi &= \left(-\frac{\hbar^2}{2m\alpha}\right)^{1/2}\qquad
 & H_c &= \left(\frac{4\pi\alpha^2}{\beta}\right)^{1/2} \\
 \kappa &= \frac{\lambda}{\xi} =
        \sqrt{\frac{\beta}{2\pi}}\frac{mc}{e\hbar}\qquad
 & \psi_0 &= \sqrt{-\alpha/\beta}\qquad
 & x &= \lambda x' \\
 H &= \sqrt{2}H_cH'\qquad
 & \mathbf{j} &= \frac{cH_c}{2\sqrt{2}\pi\lambda}\mathbf{j'}\qquad
 & \mathbf{A} &= \sqrt{2}\lambda H_c\mathbf{A}'\\
 g &= \frac{\alpha^2}{\beta} g'\qquad
 & G &= \frac{\alpha^2\lambda^3}{\beta} G'\qquad
 & t &= \frac{4\pi\sigma\lambda^2}{c^2} t'
\end{alignat*}
\ncap[Dimensionless TDGL Rescalings]{These are the rescalings of
the time-dependent \GL equations to their standard dimensionless
form.\label{tab:transform}}
\end{table}
The Onsager coefficient has become
\begin{equation}
    \gamma = \frac{1}{\alpha\Gamma t_0}
\end{equation}
where $t_0$ is the rescaling of time shown in
Table~\ref{tab:transform}, and calculations from BCS theory
determine it to be between $0.8$ and $1.0$~\cite{gorkov68}. As you
may guess, we'll take it to be unity.

The other parameter is the Ginzburg-Landau parameter,
\begin{equation}
    \kappa = \frac{\lambda}{\xi}
\end{equation}
which is a dimensionless parameter characterizing the metal under
consideration.  In this formulation of the \GL equations, it is
the only such parameter.  A metal with small $\kappa$ was once
called hard---it resists penetration by magnetic
fields.\footnote{These days \emph{hard} superconductors refer to
type-ii superconductors with lots of pinning impurities to hold
vortices.} Metals with $\kappa<1/\sqrt{2}$ are called type-i,
those with $\kappa>1/\sqrt{2}$ type-ii.

In these new variables, the current is rescaled so that the nondimensional
current
\begin{equation}
    \mathbf{J}_s = \frac{i}{2\kappa}(\psi\del\psi^*-\psi^*\del\psi)
    -|\psi^2|\mathbf{A}.
\end{equation}
The new boundary conditions are
\begin{gather}
    \left(\frac{i}{\kappa}\del\psi+\mathbf{A}\psi\right)\cdot\hat{n} = 0 \\
    \del\times\mathbf{A}\times\hat{n} = \mathbf{H}\times\hat{n}.
\end{gather}
They still express the requirement that no supercurrent flow out of the sample
and that the tangential magnetic field be continuous at the boundary.

These nondimensional TDGL equations, Eqns.~\ref{eqn:dimless1}
and~\ref{eqn:dimless2}, are also invariant under a gauge
transformation,
\begin{equation}
    \psi \rightarrow \psi e^{i\kappa\chi},\qquad \mathbf{A} \rightarrow
    \mathbf{A}+\del\chi,\qquad\text{and}\qquad
    \phi \rightarrow \phi-\pad{\chi}{t}.
\end{equation}
This indicates that, as written, the equations are not well-posed
because they do not have a single set of solutions but a whole
family of solutions joined by the function $\chi$.  While it is
possible to solve the equations analytically for a family of
solutions without specifying a specific $\chi$, we will work with
specific functions $\chi$, or specific gauges.  We will derive
those gauges as they apply in later chapters.

We have introduced the \GL equations from general properties of
the free energy and with little reference to the physics of the
electronic system of the normal or superconducting metals.  The
majority of this work will be concerned with the equations
themselves.  As just shown, the physical system informed the
derivation at two points---the choice of $SO(2)$ symmetry in the
original free energy and the length scales of the gauge variables,
$\lambda$, $\zeta$ and $\phi$.  We use those length scales to
analyze the superheating of a half-space in the next chapter.

\chapter{Superheating of Superconductors}

\section{Introduction}

A first analysis of the free energy of a metal examines the properties of
a bulk specimen large enough that surface tension associated with edges is
negligible. Below a thermodynamic critical field, $H_c$, a bulk superconductor
in equilibrium will be in a Meissner phase.  Above that critical field, a type-i superconductor
will become normal, allowing magnetic fields to penetrate, and a type-ii
superconductor will enter the Abrikosov phase, characterized by the infusion of
flux into the sample in the form of discrete vortices.

If, instead of estimating the critical field for a phase
transition just by comparing which phase has the least free
energy, we, in addition, ask at what applied field flux at the
boundary of the sample (with vacuum) can push past through the
surface currents, we see that the phase transition is delayed.  If
one were to begin with a superconducting metal and steadily
increase the magnetic field, there would not be a phase transition
at $H_c$ but for some higher value we call the superheating field,
$H_{\text{sh}}$.  (While we are not actually heating the sample,
the word ``superheating'' refers to the similar delay of a phase
transition in more familiar systems like water.)  For applied
magnetic fields well above the critical field, the bulk would
prefer to change phases, but the order parameter at the surface
decreases from its value in the bulk in order that the metal
remain superconducting.  On a microscopic scale, the depression of
the order parameter is caused when the penetrating magnetic field
aligns spins in the Cooper pairs.  The phase transition is delayed
only when the edges of the sample are considered.

If we want to understand the time-dependent collapse of
superconductivity in a superconductor, characterizing the
superheated Meissner state will tell us our initial conditions for
the collapse.  The superheating solutions to our system (described in
Eqns.~\ref{eqn:fullgi1}--\ref{eqn:gibc2}) measure energy barriers
separating local equilibrium from catastrophic collapse.  Scaling
solutions for cascading phase transitions often depend on first
integrals of initial conditions during a superheated state.  We
will not be able to produce those scaling solutions, but a later
chapter pursues a similar system found more tractable.

The matched asymptotics in this chapter were published in a
journal paper~\cite{dolgert96}.  New material appears in the
analysis of perturbations on the solutions.

\subsection{The Physical System}

The superheating field is more practically important for quite a
few reasons.  First, it is one of the most common ways to
determine the \GL parameter of a type-i
metal~\cite{burger69}.  Second, there have been proposals in the
past ten years to use superconducting granules to detect weak
electromagnetic elementary particles, including WIMPS and magnetic
monopoles~\cite{meagher97}. Here, the granules act as bubble
chambers, flipping from superconducting to normal when a particle
strikes.  A review is given in Barone~\cite{barone}.

Both measurements of the \GL parameter and detectors study small
spherical or ellipsoidal particles rather than large slabs of
superconductor.
    \begin{figure}
    \centerline{\includegraphics[width=0.6\figwidth]{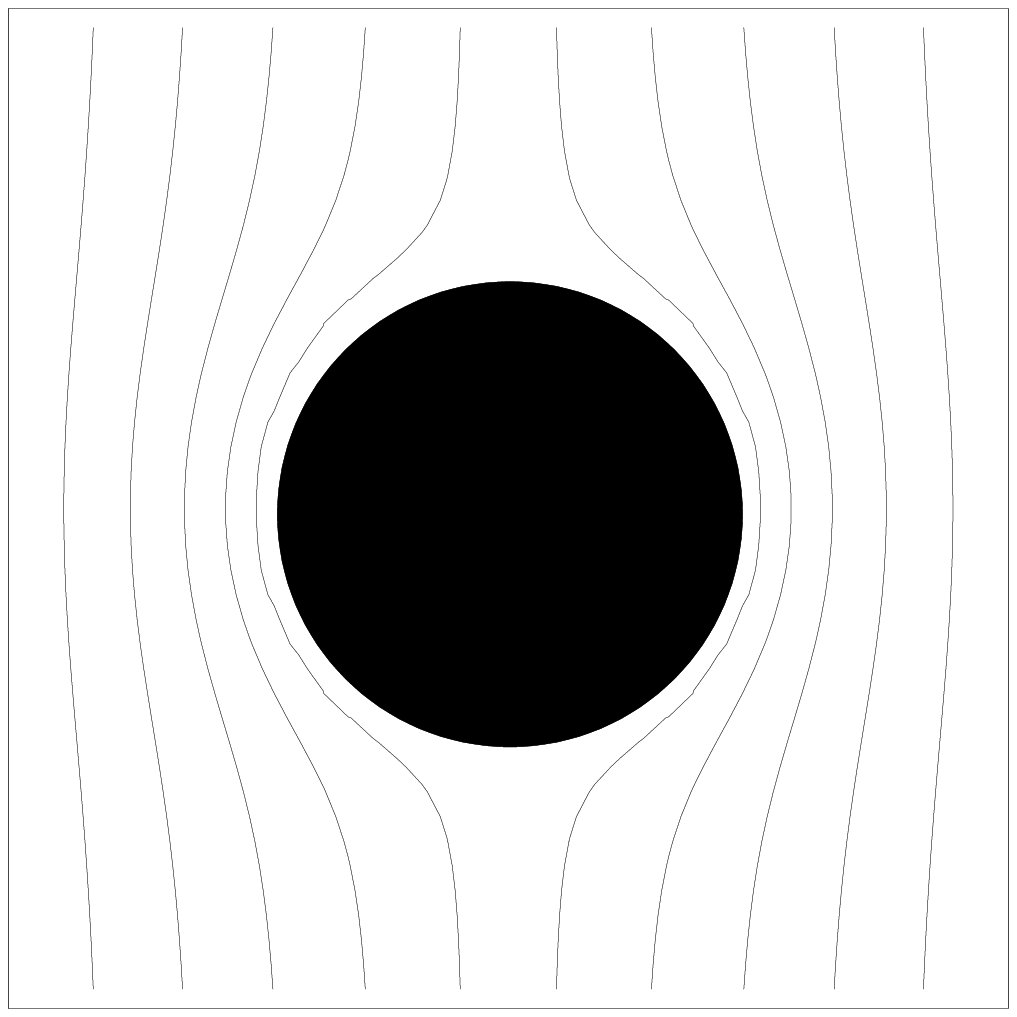}}
    \ncap[Superconducting Sphere in a Magnetic Field]{A
    superconducting sphere expels applied magnetic fields.  The
    increased field at the equator is called the demagnetizing field.
    To order $\lambda/R$, the field at the equator is $3/2$ the
    applied field.\label{sc_sphere}}
    \end{figure}
Larger samples generally show almost no
superheating due to defects.  When early theories
predicted almost no superheating~\cite{fink66}, experiments on bulk
superconductors supported the results~\cite{deblois,boato}. It was
then discovered that it is possible to make colloids of very
uniform superconducting particles ranging in size from
$10$--$100\:\text{$\mu$m}$.  Because the defects are smaller than
the order parameter coherence length, superheating is near ideal.

When calculating the superheating field of a spherical granule,
one needs to account for the demagnetization fields caused by
expulsion of flux from the granule.  For a spherical grain which
is perfectly superconducting, the field at the equator of a grain
is three halves the applied magnetic field.  Ellipsoidal grains
and penetration depths complicate the exact calculation, but the
experiment still rests on the physics of a field penetrating the
surface of a superconductor.

The master of the superheating field is Hugo Parr.  He ascribes
his lineage to Feder and McLachalan~\cite{feder69}, and in 1979,
he published his final paper of a series on the subject
\cite{parr73,parr75,parr76a,parr76b,parr76,pettersen79}.  Using
small grains of various metals, he found nearly ideal superheating
fields in both type-i and type-ii superconducting grains in
suspension.  While he did not use SQUID techniques common to later
studies on granules as detectors, his results showed that the
prediction of the large superheating fields for type-i
superconductors is accurate to within a percent.  He used the
granule system to study the temperature dependence of the
penetration depth and coherence length in detail.  He even
noted the recently rediscovered ``giant vortex state" of
type-ii granules not large enough to contain a single discrete
vortex~\cite{pettersen79,geim97}.

We want to calculate the superheating field, $H_{\text{sh}}$, as a function of
$\kappa$, the \GL parameter.  That parameter is the ratio of the
coherence length of the Cooper pairs, $\xi$, to the magnetic field penetration depth,
$\lambda$.  This is known to within $10\:\%$ for most superconductors.
    \begin{table}
    \centerline{\begin{tabular}{llll}
     & $\lambda\:[\text{\AA}]$ & $\xi\:[\text{\AA}]$ & $\kappa$
     \\ \hline
    Al & 500 & 16000 & 0.03 \\
    Sn & 500 & 2300 & 0.2 \\
    Pb & 400 & 830 & 0.5
    \end{tabular}}
    \ncap[Coherence Lengths and Penetration Depths for Some
    Materials]{These are some sample coherence lengths and
    penetration depths for a few materials~\cite{bardeen}.  The dimensionless
    constant $\kappa=\lambda/\zeta$.  The coherence length and
    penetration depth appropriate to the \GL equations differ
    slightly from the physical quantities.  A good discussion of
    this is in Feder and
    McLachlan~\protect\cite{feder69}.\label{tab:kappas}}
    \end{table}%
A few are shown in Table~\ref{tab:kappas}. We will examine the
most basic system we can, a superconducting half space.

\subsection{A Rough Estimate}

\newcommand{\ns}{\sigma_{\text{ns}}}
\newcommand{\ons}{\overline{\sigma}_{\text{ns}}}

If we focus on the first entry of flux into a superconductor, we
can think of a small nucleus at the edge of the metal.  For a
type-ii superconductor, the flux will enter as vortices typically
much smaller than the penetration depth of the magnetic field.  In
1964, Bean and Livingston estimated the critical magnetic field at
which flux would enter a type-ii superconductor by imagining a
small test flux interacting with the unperturbed
system~\cite{bean64}. The same picture of a small flux line
interacting with a nearly intact superconducting state is still
used to refine magnetization curves~\cite{burlachkov93}. For a
type-i superconductor, we can model the entry of flux as the
nucleus of a phase transition from superconducting to normal. When
the critical radius of a nucleus of normal metal becomes as small
as the typical perturbation of the system, a phase transition will
occur.  Our goal is to determine the transition field as a function of the
\GL parameter, $\kappa$.

\begin{figure}
    \centerline{\includegraphics{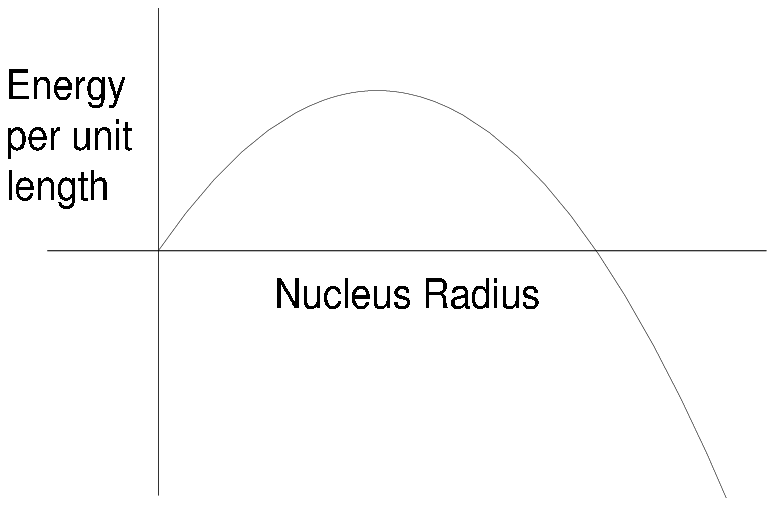}}
    \ncap[Cahn-Hilliard Estimate of Nucleus Growth]{A small
    nucleus of the normal state in a superconductor will have a
    large positive surface tension.  No nucleus can grow unless
    its size is larger than a critical radius where it is
    energetically favorable.\label{fig:cahn}}
\end{figure}
Picture a nucleus of normal metal in a superconducting cylinder.
We can look at a two-dimensional slice of an infinite cylinder so
that the nucleus has an energy per unit length determined by the
competition between volume energy and surface tension.  Surface
energy per unit volume depends on the difference between the
applied magnetic field and the critical field
\begin{equation}
    \frac{E}{\text{unit volume}} = -\frac{H_0^2-H_c^2}{4\pi}.
\end{equation}
The applied field is assumed to be larger than the critical field
for superheating. The energy cost of an interface between normal
and superconducting regions is the surface tension.  We can write
a nondimensional form of the surface tension
\begin{equation}
    \ns = \frac{H_c^2}{4\pi}\ons,
\end{equation}
where Osborn and Dorsey~\cite{osborn94} determined that, for
asymptotically small $\kappa$, $\ons$ is
\begin{equation}
    \ons = \frac{0.943}{\kappa}-\frac{0.880}{\kappa^{1/2}}.
\end{equation}
From the competition between the surface and bulk energies, we can
write the total energy of a nucleus of normal metal
\begin{equation}
    \frac{E}{\text{length}} = -\frac{H_0^2-H_c^2}{4\pi}\pi r^2 +
    \frac{H_c^2\lambda}{4\pi}\ons 2\pi r.
\end{equation}
The radius at which this energy becomes negative is the critical
radius for nucleus growth.

A graph of the energy dependence is shown in Fig.~\ref{fig:cahn}.
Smaller nuclei have a large ratio of surface to volume, so the
positive surface tension makes them shrink. Nuclei large enough
for the negative bulk energy to become dominant have radii larger
than the critical radius
\begin{equation}
    r_c = \frac{2H_c^2\lambda\ons}{H_0^2-H_c^2}.
\end{equation}
We can estimate that a phase transition must occur when the
critical radius of the nucleus is about the same size as the
penetration depth
\begin{equation}
    \frac{2H_c^2\lambda\ons}{H_0^2-H_c^2}\approx \lambda
\end{equation}
Solving this for the applied field where the system must collapse,
we find
\begin{equation}
    H_0^2 = H_c^2(1+2\ons).
\end{equation}
If we insert our asymptotic value for $\ons$, we find
\begin{equation}
    H_0/H_c \approx \frac{1.34}{\kappa^{1/2}}-0.641.
\end{equation}
The factor in front is about twice the value calculated later, but
the order of the first term is correct.  The superheating field of
a type-i superconductor will diverge as $\kappa\rightarrow 0$ with
a power of $\kappa^{-1/2}$ because the positive surface tension
diverges as $\kappa^{-1}$.

Now we look at a more mathematical approach which accounts for
surface currents and magnetic fields using the \GL equations.

\subsection{Formulation of the Problem}

The bare requirements for a superheating problem are an applied
magnetic field and a surface. We will study a superconducting
half-space where all space $x>0$ is filled with metal and all
space $x<0$ is empty with an applied magnetic field $\mathbf{H}$.
This is not a time-dependent problem so we need use only the
time-independent \GL free energy
\begin{equation}
    G = \int d^3x\left\{-|\psi|^2+\frac{1}{2}|\psi|^4+(\del\times\mathbf{A})^2
    -2\del\times\mathbf{A}\cdot\mathbf{H}+\left|\left(\frac{1}{\kappa}\del
    +i\mathbf{A}\right)\psi\right|^2\right\}.
\end{equation}
where $\kappa$ is the GL parameter, $\psi$ is the amplitude of the
superconducting order parameter, $\mathbf{A}$ is the vector
potential ($\mathbf{B}={\del\times \mathbf{A}}$), and $\mathbf{H}$
is the magnetic field applied to the side of the sample.  The
lengths are in units of the penetration depth $\lambda$ and fields
are in units of $\sqrt{2}H_c$. While supercurrents shield the
metal from applied magnetic fields, they are not dissipative so we
don't need to include any terms related to the electric field in
order to model this system.

The nondimensional free energy is invariant under a local linear
transformation of the form
\begin{equation}
    \psi \rightarrow \psi e^{i\kappa\chi}
    \qquad\text{and}\qquad
    \mathbf{A}\rightarrow\mathbf{A}+\del\chi.
\end{equation}
The gauge transformation amounts to an extra degree of freedom. It
is called local because the transformation depends on coordinates,
$\chi=\chi(x)$. While we could, in principle, perform calculations
without specifying a gauge, the easy and common practice is to fix
the gauge.  We fix the gauge by completely specifying a function
$\chi$ in terms of observables. In this case, if one writes $\psi
= fe^{i\theta}$, we can make $\psi$ real by defining
\begin{equation}
    \chi = -\theta/\kappa\qquad\Rightarrow\qquad
    \xi=\psi e^{i\theta} = fe^{i\theta}e^{-i\theta} = f.
\end{equation}
The phase of $\psi$ is eliminated without any loss of generality. The
transformed free energy is
\begin{equation}
    G = \int
    d^3x\left\{\frac{1}{\kappa^2}\del^2 f+\mathbf{Q}^2f^2-f^2+
    \frac{1}{2}f^4+(\del\times\mathbf{Q})^2 -
    2(\del\times\mathbf{Q})\cdot\mathbf{H}\right\}.
\end{equation}
where $f=|\psi|$ is the amplitude of the superconducting order
parameter and $\mathbf{Q}$ is the gauge-invariant vector
potential.  Having transformed the free energy, we take the
functional derivative in order to get the equilibrium equation
\begin{gather}
    \frac{1}{\kappa^2}\del^2f-f\mathbf{Q}^2+f-f^3 = 0 \label{eqn:fullgi1}\\
    \del\times\del\times\mathbf{Q}+f^2\mathbf{Q} = 0
    \label{eqn:fullgi}
\end{gather}
with boundary conditions
\begin{gather}
    \del f\cdot\hat{n} = 0 \label{eqn:gibc1}\\
    (\mathbf{\del\times Q}-\mathbf{H})\times\hat{n} =
    0.\label{eqn:gibc2}
\end{gather}
If we think of a classical mechanical system where $x\rightarrow
t$, these would be equations of motion for a particle in a
potential. Eq.~\ref{eqn:fullgi} describes the currents in the
system in the form
\begin{equation}
    \mathbf{J}_{\text{total}} - \mathbf{J}_{\text{superconducting}} = 0
\end{equation}
because there is no normal current.  Furthermore, the divergence
of Eq.~\ref{eqn:fullgi} demonstrates explicitly that the
supercurrent is zero.

Now we further restrict our treatment to a single dimension by treating
independent variables as constant in directions parallel to the surface.  Since
our sample is a halfspace $x>0$, we apply a magnetic field
parallel to the surface by specifying $\mathbf{H} = h_a\hat{z}$.  We represent
that magnetic field with a vector potential $\mathbf{Q} =
(0,q(x),0)$. All spatial variations are in the $\hat{x}$
direction so that the resulting equations are
\begin{gather}
    \frac{1}{\kappa^2} f'' - q^2 f + f - f^3 = 0,
    \label{GL1} \\
    q'' - f^2 q = 0 ,
    \label{GL2} \\
     h = q'.
    \label{GL3}
\end{gather}
where primes denote derivatives with respect to $x$.
These much simplified equations will suffice us the whole of the derivation of
the superheating until we need to examine its stability.

\subsection{Previous Work}
The \GL equations have been used to study the superheating field
since their inception.  All work done with them has been
approximate because these are coupled nonlinear equations and
intractable for this application.  All approaches focused either
on large or small asymptotics in $\kappa$.  We will look at both,
discussing in greater detail the work on small $\kappa$.

In 1950, Ginzburg and Landau~\cite{landau50} wrote Eqns.~\ref{GL1}
and \ref{GL2} in order to find the effect of a magnetic field
applied parallel to a superconductor.  Assuming the order
parameter varied only a little for a hard superconductor (small
$\kappa$), they wrote $f = f+\tilde{f}$ where $f=1$ was the bulk
solution and $\tilde{f}<0$ showed the depression of superconductivity at the
boundary. Second-order terms, $\tilde{f}^2$ and $q\tilde{f}$,
were negligible.  The resulting equations were
\begin{gather}
    \frac{1}{\kappa^2}\tilde{f}''-2\tilde{f}-q^2 =0\\
    q''=q.
\end{gather}
Solving these is straightforward, and the approximate order
parameter is
\begin{equation}
 f = 1+\frac{\kappa
 H^2}{\sqrt{2}(2-\kappa^2)}\left(\frac{\kappa}{\sqrt{2}} e^{-2x} -
 e^{-\sqrt{2}\kappa x}\right)\label{eqn:patsurf}
\end{equation}
While they did not further investigate the superheating field, the
above equation did establish that the order parameter at the
surface responds to applied magnetic fields according to $\kappa
H^2$.  Note also that our dimensionless units measure distance in
terms of the magnetic penetration depth so that this first
approximation shows $q\propto e^{-x}$.

In 1958, Ginzburg~\cite{ginzburg58} estimated the superheating
field for large and small $\kappa$.  For
$\kappa\rightarrow\infty$, he solved the \GL equations exactly to
find $H_{sh}=1/\sqrt{2}$ (Recall the field is normalized to
$\sqrt{2}H_c$).  He did not examine the stability of that solution
to discover the physical superheating field is actually lower.

For small $\kappa$, however, Ginzburg made fruitful conjectures
from scaling arguments: ``From [Eq.~\ref{eqn:patsurf}], and from
an analysis of [Eqs.~\ref{GL1} and~\ref{GL2}] with the introduction
of variables $\zeta=\sqrt{\kappa}x$, $\chi=f/\sqrt{\kappa}$ and
$b=\sqrt{\kappa} q$, it can be inferred that, for
$\sqrt{\kappa}\ll 1$, $H_{sh}=\text{const}/\sqrt{\kappa}$."  How
do we explain these choices?

We want to isolate and solve the relevant parts of the coupled
nonlinear differential equations.  Here, the relevant parts are
those concerning the magnetic field.  For hard superconductors,
the order parameter will remain nearly constant while the magnetic
field dies off quickly.  We also have the sense that the
superheating field for a hard superconductor is large.  A good
first step would be to rescale the vector potential so that it is
closer to being $\mathcal{O}(1)$.  Given choices like $b = \kappa
q$ or $b=\sqrt{\kappa}q$, one might look at $\kappa H^2$ in
Eqn.~\ref{eqn:patsurf} and choose the latter so that $\kappa H^2$
would become the $\mathcal{O}(1)$ $(\del\times b)^2$ in the new
variables.  There is also a mathematical expression for our
expectation that the order parameter remain steady while the
vector potential varies.  We simply change the length scale of the
problem so that the vector potential looks more like a
perturbation.  With units in terms of the penetration depth, we
have $q\propto e^{-x}$.  We can use units $\zeta=\sqrt{\kappa}x$
so that $\zeta$ remains $\mathcal{O}(1)$ when $x$ is large.  We
now have the following set of equations
\begin{gather}
    \frac{1}{\kappa}f''-\frac{1}{\kappa}b^2f+f-f^3 =
    0\\
    \kappa b''-f^2 b = 0.
\end{gather}
Since we claim it is the behavior of the vector potential that
describes this problem, we want to include the second equation. We
can do that by rescaling $\sqrt{\kappa}\chi=f$ with the effect
\begin{gather}
    \frac{1}{\kappa}\chi''-\frac{1}{\kappa}b^2\chi+\chi-\kappa\chi^3 =
    0\\
    b''-\chi^2 b = 0.\label{eqn:ginz2}
\end{gather}
From here, one can do series solutions for $\chi$ and $b$ in $\kappa$.  The zeroth
order solution is that $\chi=1/\sqrt{\kappa}$.  Just from the zeroth order solution, we can
look at Eqn.~\ref{eqn:ginz2} to find $b=-C
e^{-\zeta/\kappa}=-Ce^{-x}$ so that,
$q=b/\sqrt{\kappa}=-(C/\sqrt{\kappa})e^{-x}$.  Our rescalings
along the way sought to ensure that all proportionality constants
were $\mathcal{O}(1)$.  We conclude, therefore, that
$H=\text{const}/\sqrt{\kappa}$.  Ginzburg determined that
constant numerically to find
\begin{equation}
    H_{sh}=\frac{0.89}{\sqrt{\kappa}}H_c.
\end{equation}

The Orsay Group on Superconductivity~\cite{orsay} (including de~Gennes),
in 1966, took a similar physical
approach with a different mathematical technique.  Rather than
look at scaling, they used a variational argument.  The basis is
that the magnetic field goes rapidly to zero while the order
parameter takes its time.  The order parameter is calculated in
the absence of the applied field
\begin{equation}
    \frac{1}{\kappa^2}f''=f(f^2-1).
\end{equation}
The vector potential is calculated for a fixed order parameter so
that
\begin{equation}
    q = q(0)e^{-f(0) x}.
\end{equation}
Then they put these estimates into the free energy and
concentrate, as did Ginzburg, on the terms involving $q$, $q'$,
and $f'$.  Terms in $f$ are dropped.  They vary the free energy
with respect to $f(0)$ to find what $f(0)$ gives a free energy
minimum.  The result is a nontrivial estimation of the applied
field in terms of $f(0)$
\begin{equation}
    H^2 =
    \frac{2\sqrt{2}}{\kappa}f^2(0)(1-f^2(0)).\label{eqn:ginzh}
\end{equation}
The applied field always reaches its maximum for $f(0)=1/\sqrt{2}$
where
\begin{equation}
    H_{sh} = 2^{-1/4}\kappa^{-1/2}H_c.
\end{equation}
This exact analytical coefficient is the same as that derived
later in this chapter, but the eponymous Ginzburg would cite our
paper from 1996~\cite{dolgert96} in his 1998 work on the
thermoelectric effect~\cite{ginzburg98}. We can see in
Eqn.~\ref{eqn:ginzh} (plotted in Fig.~\ref{nose} on
page~\pageref{nose}) that the superheating field occurs where
$\pad{H}{f(0)}=0$.  It is also apparent that the order parameter
at the surface is nonzero at the maximum applied field.

More recently, Hugo Parr~\cite{parr76} combined intuitive
rescalings and a variational approach to derive the next term in
the superheating field.
\begin{equation}
    H_{\text{sh}} =
    2^{-1/4}\kappa^{-1/2}\left(1+\frac{15\sqrt{2}}{32}\kappa\right)H_c
\end{equation}
The techniques are not significantly different from those above.
The reasons for his rescaling are not clear and the calculation
for the second term is very complicated. His work is important
because the addition of the second term brings the asymptotics
quite close to the full solution of the \GL equations as shown in
Fig.~\ref{superplot} on page~\pageref{superplot}.  This permitted
him to make practical verifications of the estimated superheating
field.

Our contribution to the examination of the superheating field is
two-fold.  First, we determine the superheating field to greater
precision than previous work.  Second, and more importantly, we
use a method called matched asymptotics to do that derivation with
surety.  While matched asymptotics are informed by the physical
considerations discussed above---short penetration depth, slow
variation of order parameter---the method transforms the problem
into linear series solutions which can be integrated mechanically.

\subsection{Boundary Layer Method for a Converging Channel}
The boundary layer method is a standard technique of singular
perturbation theory, but it is much less well known than WKB, so
we will review it here.  An excellent book on the subject is
Bender and Orszag~\cite{bender}.  For physicists, a more amusing
introduction might be a simpler problem from fluid dynamics.

Landau and Lifshitz~\cite{landau59} derive the exact solution of
the equations of motion for a viscous, incompressible solution
flowing into a converging channel, but they don't find a closed
form solution.  We follow their brief derivation enough to make
sense of the main differential equation, then use the boundary
layer method to find a closed form solution for low viscosity,
high Reynolds number fluids.  This is a system with a single small
length scale, identified by the Reynolds number. It should be good
practice for the \GL equations which exhibit two or more length
scales in multiple coupled nonlinear equations.

\begin{figure}
\centerline{\includegraphics[width=0.5\figwidth]{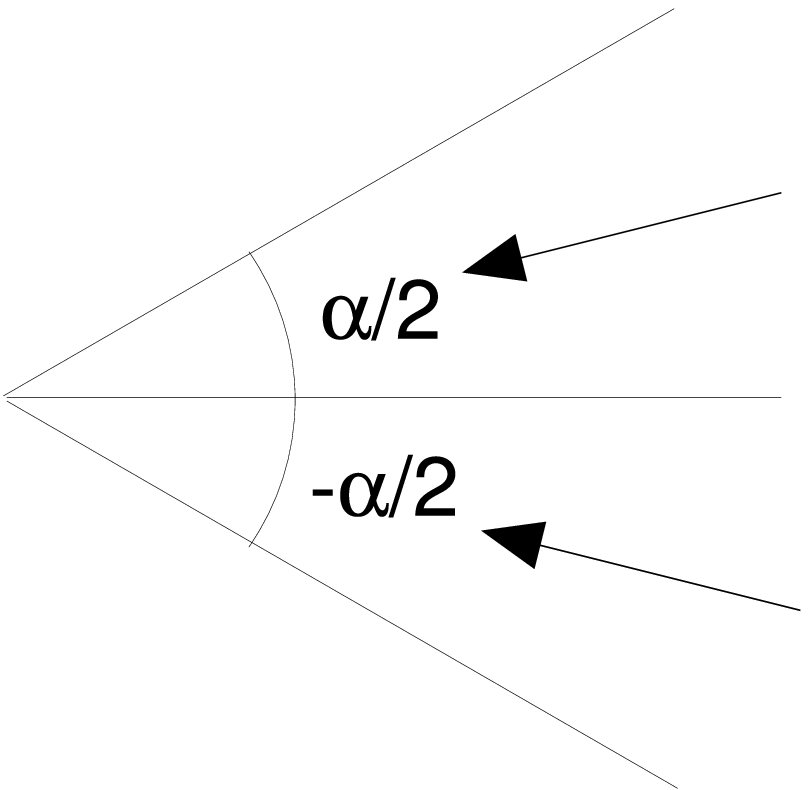}}
\ncap[Converging Channels]{Fluid flows into a converging channel
and out a pinhole at the vertex.\label{channels}}
\end{figure}
The system geometry is shown in Fig.~\ref{channels}.  The fluid
flows into the channel and out a pinhole.  We assume the flow is
entirely radial.  We could determine the equation of motion for an
inviscid liquid just from mass conservation.  Given a mass $Q$ of
liquid per unit time,
\begin{equation}
    Q = \rho\int_{-\alpha/2}^{\alpha/2}vr\:d\phi.\label{qmassconv}
\end{equation}
For an inviscid liquid, the solution would be
\begin{equation}
    Q = \frac{\alpha}{2\pi}\rho v r\qquad \Rightarrow \qquad v =
    \frac{2\pi Q}{\rho\alpha r}.\label{maxv}
\end{equation}
\begin{figure}
\centerline{\includegraphics{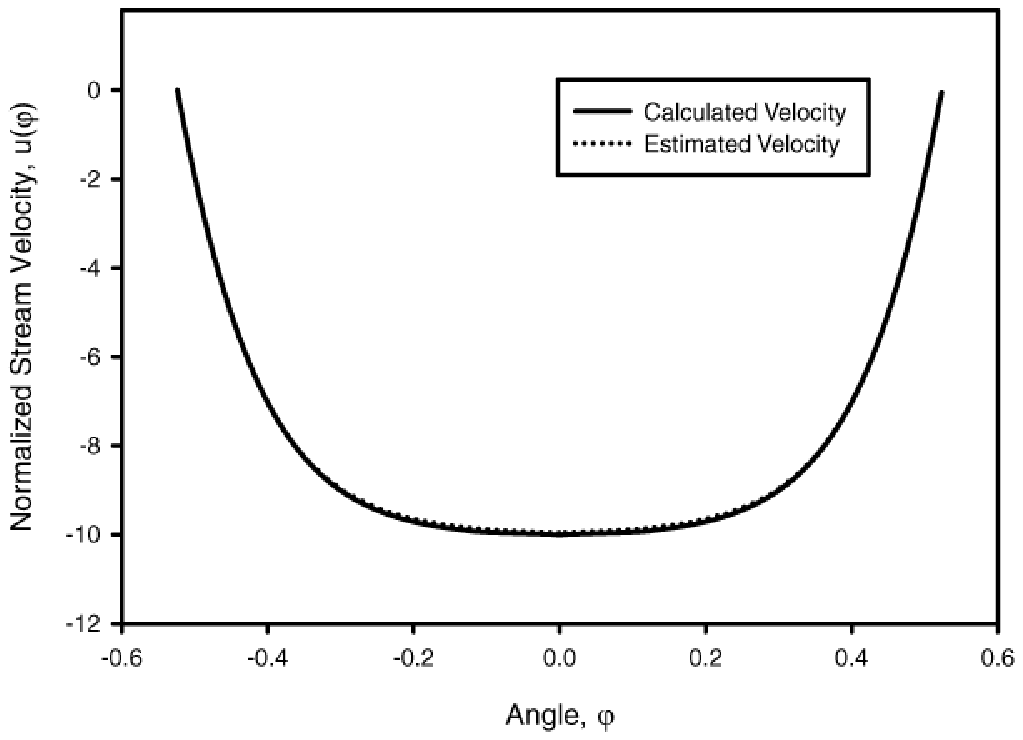}}
\ncap[Stream Velocity for Re=43]{Fluid flow is uniform in the
center of the channel.  It drops to zero only near the sides.  The
region affected by viscosity is $\mathcal{O}(1/\sqrt{\re})$.  The
estimated stream velocity for a large Reynolds number, $\re=43$,
is a very good approximation of the exact
solution.\label{channel43}}
\end{figure}
In order to include viscosity, we will use radial coordinates
where the Navier-Stokes equations are
\begin{gather}
    v\pad{v}{r}=-\frac{1}{\rho}\pad{p}{r}
        +\nu\left(\pad{^2v}{r^2}+\frac{1}{r^2}\pad{^2v}{\phi^2}+
        \frac{1}{r}\pad{v}{r}-\frac{v}{r^2}\right) \label{ns1}\\
    -\frac{1}{\rho r}\pad{p}{\phi}+\frac{2\nu}{r^2}\pad{v}{\phi} =
    0 \label{ns2}\\
    \pad{(rv)}{r} = 0.\label{ns3}
\end{gather}
These equations are still supplemented by the mass conservation
equation, Eq.~\ref{qmassconv}.  Because of the last equation,
Eq.~\ref{ns3}, we know $v\propto 1/r$, so we change variables to
\begin{equation}
    u(\phi) = \frac{rv}{6\nu}.
\end{equation}
Substituting $u$ in Eq.~\ref{ns2}, we can integrate to find the
pressure
\begin{equation}
    \frac{1}{\rho}(p-p_{\text{wall}}) = \frac{12\nu^2}{r^2}u.
\end{equation}
Substituting the pressure into Eq.~\ref{ns1}, we find
\begin{equation}
    \pad{^2u}{\phi^2}+4u+6u^2=\frac{r^3}{6\rho\nu}\pad{p_{\text{wall}}}{r}.
\end{equation}
Noticing that the left-hand side depends only on $\phi$ and the
right-hand side only on $r$, we assign each to a constant, $2c$,
and arrive at the main differential equation for this system
\begin{equation}
    u''+4u+6u^2 = 2c.\label{chanmain}
\end{equation}
Here, $u$ depends only on $\phi$ and $c$ is a constant of
integration.

Landau and Lifshitz express Eq.~\ref{chanmain} in integral form
and discuss the nature of the solutions.  We will examine the case
where the viscosity, $\nu$, is small, or Reynolds number,
$\re=|Q|/\rho\nu$, is large. Here we expect that, through most of
the channel, the liquid will move with nearly the same
velocity as the inviscid case, Eq.~\ref{maxv}.  Near the sides,
however, the velocity at the walls must drop to zero.  The
affected region in the channel will be
$\mathcal{O}(1/\sqrt{\re})$. The affected region is our boundary
layer. We can rewrite the differential equation in different
variables to see why there is a boundary layer.

First, note that the constant, $c$ is $\mathcal{O}(\re^2)$. This
could be proven by estimating $p_{\text{wall}}$ from Bernoulli's
Equation. We also know from Eq.~\ref{maxv} that the maximum $u$ is
less than
\begin{equation}
    u_{m} = \frac{Q\pi}{3\nu\rho\alpha r} = -\re\frac{\pi}{3\alpha
    r}.
\end{equation}
That tells us that $u$ is $\mathcal{O}(\re)$.  Let's first write
the main differential equation in terms of $\mathcal{O}(1)$
variables.  Substitute $U = u/\re$ and $C = c/\re^2$ to find
\begin{equation}
    \frac{1}{\re}U''+\frac{4}{\re} U+6 U^2 = 2 C.
\end{equation}
It is customary to write the small parameter as $\epsilon = 1/\re$
\begin{equation}
    \epsilon U''+4 \epsilon u+6 U^2 = 2 C.
\end{equation}
While this equation applies in the center of the channel, it is
our \emph{outer equation} because it describes the region far from
the boundary layer. For flows near the center, $U''$ is small so
we don't expect it to be important. We will need to rescale in
order to look at flows near the side. We can use a simple series
solution to solve the outer equation.  Substituting the series
\begin{align}
    U &= U_0+\epsilon U_1+\epsilon^2 U_2+\dots \\
    C &= C_0+\epsilon C_1+\epsilon^2 C_2+\dots.
\end{align}
we find
\begin{equation}
    6U_0^2 = 2C_0.
\end{equation}
The solution is $U_0 = \sqrt{C_0/3}$ in the center. We can write
down the next order differential equation, too
\begin{equation}
    U_0''+4U_0+12U_0U_1 = 2C_1 \qquad\Rightarrow\qquad
    U_1 = \frac{C_1}{6U_0}-\frac{1}{3}.
\end{equation}
We determine constants, $C_0$ and $C_1$ from matching the solution
near the boundary.  A nice side effect of the series solutions is
that each succeeding order is always a linear differential
equation.

Near the boundary, we expect the second derivative of $U$ to be
important.  We express this by rescaling the $\phi$-coordinate,
$\Phi=\phi/\sqrt{\epsilon}$.
\begin{equation}
    \pad{^2U}{\Phi^2}+4\epsilon U+6U^2 = C.
\end{equation}
We have the option of rescaling $U$, in addition.  While we expect
$U$ is smaller near the boundary, rescaling $U \rightarrow
\sqrt{\epsilon}U$ would give us $U''=C$ which could not fulfill
boundary conditions. We usually seek rescalings where more than
one term is of the highest order of $\epsilon$ because you usually
need two or more terms to find a non-trivial solution. Scaling
variables such that nontrivial solutions exist is called the
\emph{principle of dominant balance.\/}  Currently $U''$ is
balanced against $6U^2-C$ which seems like it could fulfill
boundary conditions non-trivially. If we substitute series into
this equation, we find
\begin{equation}
    U = U_{m0}(-2+3\tanh^2(\pm\sqrt{-3 U_0}(\Phi-k)).
\end{equation}
The integration constant, $k$, is determined from matching to the
outer solution.

Now that we have an inner and outer solution, we match them.  We
do that by writing the inner solution in terms of the outer
variables, $\Phi = \phi/\sqrt{\epsilon}$.
\begin{equation}
    U = U_{m0}(-2+3\tanh^2(\pm\sqrt{-3
    U_0}(\epsilon^{-1/2}\phi-k)).
\end{equation}
Now we define the constant $k$ such that $U(-\alpha/2)=0$.  The
inner equation is already set to match $U_{\text{outer}}=U_{m0}$.
We therefore see the real size of our boundary layer is
$1/\sqrt{3|U_0|} = 1/\sqrt{3|u_0|\re}$.

\begin{figure}
\centerline{\includegraphics{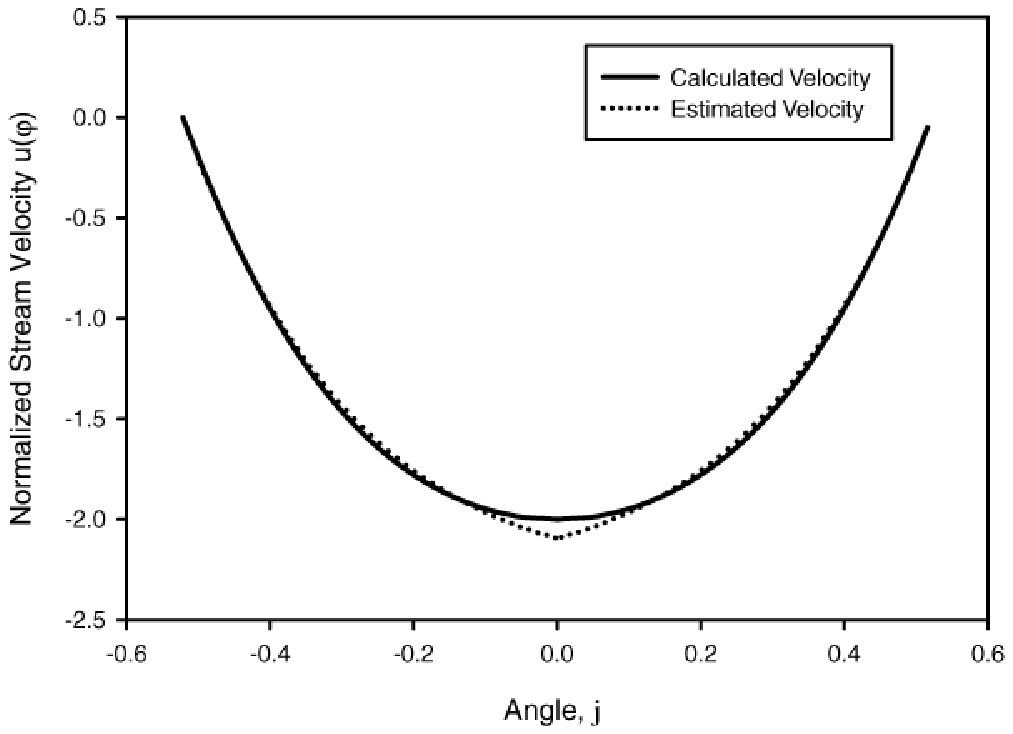}}
\ncap[Stream Velocity for Re=9]{The estimated stream velocity
for moderately large Reynolds number, $\re=9$, is a fair
approximation of the exact solution.\label{channel9}}
\end{figure}
The solution for the boundary layer meets the boundary condition,
$u=0$, at the side of the channel and meets the solution to the
inner equation as $\phi\rightarrow\infty$.  The outer solution
would not, by itself, be capable of fulfilling the boundary
conditions at the walls, but it does match to the inner solution
near the walls.  The solutions of the superconducting half-space
will have the same property.  Inner and outer solutions together
form a uniform solution to the system.

\section{Asymptotic Expansions for Small-$\kappa$}

In this section we will develop an asymptotic expansion for the
superheating field for small-$\kappa$, using the method of matched
asymptotic expansions~\cite{bender,vandyke}. For small-$\kappa$
the longer length scale is the coherence length $\xi$.  In order
to treat the problem with singular perturbation theory, we need to
write it such that the factor in front of the highest order
derivative is small.  Second order equations with a small
parameter modifying the highest derivative tend to exhibit rapid
behavior near the boundary.  In this case, the rapid behavior is
the penetration of the magnetic field.  That field is the boundary
of boundary layer theory, which we use here.

Rescaling $x$ by $\kappa$, we introduce a new dimensionless
coordinate $x'=\kappa x$ which means we now measure distance using
the coherence length. The resulting GL equations in these ``outer
variables'' are
\begin{gather}
    f'' - q^2 f + f - f^3 = 0, \label{GL4} \\ \kappa^2 q'' - f^2 q = 0, \label{GL5} \\
    h = \kappa q',
    \label{GL6}
\end{gather}
with the primes now denoting differentiations with respect to $x'$.

{\it Outer solution.}  In order to obtain the outer solutions expand
$f$, $q$, and $h$ in powers of $\kappa$:
\begin{gather}
f = f_0 + \kappa f_1 + \kappa^2 f_2 + \dots, \label{expand1}  \\ q
= q_0 + \kappa q_1 + \kappa^2 q_2 + \dots, \label{expand2}  \\ h =
h_0 + \kappa h_1 + \kappa^2 h_2 + \dots. \label{expand3}
\end{gather}
Substituting into Eqs.~(\ref{GL4})-(\ref{GL6}), at $O(1)$ we have
\begin{gather}
f_0'' - q_0^2 f_0 + f_0 - f_0^3=0, \label{outer1} \\
 - f_0^2 q_0 = 0. \label{outer2}
\end{gather}
Since we want $f\rightarrow 1$ as $x'\rightarrow \infty$, the only possible
solution to Eq.~(\ref{outer2}) is $q_0 = 0$.  We can then immediately
integrate Eq.~(\ref{outer1}),
    \begin{equation}
    f_0(x') = \tanh \left( {x' + x_0}\over \sqrt{2} \right),
    \label{outer3}
    \end{equation}
with $x_0=x_0(\kappa)$ an arbitrary constant.\footnote{Because
$x_0$ depends on $\kappa$, it should also be expanded in series.
For the orders of terms used in the following derivation, the
differences are negligible.  A letter from the Bolley \emph{et
al.\/}~\cite{bolley97} pointed out the difficulty which was
painstakingly recalculated and corrected by John Di~Bartolo in an
erratum~\cite{dolgert98}.  The affected terms are of high enough
order that they show up here only in Table~\ref{tab:shconstants}
and Eq.~\ref{pade}.} To $O(\kappa)$, the outer equations are
    \begin{gather}
    f_1'' - 2q_0f_0q_1 - q_0^2f_1 + f_1 - 3f_0^2 f_1 = 0,
    \label{outer4} \\
    - f_0^2 q_1 - 2 f_0q_0 f_1 = 0,
    \label{outer5} \\
    h_0 = 0. \label{outer6}
    \end{gather}
Once again, the only solution to Eq.~(\ref{outer5}) is $q_1 = 0$;
substituting this into Eq.~(\ref{outer4}), we find $f_1 =C_1 f_0'$,
with $C_1$ a constant:
    \begin{equation}
    f_1 = {C_1\over \sqrt{2}} {\rm sech}^2\left( { x'+x_0 \over \sqrt{2}} \right).
    \label{outer7}
    \end{equation}
We can continue in this manner; at every order $q_n= 0$, $h_n=0$,
and $f_n = C_n f^{(n)}_0$, with the $C_n$'s constants which are determined
by matching onto the inner solution.

{\it Inner solution.}  The outer solution breaks down within a boundary
layer of $O(\kappa)$ near the surface.  This suggests introducing a rescaled
inner coordinate $X=x'/\kappa$, so that $X=O(1)$ within the boundary layer.
It is also possible to rescale $f$ and $q$, with the hope that this
will lead to a tractable inner problem.  Such a rescaling must lead to
a successful matching of the inner and outer solutions; i.e., the inner
solutions as $X\rightarrow \infty$ must match onto the outer solutions as
$x'\rightarrow 0$.  Since $f_0(0) = \tanh (x_0/\sqrt{2})$, then assuming
that $x_0\neq0$ we have $f_0(0)= O(1)$, indicating
that the order parameter should not be rescaled in the inner region;
therefore we set $f(x') = F(X)$  in the inner region. However, from the
outer solution for the vector potential we see that the only constraint
on $q(X)$ in the inner region is that $q(X)\rightarrow 0 $ as
$X\rightarrow\infty$ (presumably exponentially).  Therefore, we
are free to rescale $q$ by $\kappa$ in the inner region, hopefully in
a way which simplifies the inner equations.  One possibility is
$q(x')=\kappa^{-\alpha} Q(X)$; substituting this into the GL equations,
Eqs.~(\ref{GL4})-(\ref{GL6}), we see that unless $2\alpha$ is an
integer, fractional powers of $\kappa$ will be introduced into the inner
equations, contradicting our expansion of $f$ and $q$ in integer powers of
$\kappa$ in the outer region.  Therefore, the simplest assumption is that
$\alpha=1/2$, leading to the following choice for the inner variables:
\begin{equation}
x'= \kappa X,\quad f(x') = F(X),\quad q(x') = \kappa^{-1/2} Q(X), \quad
h(x') = H(X).
\label{inner1}
\end{equation}
In these variables Eqs.~(\ref{GL4})-(\ref{GL6}) become
\begin{gather}
F'' - \kappa Q^2 F + \kappa^2 (F - F^3) = 0, \label{inner2} \\ Q''
- F^2 Q = 0, \label{inner3} \\
\kappa^{1/2} H =  Q', \label{inner4}
\end{gather}
where now the primes denote differentiation with respect to $X$.
The boundary conditions are
\begin{equation}
F'(0)=0, \qquad H(0) = H_a.
\label{inner4a}
\end{equation}

The next step is to expand the inner solutions in powers of $\kappa$:
\begin{gather}
F = F_0 + \kappa F_1 + \kappa^2 F_2 + \ldots, \label{inner5} \\ Q=
Q_0 + \kappa Q_1 + \kappa^2 Q_2 + \ldots, \label{inner6} \\
H =
\kappa^{-1/2} H_0 +  \kappa^{1/2} H_1 + \ldots. \label{inner7}
\end{gather}
Note that there is no term of $O(1)$ in the expansion for $H$, since
we would be unable to match such a term to the outer solution.
Using the boundary condition $H(0)=H_a$ leads to
\begin{equation}
H_a = \kappa^{-1/2} H_0(0) + \kappa^{1/2} H_1(0) + \ \ldots.
\label{inner7a}
\end{equation}
Substituting these expansions into Eqs.~(\ref{inner2})-(\ref{inner4}),
at $O(1)$ we obtain
\begin{equation}
F_0'' = 0, \qquad Q_0'' - F_0^2 Q_0 = 0, \qquad H_0 = Q_0'.
\label{inner9}
\end{equation}
Solving these equations subject to the boundary conditions (\ref{inner4a})
(we also need $Q_0\rightarrow 0 $ as $x\rightarrow \infty$ in order to
match onto the outer solution), we obtain
\begin{equation}
F_{0}(X)   = A_0, \qquad Q_0(X) = B_0 e^{-A_0 X},
                  \qquad H_0(0) = -A_0 B_0,
\label{inner10}
\end{equation}
with $A_0$ and $B_0$ constants.  In what follows we will assign $F_n(0) = A_n$
and $Q_n(0)=B_n$ for notational simplicity.  The $O(\kappa)$ equations are
\begin{equation}
F_1'' = Q_0^2 F_0, \qquad Q_1'' - F_0^2 Q_1 = 2 F_0 Q_0 F_1,
\qquad H_1 = Q_1'.
\label{inner11}
\end{equation}
Solving with the boundary condition $F_{1}'(0)=0$, we obtain
\begin{gather}
F_1(X) = A_1 + { B_0^2 \over 4 A_0}\left[ 2 A_0 X +   e^{-2 A_0 X}
                       - 1\right],
\label{inner12}\\
 Q_1(X) = e^{-A_0 X} \left\{ B_1 - {B_0^3\over 16 A_0^2} \biggl[
            1 - e^{-2A_0 X} \right.  \nonumber \\
      \qquad  \left. + 16 {A_0^2 A_1 \over B_0^2} X
          + 4 A_0^2 X^2 \biggr] \right\},
\label{inner13}\\
  H_1(0)= -{1\over 8} {B_0^3 \over A_0} - A_0B_1 - A_1 B_0.
\label{inner13a}
\end{gather}
Finally, to $O(\kappa^2)$ we have for $F_2$
\begin{equation}
F_2'' = - F_0 + F_0^3 + 2Q_0Q_1F_0 + Q_0^2F_1,
\label{next1}
\end{equation}
the solution of which (with $F_2'(0) = 0$) is
\begin{eqnarray}
F_2(X) &=& {17 \over 128}{B_0^4\over A_0^3} + {1\over 4} {B_0^2 A_1 \over A_0^2}
      - {1\over 2} {B_0 B_1 \over A_0} + A_2  + \left( B_0B_1 - {3\over 32}
      {B_0^4 \over A_0^2} \right) X - {1\over 2} A_0(1-A_0^2) X^2 \nonumber \\
 & & \quad + \left[ {1\over2} {B_0 B_1 \over A_0}
  - {1\over 4}{B_0^2 A_1\over A_0^2}- {5\over 32}{B_0^4\over A_0^3}
  - \left( {1\over 8}{B_0^4\over A_0^2} +{1\over 2} {B_0^2 A_1\over A_0}\right)X
     - {1\over 8} {B_0^4\over A_0} X^2\right] e^{-2A_0 X} \nonumber \\
 & & \qquad + {3\over 128} {B_0^4 \over A_0^3} e^{-4A_0 X}.
\label{inner14}
\end{eqnarray}
The expression for $Q_2$ is even more unwieldy, and is not needed
in what follows.

{\it Matching.}  To determine the various integration constants
which have been introduced we must match the inner solution to the
outer solution. Since the outer solution for $q$ is simply $q=0$,
and all of our inner solutions decay exponentially for large $X$,
the matching is automatically satisfied for $q$, as well as for
the magnetic field $h$. To match the inner and outer solutions for
the order parameter, we are guided by the \emph{van Dyke matching
principle\/}~\cite{vandyke}, which states that the $m$ term inner
expansion of the $n$ term outer solution should match onto the $n$
term outer expansion of the $m$ term inner solution.  While the
matching principle works for any pair, $(m,n)$, experience shows
which choices yield meaningful results at a particular order of
the computation.  In our case we will take $m=3$ and $n=2$.
Therefore, write the two term outer solution $f_0(x') + \kappa
f_1(x')$ in terms of the inner variable $X$, and expand for small
$\kappa$, keeping the first three terms in the expansion in powers
of $\kappa$:
\begin{eqnarray}
f_0(\kappa X) + \kappa f_1(\kappa X) & \sim&
       \tanh \left({x_0\over \sqrt{2}}\right)
        + \kappa\,  {\rm sech}^2 \left({x_0 \over \sqrt{2}}\right)
              {1\over \sqrt{2}}\left[ C_1 + X \right] \nonumber \\
  & & \qquad  + \kappa^2 {\rm sech}^2 \left({x_0 \over \sqrt{2}}\right)
      \tanh \left({x_0\over \sqrt{2}}\right)\left[- C_1 X
         - {X^2\over 2}\right].
\label{match1}
\end{eqnarray}
Next, write the three term inner solution
$F_0(X) + \kappa F_1(X)+\kappa^2 F_2(X)$ in terms
of the outer variable $x'$, and expand for small $\kappa$, this time keeping
the first two terms of the expansion:
\begin{eqnarray}
F_0(x'/\kappa) + \kappa F_1(x'/\kappa) + \kappa^2 F_2(x'/\kappa) & \sim&
  A_0 + {B_0^2 \over 2} x'  - {1\over 2} A_0(1-A_0^2) x^{'2} \nonumber \\
& & \qquad+ \kappa \left[ A_1-{B_0^2 \over 4 A_0} + \left( B_0 B_1 - {3\over 32}
       {B_0^4 \over A_0^2}\right)x'  \right].
\label{match2}
\end{eqnarray}
By writing both expressions in terms of $x'$, and equating the various
coefficients of $x'$ and $\kappa$, we see that the expansions do indeed
match if we choose
\begin{gather}
A_0 = \tanh\left({x_0\over \sqrt{2}}\right), \label{match3} \\
\qquad B_0 =- 2^{1/4} {\rm sech}\left({x_0 \over \sqrt{2}}\right)
            = -2^{1/4} ( 1- A_0^2)^{1/2},
\label{match4} \\
A_1 = {B_0^2 \over 4A_0} + {\rm sech}^2\left({x_0
\over \sqrt{2}}\right)
           {C_1 \over \sqrt{2}}
     = {\sqrt{2}\over 4} {1 - A_0^2 \over A_0} + (1-A_0^2) {C_1\over \sqrt{2}},
\label{match5} \\
B_1 = {3\over 32} {B_0^3 \over A_0^2} - {\sqrt{2}
A_0(1-A_0^2) \over B_0}
                 {C_1\over \sqrt{2}}.
\label{match6}
\end{gather}
Eliminating $C_1$,
\begin{equation}
B_1 = -{\sqrt{2} A_0 A_1 \over B_0} + {3\over 32} {B_0^3\over A_0^2}
      + {1\over 2} {1-A_0^2 \over B_0}.
\label{match7}
\end{equation}
Substituting into our expressions for $H_0(0)$ and $H_1(0)$ from
Eqs.~(\ref{inner10}) and (\ref{inner13a}), we obtain
\begin{gather}
 H_0(0) = 2^{1/4} A_0 ( 1- A_0^2)^{1/2},
\label{match8} \\ H_1(0) = {2^{3/4} \over 64} {(2A_0^2
+14)(1-A_0^2)^{1/2} \over A_0}
        - {2^{1/4}(2A_0^2 -1) \over (1-A_0^2)^{1/2}} A_1.
\label{match9}
\end{gather}
In order to calculate the superheating field (or, more correctly, the
\emph{maximum\/} superheating field), we need to maximize $H_0(0)$ and
$H_1(0)$ with respect to $A_0$ and $A_1$.
Maximizing $H_0(0)$ with respect to $A_0$, we find that the maximum occurs at
$A_0^* = 1/\sqrt{2}$, $B_0^* = -2^{-1/4}$, so that $H_0 (0) = 2^{-3/4}$.
Substituting this
result into $H_1(0)$, we find the surprising result that the coefficient
of $A_1$ is zero, and $H_1(0) = 2^{3/4} 15/64$.
Our superheating field is then
\begin{equation}
H_{\rm sh} = 2^{-3/4}\kappa^{-1/2} \left[ 1 + {15 \sqrt{2} \over 32} \kappa
            + O(\kappa^2) \right].
\label{final1}
\end{equation}

\begin{table}
    \centerline{\begin{tabular}{ccccc}
    $n$ & $A_n$ & $B_n$ & $x_n$ & $H_n(0)$ \\ \hline
    0 & $2^{-1/2}$ & $-2^{-1/4}$ & $2^{1/2}\tanh^{-1}(2^{-1/2})$ &
        $2^{-3/4}$ \\
    1 & $-7/32$ & $-(9/16)2^{1/4}$ & $-(15/16)2^{1/2}$ &
        $(15/64)2^{3/4}$ \\
    2 & $(395/2048)2^{1/2}$ & $(147/512)$ & $429/512$ &
        $-(325/2048)2^{1/4}$ \\
    3 & ? & ? & ? & $(14191/65539)2^{3/4}$ \\
    4 & ? & ? & ? & $-(67453267/62914560)2^{1/4}$
    \end{tabular}}
    \ncap[Integration Constants for Small-$\kappa$
    Asymptotics]{The lower order integration constants are shown
    in the text.  Higher orders required symbolic mathematics
    software.  Miraculous cancellations allowed derivation of the
    first six terms of the superheating field without knowledge of
    most third- and fourth-order constants.\label{tab:shconstants}}
\end{table}

\begin{figure}
    \centerline{\includegraphics[width=\figwidth]{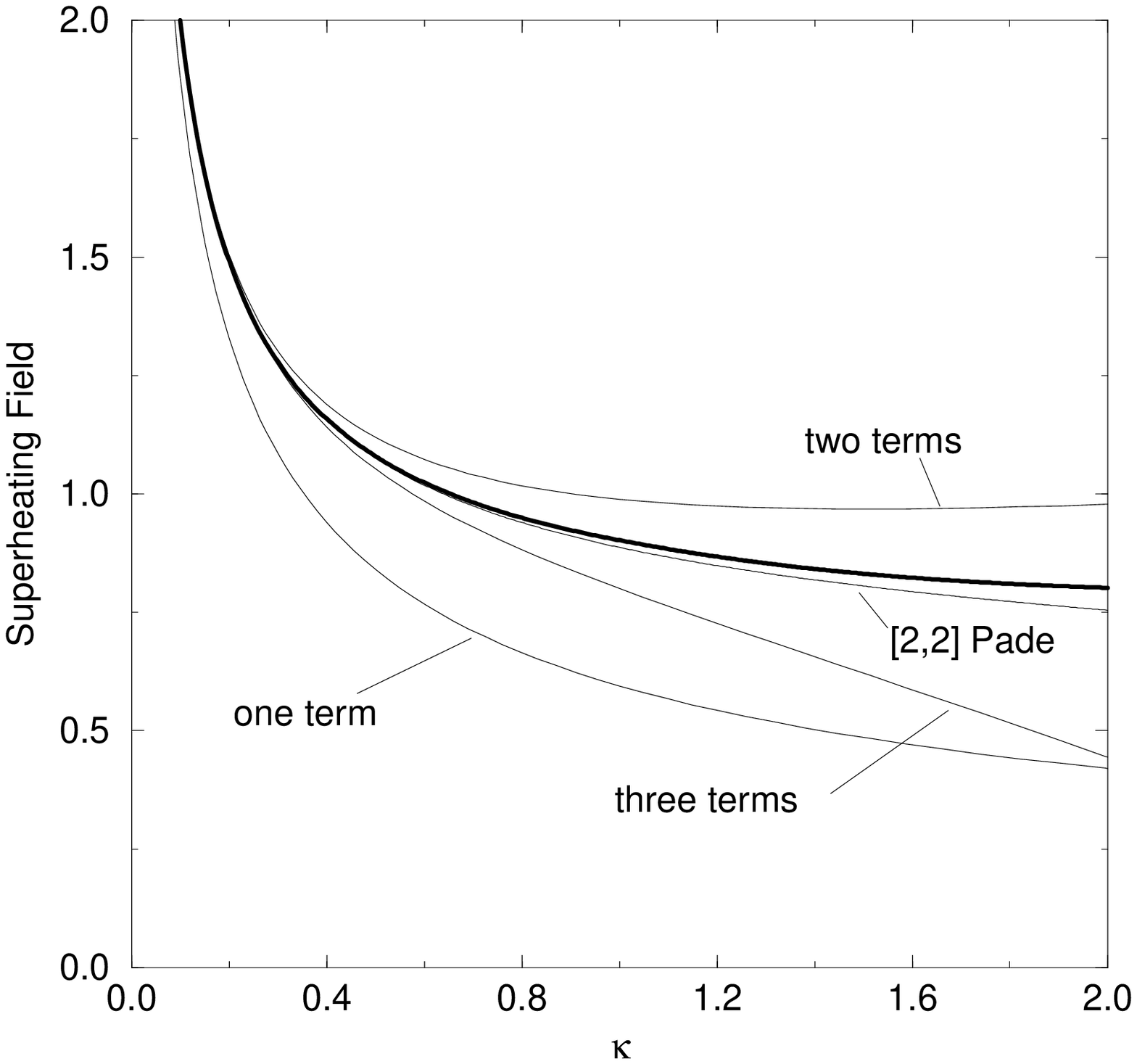}}
    \ncap[Comparison of Superheating Field with Expansions]{\label{superplot}A comparison of the numerically
    calculated superheating field $H_{\rm sh}$ (heavy line) with the
    three term asymptotic expansion for small-$\kappa$, and the
    $[2,2]$ Pad\'e approximant.  The one-term expansion due to the
    Orsay group deviates systematically from the calculated
    superheating field.  The two- and three-term expansions provide a
    marked improvement over the one-term expansion.}
    \end{figure}

In order to determine $A_1$ we need to proceed to a higher order
calculation.  The method is the same as before, although the
algebra quickly becomes tedious; we have used the computer algebra
systems \emph{Maple~V\/} and \emph{Mathematica\/} to organize the
expansion. The results from a six term inner expansion are
summarized in Table~\ref{tab:shconstants}.  Including the next
order term in the expansion in the superheating field, we have
\begin{equation}
H_{\rm sh} = 2^{-3/4}\kappa^{-1/2} \left[ 1 + {15 \sqrt{2} \over 32} \kappa
             - {325\over 1024} \kappa^2 + O(\kappa^3) \right].
\label{final2}
\end{equation}
The first term is exactly the result obtained by the Orsay
group~\cite{ginzburg58,orsay}, who used a variational argument to
obtain their result. The second term is identical to the result
obtained by Parr~\cite{parr76}.  Parr combined an inspired guess
for the behavior of the order parameter near the surface with a
variational calculation in order to obtain his result.  It is
interesting to note that our result for $A_1$ also agrees with
Parr's result.  The advantage of the method of matched asymptotic
expansions is that we can make this expansion systematic, and
therefore in principle carry out this expansion as far as we wish.
The third term in Eq.~(\ref{final2}) is one of the new results of
this chapter; the fourth and fifth terms are included in
Table~\ref{tab:shconstants}.  With the five-term expansion for $H_{\text{sh}}$
it is possible to employ resummation techniques to improve the
expansion.  For instance, the $[2,2]$ Pad\'e
approximant~\cite{bender} is
\begin{equation}
    H_{\text{sh}}^{\text{Pad{\'e}}} =  2^{-3/4}\kappa^{-1/2}
  { 1 + 4.6825120\,\kappa + 3.3478315\,\kappa^2 \over
    1 + 4.0195994\, \kappa + 1.0005712\, \kappa^2}.
\label{pade}
\end{equation}
In Fig.~\ref{superplot}
we compare the numerically calculated superheating
field against the one, two, and three term asymptotic expansions.
The one term (i.e., the Orsay group) result never seems particularly
accurate. There is a marked improvement with the two term expansion, with
the three term expansion offering only a modest additional improvement.
The $[2,2]$~Pad\'e approximant agrees with the numerical data to within about
$1\:\%$ all the way to $\kappa=1$.

{\it Uniform solutions.}   From the inner and outer expansions it is
possible to construct {\it uniform} solutions, which are asymptotically
correct for all $x$ as $\kappa\rightarrow 0$.  To do this we simply add
the inner and outer solutions of a given order, which guarantees the correct
behavior in the outer region as well as in the boundary layer.  However,
this would produce a result which was $2 f_{\rm match}$  in the
matching region, so we need to subtract $f_{\rm match}$ in order to obtain
the correct behavior in this region.  As an example, we will construct the
2-term uniform solution for the order parameter.  Adding the two-term
outer solution, $f_0(x')+ \kappa f_1(x')$, to the two-term inner solution,
$F_0(X) + \kappa F_1(X)$, subtracting the solution in the matching region,
which is $1/\sqrt{2} + (\sqrt{2}/4)\kappa X - (15/32)\kappa$, and writing
the entire combination in terms of the original variable $x$ (which is the
same as $X$), we obtain
\begin{equation}
f_{\rm unif,2}(x) = \tanh\left( {\kappa x + x_0 \over \sqrt{2}}\right)
- {15\over 16} \kappa\, {\rm sech}^2\left( {\kappa x + x_0 \over \sqrt{2}}\right)
 + {\kappa \over 4} e^{-\sqrt{2} x}.
\label{unif}
\end{equation}
As $x\rightarrow \infty$, $f_{\rm unif,2}(x)\rightarrow 1$; also,
$f_{\rm unif,2}(0) = 1/\sqrt{2} - (7/32)\kappa$, as we expect.  However,
$f_{\rm unif,2}'(0) = (15/64)\kappa^2$, so that the zero-derivative boundary
condition is only satisfied to $O(\kappa)$.

\begin{figure}
    \centerline{\includegraphics[width=\figwidth]{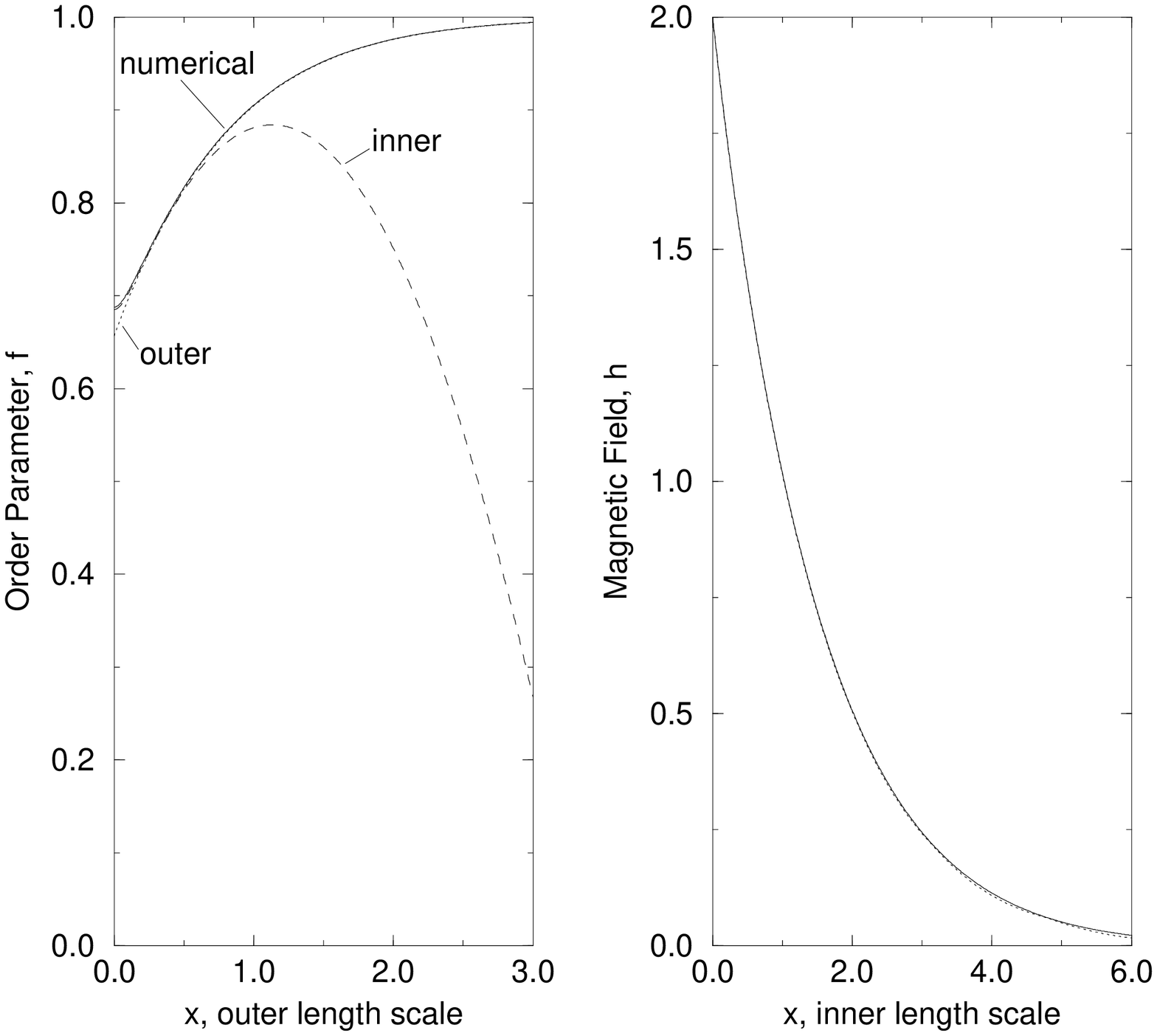}}
    \ncap[Order Parameter for $\kappa=0.1$]{A comparison of the three term inner and outer solutions for the
    order parameter and the magnetic field
    with the numerical solution for $\kappa=0.1$.  The asymptotic solutions
    approximate the computed values only in the appropriate regions.
    The matching region where the inner and outer meet is
    $O(\kappa)$ as can be estimated from the inner solution for $f$.\label{matching_fig}}
    \end{figure}

\begin{figure}
    \centerline{\includegraphics[width=\figwidth]{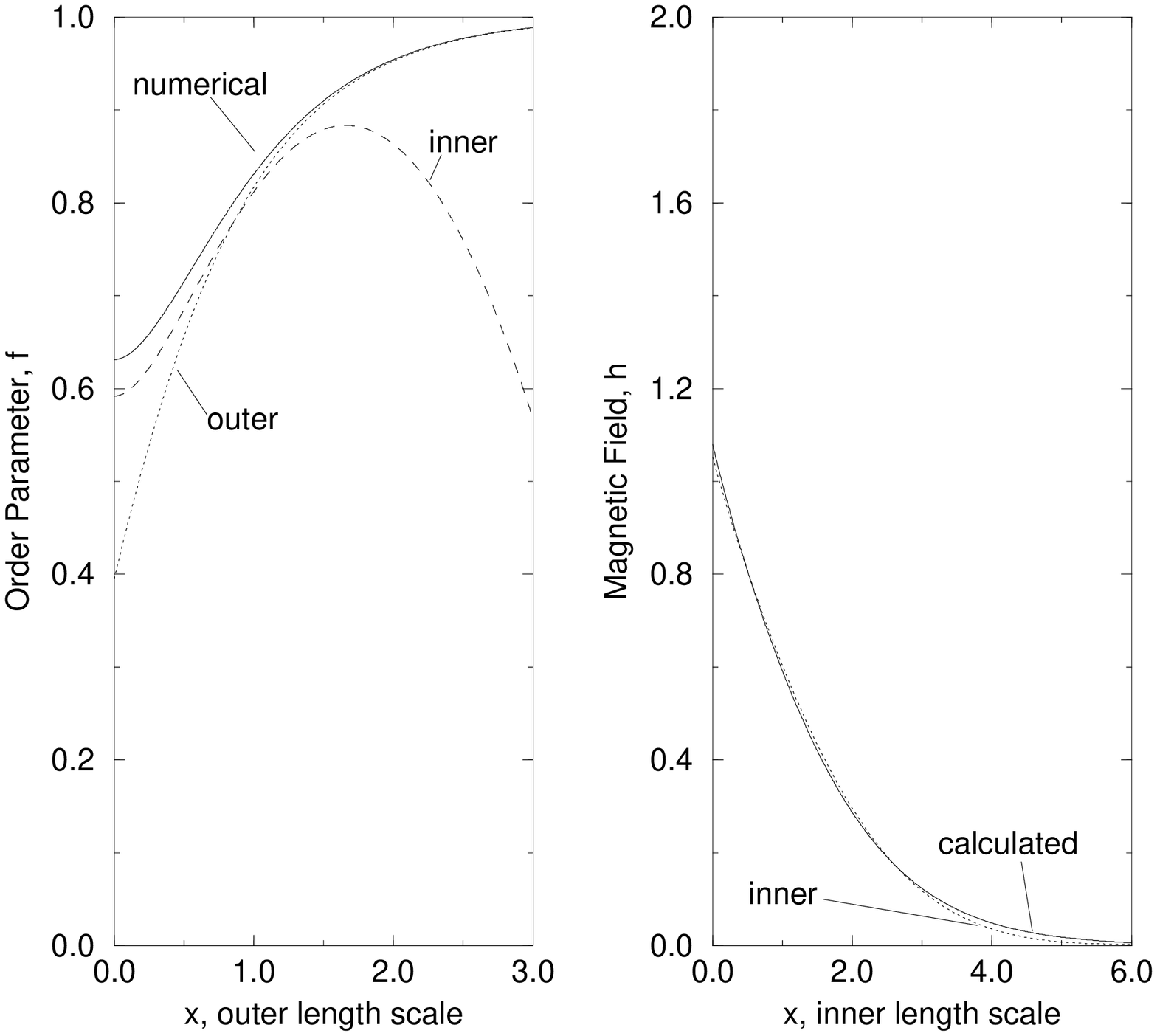}}
    \ncap[Order Parameter for $\kappa=0.5$]{The same as Fig. \ref{matching_fig} for $\kappa=0.5$.\label{matching_fig2}}
    \end{figure}

\begin{figure}
    \centerline{\includegraphics[width=\figwidth]{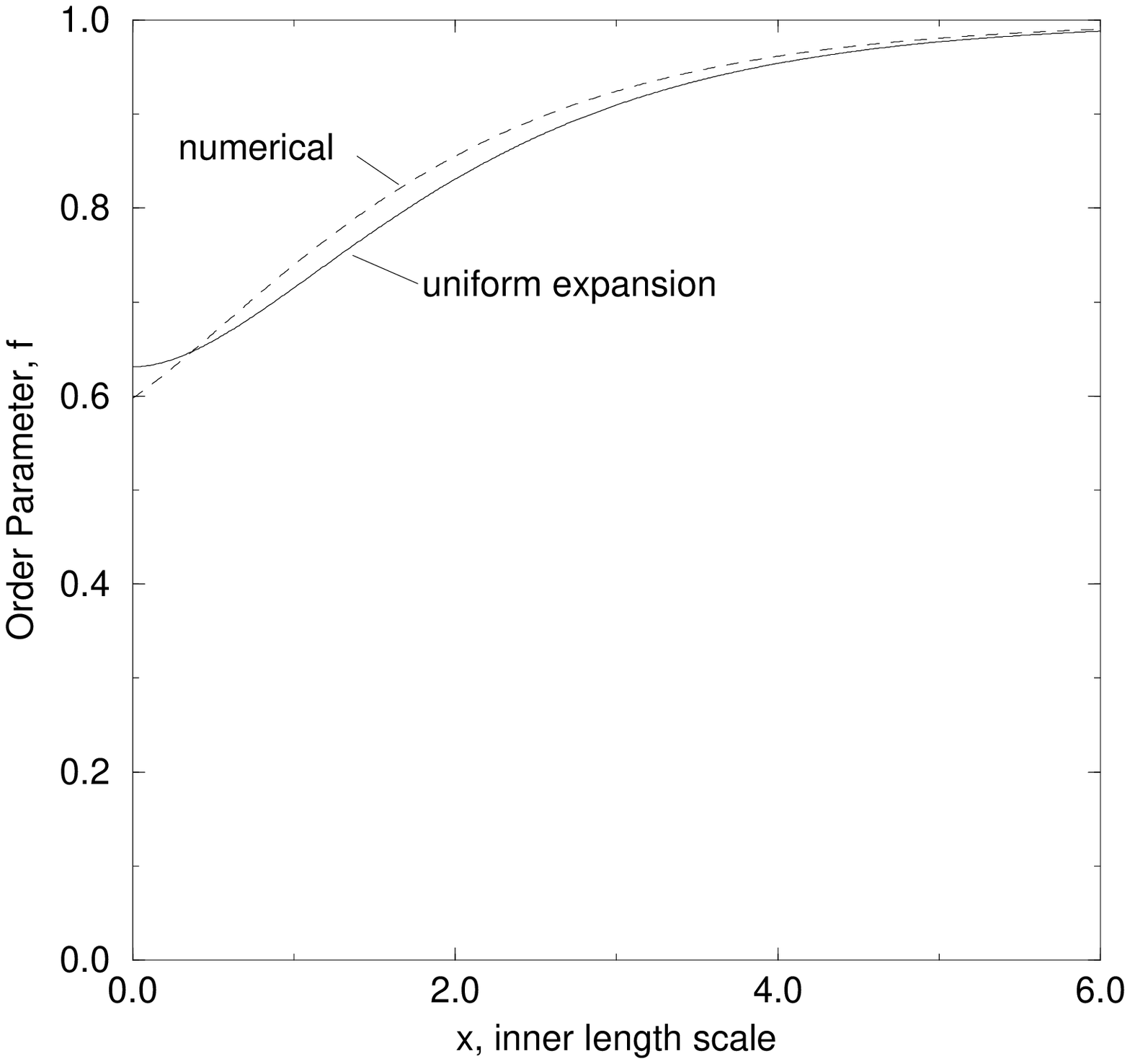}}
    \ncap[Uniform Solution of the Order Parameter at $\kappa=0.5$]{A comparison of the two-term uniform solution for the
    order parameter, $f_{\rm unif,2}(x)$ (dashed line),
    with the numerical solution (solid line) at
    $\kappa= 0.5$.  The disagreement of the uniform solution with the boundary
    condition at $x=0$ is of order $\kappa^2$.\label{uniform_fig}}
    \end{figure}

In Fig.~\ref{matching_fig} and Fig.~\ref{matching_fig2}
we compare the
numerically calculated order parameter and magnetic field with the
two term outer solutions and the three term inner solutions.
The agreement is quite good for $\kappa=0.1$, with deviations appearing
at $\kappa=0.5$.  These figures also illustrate the existence of a
matching region where the inner and outer solutions overlap; this
region grows as $\kappa\rightarrow 0$.
Lastly, we show in Fig.~\ref{uniform_fig}
how the two term uniform expansion
constructed earlier supplies a uniform approximation to the order parameter and
magnetic field over the whole region.

\section{Stability Analysis of the Solutions}

\subsection{Derivation of Equations}

We test the stability of the solutions by solving the second
variation of the free energy to see whether the \GL solutions sit
in a minimum, maximum or saddle point of the free energy. The
second variation of the Free Energy can be found any one of a
number of ways.  The first solution shown is equivalent to using
Calculus of Variations and the second is known as a time-dependent
formulation.

If we perturb the extremal solution $(f,q)$ of the GL equations by
allowing $f(x)\rightarrow f(x)+\tilde{f}(x)$ and $q(x)\rightarrow
q(x)+\tilde{q}(x)$, then the second variation of the free energy
functional is
\begin{equation}
  \delta^2 {\cal F} =
  \int_0^{\infty}dx\left[\frac{1}{\kappa^2}\tilde{f}^{'2}+(3f^2+q^2-1)
  \tilde{f}^2+4fq\tilde{f}\tilde{q}+f^2\tilde{q}^2+
  \tilde{q}^{'2}\right].
\label{stab1}
\end{equation}
The boundary conditions on $\tilde{f}$ and $\tilde{q}$ should be
chosen so as to not perturb $f$ and $h$ at the surface, so that
\begin{equation}
  \tilde{f}'(0) = \tilde{q}'(0)=0, \qquad
          \tilde{f}(\infty)=\tilde{q}(\infty)=0.
\label{stab_boundary}
\end{equation}
We can then integrate Eq.~(\ref{stab1}) by parts to obtain
\begin{equation}
  \delta^2 {\cal F} = \int_0^{\infty}dx\left[\tilde{f}\left(
   -\frac{1}{\kappa^2}\frac{d^2}{dx^2} +
           q^2+3f^2-1\right)\tilde{f} +
  \tilde{q}\left(-\frac{d^2}{dx^2}+
  f^2\right)\tilde{q}+4qf\tilde{q}\tilde{f}\right].
\end{equation}
This quadratic form can be conveniently written as
\begin{equation}
  \delta^2{\cal F} = \int_0^{\infty}dx\, (\tilde{f},\tilde{q})\hat{L}_1
  \left(\begin{array}{c}\tilde{f}\\ \tilde{q}\end{array}\right)
\end{equation}
where $\hat{L}_1$ is the self-adjoint linear operator
\begin{equation}
  \hat{L}_1\left(\begin{array}{c}\tilde{f}\\ \tilde{q}\end{array}\right) =
  \left(\begin{array}{cc}
    -\frac{1}{\kappa^2}\frac{d^2}{dx^2}+q^2+3f^2-1 & 2fq \\
    2fq & -\frac{d^2}{dx^2} +f^2\end{array}\right)
  \left(\begin{array}{c}\tilde{f}\\ \tilde{q}\end{array}\right).
\end{equation}
In order to analyze the stability, expand $\tilde{f}$ and
$\tilde{q}$ as
\begin{equation}
\left(\begin{array}{c}\tilde{f}\\ \tilde{q}\end{array}\right) =
\sum_{n} c_n \left(\begin{array}{c}\tilde{f}_n\\
\tilde{q}_n\end{array}\right), \label{eigen1}
\end{equation}
where the $c_n$'s are real constants, and
$(\tilde{f}_n,\tilde{q}_n)$ is a normalized eigenfunction of
$\hat{L}_1$ with eigenvalue $E_n$:
\begin{equation}
  \hat{L}_1\left(\begin{array}{c}\tilde{f}_n\\ \tilde{q}_n\end{array}\right) =
  E_n\left(\begin{array}{c}\tilde{f}_n\\
  \tilde{q}_n\end{array}\right).\label{matrixEigen}
\end{equation}
Then
\begin{equation}
  \delta^2{\cal F} = \sum_n E_n c_n^2.
\end{equation}
The second variation $\delta^2 {\cal F}$ ceases to be
positive-definite when the lowest eigenvalue first becomes
negative, indicating that the corresponding solutions $(f,q)$ of
the GL equations are unstable. Therefore the entire issue of the
stability of the solutions has been reduced to finding the
eigenvalue spectrum of the linear stability operator $\hat{L}_1$,
which in the $\kappa\rightarrow 0$ limit can be studied using
matched asymptotic expansions.

The condition that the second variation be positive definite is
identical to the traditional requirements of stability
calculations using Calculus of Variations.  It measures stability
of the system with respect to infinitesimal perturbations.  We
also know, either from theorems of Calculus of Variations or from
equivalence with the Schr\"odinger equation, that the solutions
with lowest eigenvalues will never cross zero---they will be
lowest energy bound states.

Another way to analyze the stability of solutions to the GL
equations is to substitute perturbations into the time-dependent
GL equations and find whether they remain bounded over time.  This
is often presented as a second way to measure the same stability
as that of the Calculus of Variations, but it is essentially
different.  Time-dependent stability analysis is dynamical.  Let's
look at the derivation of equations to see the difference.

In many systems, one can determine the stability of the solutions
by substituting time-dependent perturbations
\begin{eqnarray}
    f(x) & \rightarrow & f(x)+\tilde{f}(x)e^{-Et} \label{eqn:tdpert1}\\
    q(x) & \rightarrow & q(x)+\tilde{q}(x)e^{-Et} \label{eqn:tdpert2}
\end{eqnarray}
into the time-dependent equations.  Here $(f,q)$ are the exact
solutions. Then linearize the resulting equations in order to find
linear stability.  As above, positive eigenvalue, $E$, indicates a
stable system.

In the gauge where $\psi$ is real, the set of time-dependent differential
equations which would yield the exact same results as the Calculus
of Variations approach are
\begin{gather}
    \pad{f}{t}-\frac{1}{\kappa^2}\del^2f+f\mathbf{Q}^2-f+f^3 = 0 \\
    \pad{\mathbf{Q}}{t}+\del\times\del\times\mathbf{Q}+f^2\mathbf{Q} =
    0.
    \label{eqn:tdglpart}
\end{gather}
We just inserted time-dependence of the correct sign to look like
a diffusion equation.  If we put the perturbations into the above
equations (Eqns.~\ref{eqn:tdpert1} and~\ref{eqn:tdpert2}), the
resulting equations are
\begin{eqnarray}
    -\frac{1}{\kappa^2}\tilde{f}''+
        (3f^2+q^2-1)\tilde{f}+2fq\tilde{q} & = & E\tilde{f} \\
    -\tilde{q}'' + f^2\tilde{q}+2fq\tilde{f} & = & E\tilde{q}.
\end{eqnarray}
You see that the time dependence cancels when we linearize in
$\tilde{f}$ and $\tilde{q}$ leaving the eigenvalue $E$.  When
$E>0$, the solutions $(f,q)$ are stable with respect to
infinitesimal perturbations.  These equations are formally
identical to Eqns.~\ref{matrixEigen}.

The Eqs.~\ref{eqn:tdglpart} are almost, but not quite, the full
time-dependent \GL equations.  We haven't yet seen the TDGL in the
gauge where $\psi$ is real, so let's write them now.  If we make a
gauge transformation to Eqs.~\ref{eqn:dimless1} and~\ref{eqn:dimless2} of the form
\begin{gather}
\zeta = \psi e^{i\kappa\chi}\qquad \mathbf{Q} =
\mathbf{A}+\del\chi\qquad \Theta = \phi-\pad{\chi}{t}
\end{gather}
where we specify
\begin{equation}
    \kappa\chi = -\theta \quad\text{for}\quad \psi=f e^{i\theta},
\end{equation}
we find the TDGL in the gauge where $\psi$ is real
\begin{gather}
    \gamma\pad{f}{t}-\frac{1}{\kappa^2}\del^2f
    +\mathbf{A}^2f - f + f^3 = 0 \label{eqn:tdglf1}\\
    \gamma\kappa^2\phi f+\del\cdot (\mathbf{Q}^2\psi) = 0 \label{eqn:tdglf2}\\
    \pad{\mathbf{Q}}{t}+\del\phi+\del\times\del\times\mathbf{Q} +
    f^2\mathbf{Q}-\del\times\mathbf{H} = 0.\label{eqn:tdglf3}
\end{gather}

Given our initial solution in one dimension is of the form
\begin{gather}
    \mathbf{Q} = (0,q_y(x),0) \\
    f = f(x)
\end{gather}
we know the divergence on the right of the second equation,
Eq.~\ref{eqn:tdglf2}, is zero,
\begin{equation}
    2f\del\cdot\mathbf{Q} = \mathbf{Q}\cdot\del f = 0
\end{equation}
which makes $\phi=0$.  The full TDGL then reduce to
Eqs.~\ref{eqn:tdglpart}.  This is true of our initial solution,
but not true for general one-dimensional systems.

In two dimensions, however,
Eqns.~\ref{eqn:tdglf1}--\ref{eqn:tdglf3} cannot be reduced to
Eqn.~\ref{eqn:tdglpart}.  They include an extra degree of freedom
in $\phi$.  This extra degree of freedom allows them to include
normal currents, which are dissipative.  Recall that
the Calculus of Variations perturbations do not include
dissipation.  Because the equations in one
dimension are the same as the TDGL, they do describe dynamic
perturbations for this system in one dimension.  In two
dimensions, however, the Calculus of Variations perturbations
describe only non-dissipative perturbations.

The physical perturbations relevant to this system are incident
charged particles and thermal fluctuations.  They are not
infinitesimal and certainly are dissipative.  It would be more
appropriate to allow perturbations in the normal current for
infinitesimal calculations, and any finite perturbations
calculated for two dimensions must use the TDGL in order to be
relevant.  This is the heart of the debate by Kramer et al.\ about
whether perturbations on superheating can describe the first
introduction of vortices into a superconductor.  Just using the
\GL Hamiltonian is not appropriate to what would seem an
immediately relevant problem.

Perturbations using the equations of the Calculus of Variations
approach should at least be close to the correct results.  It is
possible that the least stable subspace of dissipative
perturbations is the space of stable perturbations.  More ardent
numerical work is required to test the dissipative perturbation
equations which are of higher order than the conservative ones.
We proceed with the conservative \GL perturbations and their
boundary conditions.

The boundary conditions on the differential equations for the
perturbations have to leave the applied fields unchanged. Given GL
solutions like those shown in Figure~\ref{matching_fig}, the
applied field, $\mathbf{h}(o)=q'(0)=\mathbf{H}_{\text{applied}}$
and $f'(0) = 0$ must not change. Since the perturbation equations
are linear, only the relative magnitude of the solutions is
important.

The eigenvalue, $E$, however, is nonlinearly coupled into the
equation.  However we scale the perturbations, the size of the
eigenvalue will remain the same. It will depend only on the
parameters $(\kappa, H_a)$ of the GL equation.  In fact, the
magnitude of the stability eigenvalue is another of the new
results of this work.  Knowledge of the eigenvalues as a function
of perturbation wavelengths are useful in calculating higher-order
stability.  We begin as before with matched asymptotic analysis in
inner and outer regions.

\subsection{One-dimensional Perturbations}

{\it Outer solution.} The outer equations for $(\tilde{f},\tilde{q})$
are rescaled with
$x'=\kappa x$ as before to yield (we will drop the subscript $n$
for notational convenience)
\begin{equation}
  -\tilde{f}''+(3f^2+q^2-1)\tilde{f}+2fq\tilde{q} = E\tilde{f},
\label{2d_perturb1}
\end{equation}
\begin{equation}
         -\kappa^2\tilde{q}''+ f^2\tilde{q} + 2fq\tilde{f} = E\tilde{q}.
\label{2d_perturb2}
\end{equation}
Expanding $\tilde{f}$, $\tilde{q}$, and $E$ in powers of $\kappa$, and
recalling that $q=0$ to all orders in $\kappa$ in the outer region,
we have at leading order
\begin{equation}
  -\tilde{f}_0''+(3f_0^2-1)\tilde{f}_0 = E_0\tilde{f}_0,
\label{perturb1}
\end{equation}
where $f_0 = \tanh\left(\frac{x'+x_0}{\sqrt{2}}\right)$.
By changing variables to $y = \tanh\left(\frac{x'+x_0}{\sqrt{2}}\right)$
we see that the solution of Eq.~(\ref{perturb1}) is the associated
Legendre function of the first kind:
\begin{equation}
  \tilde{f}_0(x') = c_0 P_2^{\mu}\left[\tanh\left(\frac{x'+x_0}{\sqrt{2}}
             \right)\right],
\label{legendre}
\end{equation}
where $\mu = -\sqrt{2(2-E_0)}$ and $c_0$ is a constant.
The leading order solution for $\tilde{q}$ is $\tilde{q}_0=0$.

{\it Inner solution.} To obtain the inner equations, we rescale as in
Eq.~(\ref{inner1}), with the perturbations rescaled as
\begin{equation}
\tilde{f}(x') = \tilde{F}(X),\quad \tilde{q}(x') = \kappa^{-1/2}\tilde{Q}(X),
\label{perturb2}
\end{equation}
such that
\begin{eqnarray}
  -\frac{1}{\kappa^2}\tilde{F}''+
  (3F^2+\frac{1}{\kappa}Q^2-1)\tilde{F}+\frac{1}{\kappa}2 FQ
\tilde{Q} = E\tilde{F}, \\
         -\tilde{Q}''+ F^2\tilde{Q} + 2FQ\tilde{F} = E\tilde{Q}.
  \end{eqnarray}
To leading order, $\tilde{F_0}'' = 0$, so that $\tilde{F_0} = a_0$, with
$a_0$ a constant.  The leading order equation for
the variation in $Q$ is
\begin{equation}
  -\tilde{Q}_0'' + 2F_0Q_0\tilde{F}_0+(F_0^2-E_0)\tilde{Q}_0 = 0.
\end{equation}
The solution which satisfies the boundary condition $\tilde{Q}'(0)=0$ is
\begin{equation}
  \tilde{Q}_0(X) =
  \frac{2a_0A_0B_0}{E_0}\left(e^{-A_0X}-\frac{A_0}{\sqrt{A_0^2-E_0}}
e^{-\sqrt{A_0^2-E_0}X}\right).
\end{equation}
At $O(\kappa)$ we find
\begin{equation}
  \tilde{F}_1'' = Q_0^2\tilde{F}_0+2F_0Q_0\tilde{Q}_0,
\end{equation}
with the solution
\begin{equation}
  \tilde{F}_1(X) =\displaystyle
  a_1+ a_0B_0^2\left[\frac{E_0+4A_0^2}{4A_0^2E_0}e^{-2A_0X}-
    \frac{4A_0^2}{E_0}\frac{e^{-(A_0+\sqrt{A_0^2-E_0})X}}
            {(A_0+\sqrt{A_0^2-E_0})^2\sqrt{A_0^2-E_0}}\right]
   \qquad\atop\quad\displaystyle
     +a_0B_0^2\left[\frac{E_0+4A_0^2}{2A_0E_0}-
            \frac{4A_0^3/E_0}{(A_0+\sqrt{A_0^2-E_0})\sqrt{A_0^2-E_0}}\right]X.
\end{equation}
We now have enough terms in the inner and outer region for a nontrivial match.

\begin{figure}\centerline{\includegraphics[width=\figwidth]{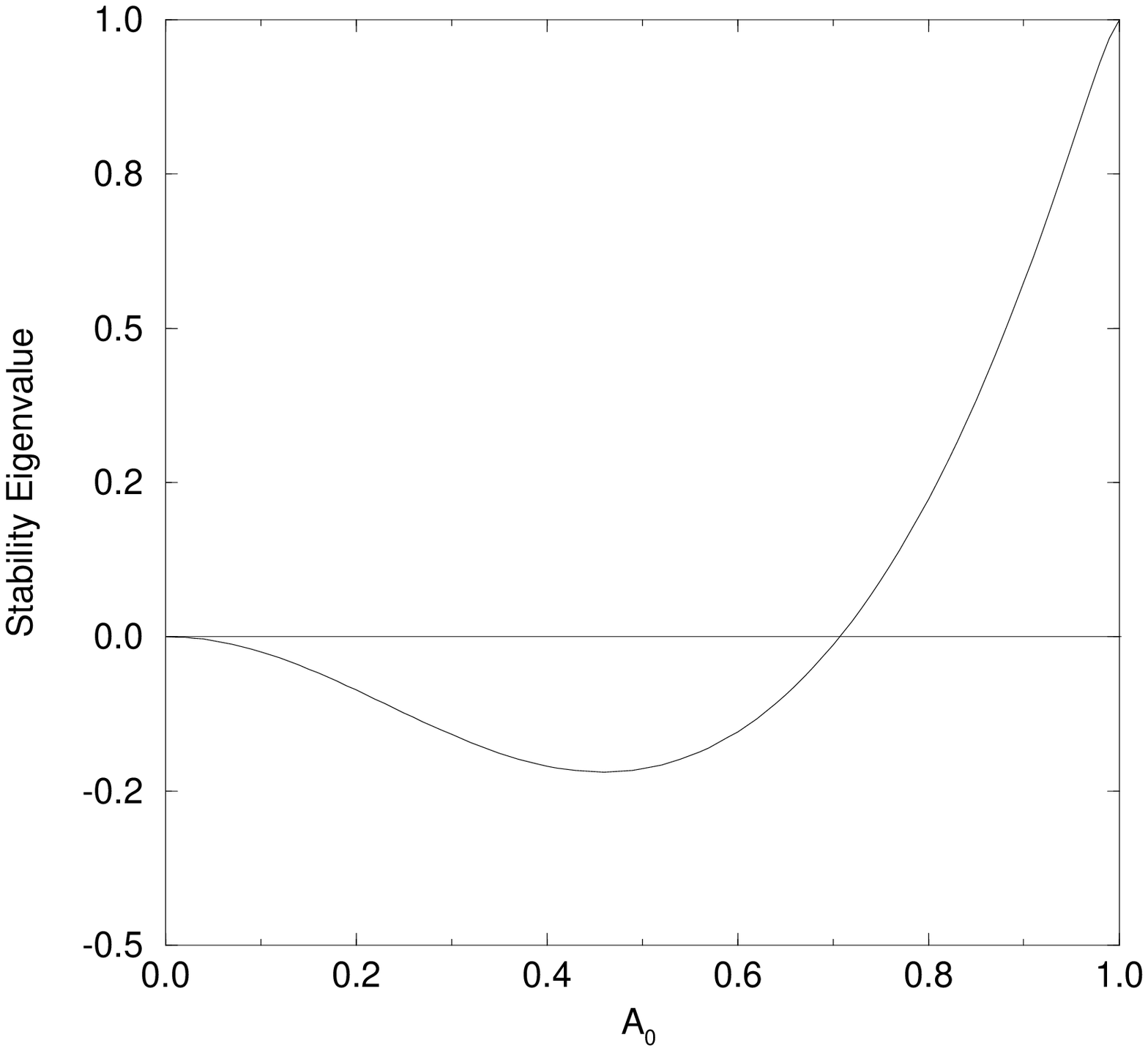}}
    \ncap[Dependence of Stability Eigenvalue on Order Parameter]{The stability
    eigenvalue $E(A_0)$, with $A_0$ the value of the
    order parameter at the surface at leading order.  We see that $E>0$
    for $A_0> 1/\protect\sqrt{2}$, indicating locally stable solutions.\label{stable_plot}}
    \end{figure}

\begin{figure}\centerline{\includegraphics[width=\figwidth]{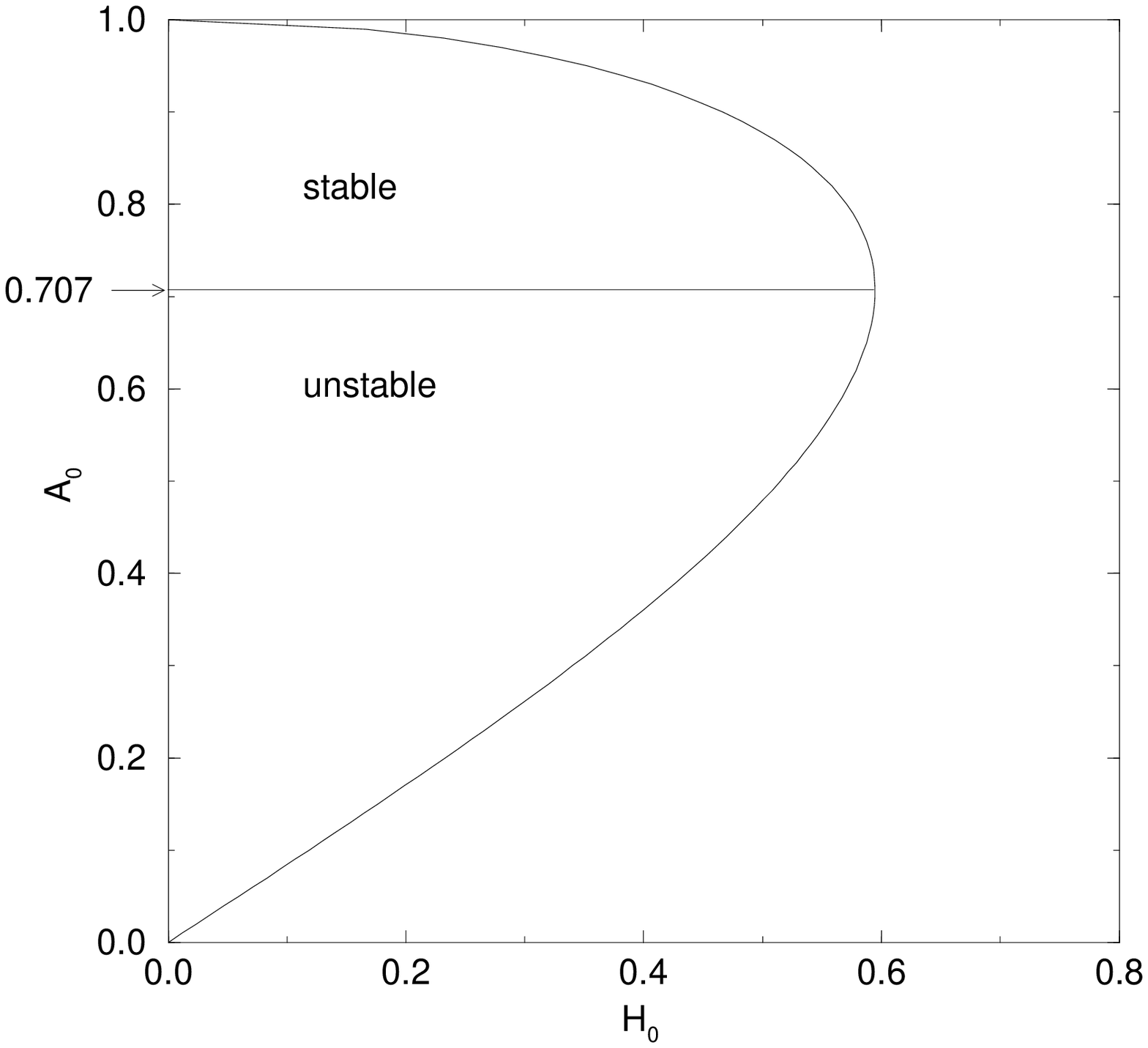}}
    \ncap[Dependence of Order Parameter on Applied Field]{The
    order parameter at the surface, $A_0$, as a function of the
    field at the surface, $H_0$, at leading order.  The stability analysis
    shows that only the upper branch corresponds to locally stable solutions.
    The field at the ``nose'' is the limit of stability, and corresponds
    to the superheating field $H_0 = 2^{-3/4}=0.595$.\label{nose}}
    \end{figure}

\emph{Matching.} We complete the matching of the inner and outer perturbations
to obtain the eigenvalue, $E_0$.  Performing a two term inner expansion of
the one term outer solution, we have
\begin{equation}
  \tilde{f}_0(\kappa X) \sim c_0 \left[ P_2^{\mu}(A_0) +
  \frac{1}{\sqrt{2}}\mbox{sech}^2(x_0/\sqrt{2})
   \frac{dP_2^{\mu}(A_0)}{d A_0 } \kappa X \right],
\end{equation}
where we have used $\tanh(x_0/\sqrt{2}) = A_0$.
Next, the one term outer expansion of the two term inner solution is
\begin{equation}
  \tilde{F}_0(x'/\kappa) + \kappa \tilde{F}_1(x'/\kappa) \sim
 a_0+\frac{a_0 2^{1/2} (1-A_0^2)}{E_0}\left[
\frac{E_0+4A_0^2}{2A_0}
      -\frac{4A_0^3}{(A_0+\sqrt{A_0^2-E_0})\sqrt{A_0^2-E_0}}\right] x',
\end{equation}
where we have used $B_0 = -2^{1/4}(1-A_0^2)^{1/2}$.
Matching the two expansions using the van Dyke matching principle yields
\begin{equation}
  c_0 = \frac{a_0}{P_2^{\mu}(A_0)},
\end{equation}
\begin{equation}
  \frac{1}{P_2^{\mu}(A_0)}\frac{dP_2^{\mu}(A_0)}{dA_0} =
  \frac{2}{E_0}\left[
    \frac{E_0+4A_0^2}{2A_0}
  -\frac{4A_0^3}{(A_0+\sqrt{A_0^2-E_0})\sqrt{A_0^2-E_0}}\right].
\label{metastable}
\end{equation}
The last equation is a rather complicated implicit equation for the
eigenvalue $E_0(A_0)$, which generally must be solved numerically.
However, when $A_0=1/\sqrt{2}$ we find $E_0=0$, corresponding to the
critical case, with $E>0$ for $A_0>1/\sqrt{2}$.  The numerical evaluation
of Eq.~(\ref{metastable}) is shown in Fig.~\ref{stable_plot}.
Therefore, we
see that our maximum superheating field (at lowest order) corresponds to
the limit of metastability for these one-dimensional perturbations.
In Fig.~\ref{nose}
we show $A_0$ as a function of the lowest order
magnetic field at the surface, $H_0$, from Eq.~(\ref{match8}).
The stability analysis of this section shows that only the upper
branch of this double valued function corresponds to solutions which
are locally stable, with the field at the ``nose'' being the superheating
field.

\subsection{Two-dimensional perturbations}

We next turn to the stability of the solutions with respect to two
dimensional perturbations. It is very likely that there may be
solutions stable with respect to one-dimensional perturbations but
not two.  The GL solutions are minimizers of the free energy and
we expect them to usually sit in the well of free energy
potential.  They will likely always be stable with respect to
infinitesimal one-dimensional perturbations.  However, we can
imagine that if we allow the free energy schematic a second
direction, the GL free energy minimizer may be either the minimum
or maximum of a parabola in the $\hat{y}$ direction.  In other
words, we are searching for the applied field at which the GL
solution becomes a saddle point in the free energy.

If we perturb the extremal solution $(f,{\bf q})$
of the GL equations by allowing
$f\rightarrow f+\delta f$ and ${\bf q}\rightarrow {\bf q}+\delta {\bf q}$,
then the second variation of the free energy functional is
\begin{equation}
  \delta^2 {\cal F} = \int dx\,dy \left[\frac{1}{\kappa^2}
({\bf \nabla}\delta f)^2+
  4 f (\delta f) {\bf q}\cdot\delta {\bf q}+f^2(\delta {\bf q})^2+
  (3 f^2+{\bf q}^2-1)(\delta f)^2 + ({\bf\nabla}\times\delta {\bf
  q})^2\right].
\label{2d}
\end{equation}
We neglect perturbations along the $\hat{z}$-direction because Fink and
Presson~\cite{fink69} showed that terms in $\hat{z}$ are purely positive
definite and thus any variation in $z$ only increases the free
energy. Expanding in Fourier modes with respect to $y$~\cite{kramer68},
\begin{equation}
  \delta f(x,y) = \tilde{f}(x)\cos ky, \quad
  \delta q_x(x,y) = \tilde{q}_x(x)\sin ky,\quad
  \delta q_y (x,y) = \tilde{q}_y(x)\cos ky,
\end{equation}
substituting into Eq.~(\ref{2d}), recalling that ${\bf q} = (0,q(x),0)$,
and integrating over $y$, we obtain (up to a multiplicative constant)
\begin{equation}
  \delta^2{\cal F} =
  \int_0^{\infty}dx\left[\frac{1}{\kappa^2}\tilde{f}^{'2}+(3f^2+q^2
    +\frac{1}{\kappa^2}k^2-1)
  \tilde{f}^2+4fq\tilde{f}\tilde{q}_y+f^2(\tilde{q}_x^2+\tilde{q}_y^2)+
  (\tilde{q}_y'-k\tilde{q}_x)^2\right].
\end{equation}
By integrating by parts and using the boundary conditions,
Eq.~(\ref{stab_boundary}), we can cast this functional into the form
\begin{equation}
  \delta^2{\cal F} = \int_0^{\infty}dx\, (\tilde{f},\tilde{q}_y,\tilde{q}_x)
\hat{L}_2 \left(\begin{array}{c}\tilde{f}\\ \tilde{q}_y\\ \tilde{q}_x\end{array}
  \right),
\end{equation}
where the self-adjoint linear operator $\hat{L}_2$ is given by
\begin{equation}
  \hat{L}_2\left(\begin{array}{c}\tilde{f}\\ \tilde{q}_y\\ \tilde{q}_x
 \end{array}\right) =
  \left(\begin{array}{ccc}
    -\frac{1}{\kappa^2}\frac{d^2}{dx^2}+q^2+3f^2+ k^2/\kappa^2-1 & 2fq & 0\\
    2fq & -\frac{d^2}{dx^2} +f^2 & - k \frac{d}{dx}\\
   0 &  k \frac{d}{dx} & f^2 + k^2 \end{array}\right)
\left(\begin{array}{c}\tilde{f}\\ \tilde{q}_y \\ \tilde{q}_x \end{array}\right).
\end{equation}
That the operator is self-adjoint shows that it represents
perturbations on a conservative hamiltonian. As in the previous
section, we want to determine the eigenvalue spectrum of this
operator. We are primarily interested in the effects of
long-wavelength perturbations (i.e., $k\rightarrow 0$), so we
rescale $k$ as $k=\kappa k'$. Then the eigenvalue equations in
terms of the outer coordinate $x'=\kappa x$ are (dropping the
prime on $k$ from now on)
\begin{gather}
  -\tilde{f}''+(3f^2+q^2-1+k^2)\tilde{f}+2fq\tilde{q} = E\tilde{f},
\label{2d1} \\
 - \kappa^2 \tilde{q}_y'' + f^2 \tilde{q}_y + 2 fq\tilde{f}
            - \kappa^2 k \tilde{q}_x' = E\tilde{q}_y,
\label{2d2} \\
 \kappa^2 k \tilde{q}_y' + (f^2 + \kappa^2 k^2) \tilde{q}_x = E \tilde{q}_x.
\label{2d3}
\end{gather}
By using the last equation we may eliminate $\tilde{q}_x$ from Eq.~(\ref{2d2}),
which becomes
\begin{equation}
- \kappa^2 {d \over dx}\left[ {f^2 - E \over f^2 + \kappa^2 k^2 - E}
 \tilde{q}_y'\right] + f^2\tilde{q}_y + 2fq\tilde{f} = E \tilde{q}_y.
\label{2d4}
\end{equation}
For $k=0$ Eqs.~(\ref{2d1}) and (\ref{2d4}) reduce to the one-dimensional
perturbation equations of the last section, Eqs.~(\ref{2d_perturb1})
and (\ref{2d_perturb2});  for $E=0$ they reduce to
the Euler-Lagrange equations derived by Kramer~\cite{kramer68}.

\begin{figure}\centerline{\includegraphics[width=\figwidth]{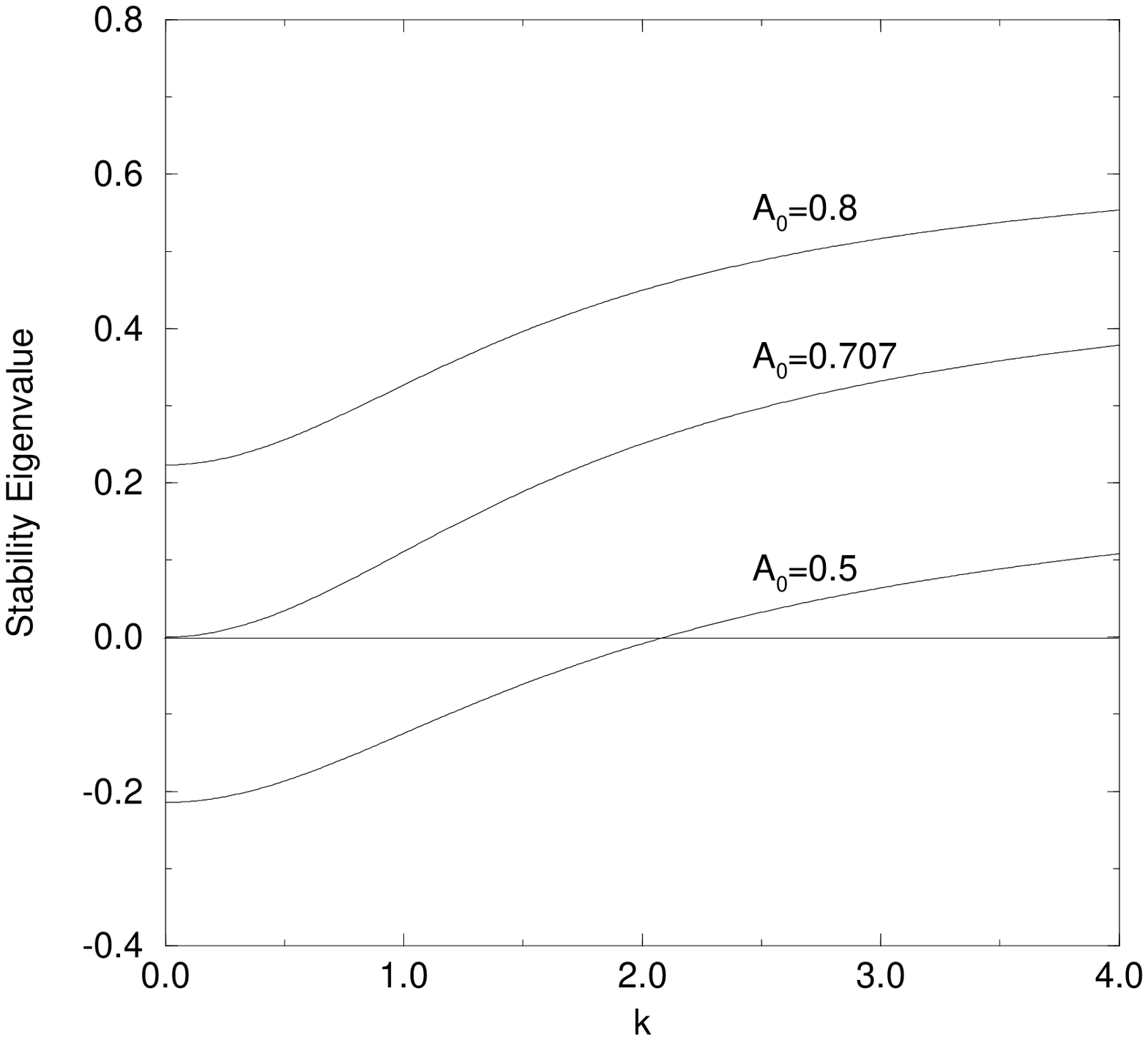}}
    \ncap[Two-dimensional Stability Eigenvalues]{The
    stability eigenvalue $E(k)$ for two-dimensional perturbations
    of wavenumber $k$, for several different values of $A_0$. For
    $A_0> 1/\protect\sqrt{2}$ the eigenvalue is stable for all wavenumbers,
    while for $A_0< 1/\protect\sqrt{2}$ there exists a band of wavenumbers
    for which the solution is unstable.\label{stable_plot2}}
    \end{figure}

The perturbation equations (\ref{2d1}) and (\ref{2d4}) may be solved
by the method of matched asymptotic expansions, just as in the
one-dimensional case.  The derivation of the eigenvalue condition
is essentially identical, with the final result that
\begin{equation}
  \frac{1}{P_2^{\mu}(A_0)}\frac{dP_2^{\mu}(A_0)}{dA_0} =
  \frac{2}{E_0}\left[
    \frac{E_0+4A_0^2}{2A_0}
  -\frac{4A_0^3}{(A_0+\sqrt{A_0^2-E_0})\sqrt{A_0^2-E_0}}\right],
\end{equation}
where now $\mu = - \sqrt{2(2+E_0-k^2)}$. The eigenvalue $E_0(k)$ is plotted
in Fig. \ref{stable_plot2}
for several different values of $A_0$.
For $A_0>1/\sqrt{2}$, $E_0(k)>0$ for all $k$, while for $A_0<1/\sqrt{2}$
there exists a band of long-wavelength perturbations for which
$E_0(k)<0$.  In all cases the most unstable modes are at $k=0$, i.e.,
the one-dimensional perturbations are the least stable.  This is
in contrast to the large-$\kappa$ limit, where the most unstable mode
occurs for $k\neq 0$~\cite{galaiko66,kramer68,chapman95}.

\subsection{Large-$\kappa$ two-dimensional Stability}
The exact numerical solution of the \GL equations shows that the
system is unstable with respect to two-dimensional perturbations
at an applied field, $H_{\text{2D}}<H_{\text{sh}}$.  Fink and
Presson~\cite{fink69} estimate that the $H_{\text{2D}}$ separates from
$H_{\text{sh}}$ at $\kappa\approx 1.10$ or $\kappa\approx 1.13$.
Our calculations show the crossover occurs at $1.16<\kappa<1.17$.
At the bifurcation, the least stable mode lifts from $k=0$ to
steadily higher wavenumbers.

\begin{figure}
    \centerline{\includegraphics[width=\figwidth]{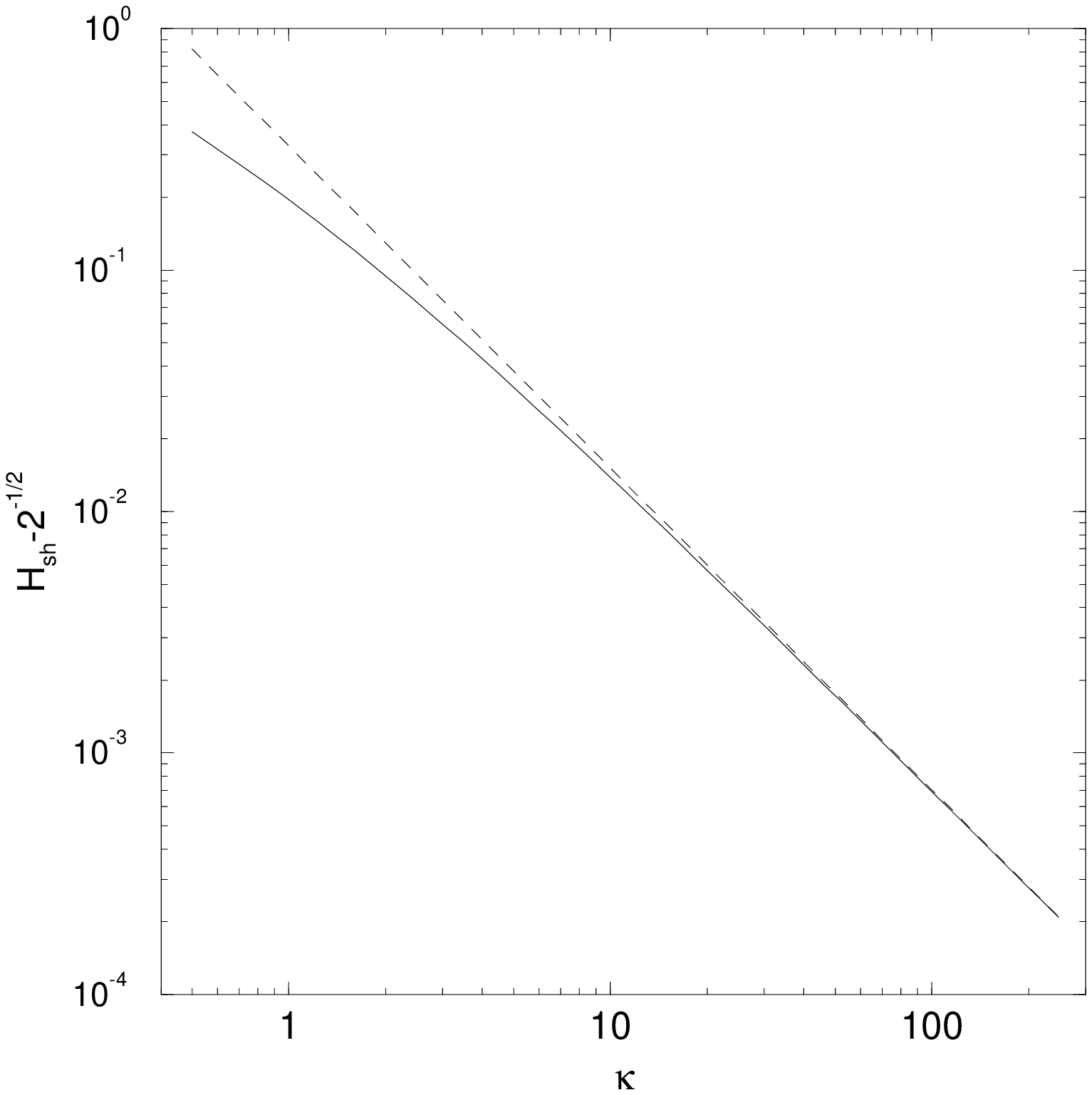}}
    \ncap[Asymptotic Dependence of Large-$\kappa$ Superheating Field]{\label{chapfig}This
    figure shows the numerically
    calculated superheating field for large $\kappa$ (solid line)
    compared with the two-term asyptotic expansion derived by
    Chapman (dashed line).  The slope of the dashed line is
    $-4/3$.}
    \end{figure}

Chapman~\cite{chapman95} recently used matched asymptotics to
examine the \GL equations in the high-$\kappa$ limit.  His final
result for the superheating field is
\begin{equation}
    H_{\text{sh}} = \frac{1}{\sqrt{2}}+
    C\kappa^{-4/3}+O(\kappa^{-6/3})
\end{equation}
where the constant $C$ is determined from the solution of the
second Painlev\'e transcendent;  a numerical evaluation yields
C=0.326~\cite{dolgert}.  The first term was originally derived by
Ginzburg~\cite{ginzburg58}, and the second term with the unusual
dependence on $\kappa$ is new.  We verify the second term in
Fig.~\ref{chapfig}.
The dependence is correct but does not converge as rapidly as the
small-$\kappa$ solution.

\section{Numerical methods}

We used two separate algorithms to evaluate our solutions of the
superheating field.  First, we calculated solutions to the \GL
equations for the half-space as a function of
$(\kappa,H_{\text{applied}})$.  Second, we calculated
perturbations to those solutions as a function of wavenumber, $k$.

Finding solutions to the \GL equations themselves was straightforward.  To
ensure that no current passes through the boundary at $x=0$ and that
the sample is totally superconducting infinitely far from the surface,
we impose the boundary conditions
\begin{equation}
f'(0) = 0,  \qquad f(x)\rightarrow 1\ \ {\rm as} \ x\rightarrow \infty.
\label{BC1}
\end{equation}
Since the field at the surface must equal the applied field $H_a$,
and there must be no field infinitely far from the surface, we
impose the boundary conditions
\begin{equation}
h(0) = H_a, \qquad q(x) \rightarrow 0 \ \ {\rm as} \ x\rightarrow \infty.
\label{BC2}
\end{equation}
The discretization used a finite domain, so the boundary
conditions at infinity were generally enforced at a coordinate
large enough not to change the shape of the solutions of the
independent variables $(f,q)$ as measured by the norm of the
difference between successive solutions,
\begin{equation}
    \int \left(|f_1(x)-f_2(x)|+|q_1(x)+q_2(x)|\right)dx.
\end{equation}
Later investigations instead used boundary conditions which were
the analytically derived asymptotics of the \GL equations.  They
had no measurable affect on the independent variables.

For $\kappa\rightarrow0$, we rescale the equations as $x'=\kappa
x$ making the new unit of length the correlation length~$\xi$.
Since $\xi\gg\lambda$ in this limit, a numerical solution over a
domain much larger than $\xi$ would ensure that the regions of
rapid change for $f$ and $h$ would be included.  (For small
$\kappa$, we find that solving for $x'<500$ is sufficient.)  In
the large $\kappa$ limit, we use the rescaled equations again, but
we increase the size of the domain depending on the value of
$\kappa$.  (The equations must be solved for domains as large as
$x'<10^4$ for values of $\kappa\sim10^3$.)

The equations can be solved using the relaxation
method~\cite{press}. By replacing these ordinary differential
equations with  finite difference equations, one can start with a
guess to the solution and iterate using a multi-dimensional
Newton's method until it relaxes to the true solution. In order to
more accurately pick up the detail near the boundary, we choose a
grid of discrete points with a higher density near $x=0$.  In
particular we choose a density which roughly varies as the inverse
of the distance from the boundary.  (For low $\kappa$ our density,
in units of mesh points per coherence length, varies approximately
from $10^7$ near the boundary to $10^3$ at the farthest point from
the boundary, while for high $\kappa$ it varies from $10^5$ to
$10^{-2}$.)

$H_{\rm sh}$ can be found in the following way.  For a given value
of $\kappa$ an initial guess is made where there is no applied
field and the sample is completely superconducting ($f\equiv 1,
q\equiv 0, h\equiv 0$). The field $H_a$ is then increased in
small increments.  For each value of $H_a$ a solution is sought
using the result from the previous lower field solution as an
initial guess.  Eventually a maximum value for $H_a$ is reached,
above which one of two things happens: our algorithm fails to
converge to a solution or it converges to the normal
(nonsuperconducting solution).  This maximum value of $H_a$ is the
numerical result for $H_{\rm sh}$. Using this algorithm, $H_{\rm
sh}(\kappa)$ can be found for a wide range of $\kappa$'s.

It is possible to imagine a situation in which this algorithm
might not work.  Suppose that you have a solution at $H_{a1}$ and
fail to converge on a solution at $H_{a2}=H_{a1}+0.01$.  It is
possible that your initial guess was just not close enough.  For
instance, a smaller stepsize would permit one to find a solution
first at $H_{a1}+0.05$, and then that solution would be close
enough to find the solution at $H_{a2}=H_{a1}+0.01$.  The
superheating solution could creep away as quickly as you could
approach it.  Using a variation on the Contracting Mapping
Theorem, Herbert Keller~\cite{keller} showed that Euler's Method,
in particular, guarantees that there exists a finite neighborhood
of the solution which will always converge for well-behaved
systems such as ours.

Each run (for a given $\kappa$) takes about 10 seconds on a
Pentium~II~300. We find it sufficient to deal with superheating
field values for $10^{-3}<\kappa<10^3$.

More interesting is the study of the perturbations.  These require
a third parameter, the wavenumber $k$.  The solutions are in a
three-parameter space $(\kappa, H_a, k)$ and the solution of each
perturbation requires calculation of the initial \GL solution.
Because the two-dimensional perturbations were of interest for
large $\kappa$ where there are not readily available analytic
solutions, even asymptotics, the initial conditions for each run
(of wavenumbers) are finicky at best.

The primary objective of solving the two-dimensional perturbations
is to find the $H_{\text{2D}}$ line, but solutions to our
equations yield, in addition, the dependence of the stability
eigenvalue on the wavenumber.  That is more information than is
relevant to the superheating field, so it is not included here.

Calculations of the two-dimensional stability required
significantly more computer resources, both in CPU time and
storage space.  The algorithms were written to function on a
cluster of a dozen systems and save only relevant data in indexed,
binary files on a central server.  Time-critical sections were
re-written as Fortran90 subroutines to C$^{++}$ control
structures.  Finding the bifurcation point at $\kappa\approx 1.17$
required about a week on our Pentium~II cluster or three months of computer
time on a single machine.

\section{Discussion}

We have solved the \GL equations both analytically and numerically
for a superconducting half-space.  The asymptotic methods depend
on disparity between coherence length and penetration depth, but
the solutions remain relevant even where they are equal.  The
resulting expansions for the superheating field should be
immediately useful for the reverse operation---calculating the
\GL parameter from the superheating field.

The same techniques were also effective for deriving perturbations
on the superheating field.  These were previously deemed
complicated enough to be beyond analysis by most authors.  We not
only calculated the exact two-dimensional perturbations but also
elucidated a vexing question last posed by Kramer~\cite{kramer73}
about whether perturbation solutions can represent vortex
nucleation.

\chapter{Phase Transition in a Current-carrying Wire}

\section{Introduction}

When a superconductor is placed in a magnetic field equal to its
critical field $H_c$, the normal and superconducting phases can
coexist in a state of equilibrium with the two phases separated by
normal-superconducting (NS) interfaces. The dynamics of such
interfaces is important for various nonequilibrium phenomena. For
instance, if the applied magnetic field is quenched below $H_c$,
these interfaces move through the sample, expelling the magnetic
flux so as to establish the Meissner phase
\cite{frahm91,liu91,dibartolo96,dorsey94,osborn94,chapman95a,%
pippard50,chapman95b,goldstein96}. Just as superconductivity can
be destroyed by applying a magnetic field exceeding $H_c$, it can
also be destroyed by applying a current exceeding the critical
depairing current $J_c$. Thus by analogy with the magnetic field
case, one might expect the competition between the superconducting
sample and the applied field to stabilize an NS interface in a
current-driven system \cite{likharev74}. In contrast to the
magnetic field induced NS interfaces, these current-induced NS
interfaces are intrinsically nonequilibrium entities, and their
structure depends upon the {\it dynamics} of the order parameter
and magnetic field. The evolution and dynamics of these
nonequilibrium interfaces is the subject of this chapter.

The current-induced NS interfaces arise in several contexts.
First, they are known to be important in understanding the
dynamics of the ``resistive state'' in superconducting wires and
films (for a review see \cite{ivlev84} or~\cite{tidecks90}), and
in determining the global stability of the normal and
superconducting phases in the presence of a
current~\cite{kramer77}. Second, Aranson {\it et
al.}~\cite{aranson96} have recently used simulations to study the
nucleation of the normal phase in thin type-II superconducting
strips in the presence of both a magnetic field and a transport
current. They found that a sufficient current produced large
normal droplets containing multiple flux quanta. Without a current
one finds stationary,  singly quantized vortices, with a larger
amount of NS interface per flux quantum than a multiply quantized
droplet. They conclude that the current produces an effective
surface tension for the NS interface which is positive,
stabilizing the interface and producing larger droplets with
smaller surface area. In their simulations, topological
singularities of multiple flux lasted for the duration of
simulations of flux entry from demagnetizing fields.  Motivated in
part by its role in this phenomenon we wanted to re-examine the
nonequilibrium stabilizing effects of current.

Even when the superconducting phase is ostensibly the equilibrium
phase, a current makes the normal phase metastable, i.e., linearly
stable to infinitesimal superconducting perturbations. A localized
superconducting perturbation of finite amplitude, on the other
hand, has one of two fates:  (1)~its amplitude may ultimately
shrink to zero restoring the normal phase (undercritical) or
(2)~it may grow eventually establishing the superconducting state
(overcritical). Separating these two possibilities are the {\it
critical nuclei\/} or {\it threshold perturbations}, for present
purposes stationary solutions of the time-dependent
Ginzburg-Landau (TDGL) equations localized around the normal
state. As one raises the current, the amplitude of the threshold
solution grows, implying that the normal phase becomes
increasingly stable.
    \begin{table}
    \centerline{\begin{tabular}{lcc}
        & normal &  superconducting \\
        \hline
    $J$ &  & \\
        & globally stable & metastable \\
    $J^*$ & & \\
        & metastable  & globally stable \\
    $J=0$ &  & \\ \hline
    \end{tabular}}
    \ncap[Simple Stability Diagram of a Current-carrying Wire]{We
    examine transitions between the two homogeneous
    states of a current-carrying superconductor.  While the system
    does not conserve energy, we can define metastability with
    respect to small perturbations.\label{simplephasetab}}
    \end{table}

At very low currents, the widths of the critical nuclei shrink as
the current is increased, but eventually this trend is reversed
and the width grows as the current is increased further. In fact,
as the current approaches a particular value, $J^*$ (the ``stall
current"~\cite{aranson96}), the width diverges resulting in two
well-separated, stationary NS interfaces.  Above $J^*$, no nucleus
solutions exist in the TDGL.  The absence of a critical nucleus
for the superconducting state defines the normal state above $J^*$
to be globally stable~\cite{kramer77,esss79}.

When a thermal fluctuation is larger than the critical nucleus for
a particular critical current, that perturbation will grow first
to locally saturate the order parameter, then form separated NS
interfaces which move at constant velocities controlled by the
applied current.  The transformation of random thermal
fluctuations into NS interfaces of fixed form is phase ordering.
Once an interface forms, it will travel towards the normal phase
if $J<J^*$ and towards the superconducting phase if $J>J^*$.

The interface solutions have been studied numerically by
Likharev~\cite{likharev74}, who found that the interfaces were
stationary at $J^* \approx 0.335$ for $u=5.79$, where $u$
characterizes the material and is $5.79$ for nonmagnetic
impurities~\cite{schmid66}. They were also studied by Kramer and
Baratoff~\cite{kramer77}, who found $J^* \approx 0.291$ for $u=12$
(corresponding to paramagnetic impurities~\cite{gorkov68}).
However, we know of no systematic study of the dependence of $J^*$
upon $u$. In this work we remedy this situation by using a
combination of numerical methods and analysis including matched
asymptotic expansions~\cite{bender,vandyke}. We show that $J^*
\sim u^{-1/4}$ for large $u$ in contrast to a previous
conjecture~\cite{likharev74}, and we find how $J^*$ approaches
$J_c$ in the small-$u$ limit.

At currents close to $J^*$, we can treat the interface velocity as
proportional to $(J-J^*)$ and calculate a kind of susceptibility.
Likharev \cite{likharev74} defined the constant of
proportionality, $\eta=(dc/dJ)^{-1}|_{J=J^*}$, where $c$ is the
interface speed; he found $\eta \approx 0.7$ for $u=5.79$. In the
extreme limits, $J \rightarrow 0$ and $J \rightarrow J_c$ Likharev
predicted that the speed $c$ diverges. We find $c$ to be bounded
in both cases and provide an analytic expression for it as $J
\rightarrow 0$.

The results of this work are summarized in Table~\ref{table1}. The
rest of the chapter is organized as follows. After briefly
reviewing the TDGL equations and the approximations used in this
work (section~\ref{TDGLsection}), we study the critical nuclei
focusing on their size and shape in the limit $J \rightarrow 0$
(section~\ref{nuclei}). We then move on to consider the stationary
interface solutions; in particular we map out the dependence of
the stall current $J^*$ on $u$ and supplement the numerical work
with analysis of the $u \rightarrow \infty$ and $u \rightarrow 0$
limits (section~\ref{stationary}). Next, we examine moving
interfaces first in the linear response regime and then in the
limits $J \rightarrow 0$ and $J \rightarrow J_c$
(section~\ref{moving}). Appendix~\ref{appendix} contains a
calculation of the amplitudes of the critical nuclei in the $J
\rightarrow 0$ limit.

\begin{table}
\ncap[Primary Results for the Chapter]{Summary of the primary
results.\label{table1}} \centerline{
\begin{tabular}{lll}
I. Critical nucleus & \\
\ \ \ \ \ Small-$J$ width & $W \sim (uJ)^{-1/2}$ & Sec.
\ref{nuc-smallj} \\
\ \ \ \ \ Small-$J$ amplitude  & $\psi_0 \sim {\rm exp}\{-A/uJ \}$ &
Sec. \ref{nuc-smallj} \\
II. Stall current $J^*$            &     &   \\
\ \ \ \ \ Large-$u$ & $J^* = 0.584491 ~u^{-1/4}$ &
Sec. \ref{stat-largeu} \\
\ \ \ \ \ Small-$u$ & $J^* =  J_c
{\textstyle (1-u/8)^{1/2}\over \textstyle (1-u/24)^{3/2}}$ &
Sec. \ref{stat-smallu} \\
III. Kinetic coeff. $\eta$ & & \\
\ \ \ \ \ Large-$u$ & $\eta  = 0.797 ~u^{3/4}$ & Sec. \ref{moving} \\
\ \ \ \ \ Small-$u$ & $\eta \sim u^{3/2}$ & Sec. \ref{moving} \\
IV. Interface speed & & \\
\ \ \ \ \ $J \rightarrow 0$ & $c \rightarrow 2/u$ &
Sec. \ref{moving} \\
\ \ \ \ \ $J \rightarrow J_c$ & $c \sim u^{1/2}$ \ \ \ \
$(u \rightarrow 0)$  & Sec. \ref{moving} \\
\ \ \ \ \  & $c \sim u^{-0.85}$ \ \ $(u \rightarrow \infty)$
& Sec. \ref{moving}
\end{tabular}}
\end{table}

\section{The Physical System}
Our simplified picture of a one-dimensional superconductor
carrying a current clarifies some of the essential physics of more
complex phenomena in experiments on thin whiskers or strips of
superconducting material.  A general review of current-induced
phenomena in one-dimensional superconductors can be found in
Tidecks~\cite{tidecks90}.  In this section, we will discuss
properties of samples (dimensions, materials, contacts) and give a
brief taxonomy of behaviors.

Systems are quasi-one-dimensional because the coherence length is
larger than the lateral dimensions of the superconducting
material. Both whiskers of single crystal and etched film
depositions can meet this criterion easily.  The point is that the
material is thin and skinny enough that variations in the order
parameter and its phase are not significant across the width of
the superconductor. A sample of YBCO from Jelila et
al.~\cite{jelila98} was $200\:\text{$\mu$m}$ long,
$20\:\text{$\mu$m}$ wide, and $90\:\text{$\mu$m}$ thick.  This
precludes the formation of vortices in the sample by self-induced
magnetization from applied currents.

We have discussed already the metastability of the two uniform
states, the normal and superconducting states.  Measurements in
steady state rarely show a transition from normal to
superconducting or vice versa. Instead, the system passes into an
intermediate state called the resistive state where the sample is
mostly superconducting but there are occasional slips in the order
parameter.  These slips are oscillatory regions where the order
parameter decreases and more of the current is carried by both an
increase in the change of phase of the order parameter ($J_s
\propto \psi^*\del\psi-\psi\del\psi^*$) and by a localized normal
current.
    \begin{figure}\centerline{\includegraphics{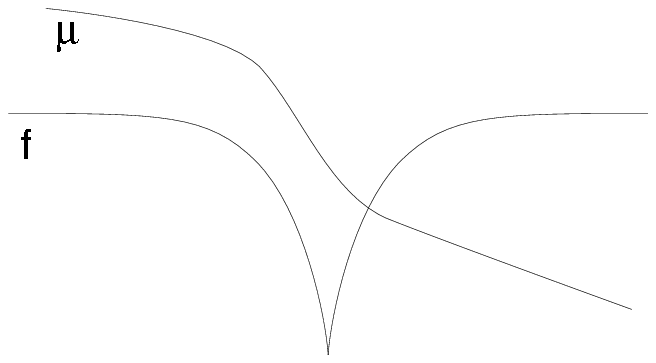}}
    \ncap[Sketch of a Phase-Slip Center]{A phase-slip center is formed when a local
    perturbation of the current causes the order parameter to drop
    to zero.  When the order parameter is small, the phase of the
    order parameter twists another $2\pi$.  The order parameter
    then heals slightly leaving a localized area with higher
    normal current and a supercurrent decreased to about $25\:\%$
    of the total current.\label{pscfig}}
    \end{figure}%
Figure~\ref{pscfig} shows a single phase slip.  At currents above
$J^*$, the resistive state consists of a periodic array of these
oscillatory regions, called phase slip centers, PSC.  The PSC
state allows the strip to remain superconducting above its
critical current, $J_c$.

\begin{figure}\centerline{\includegraphics{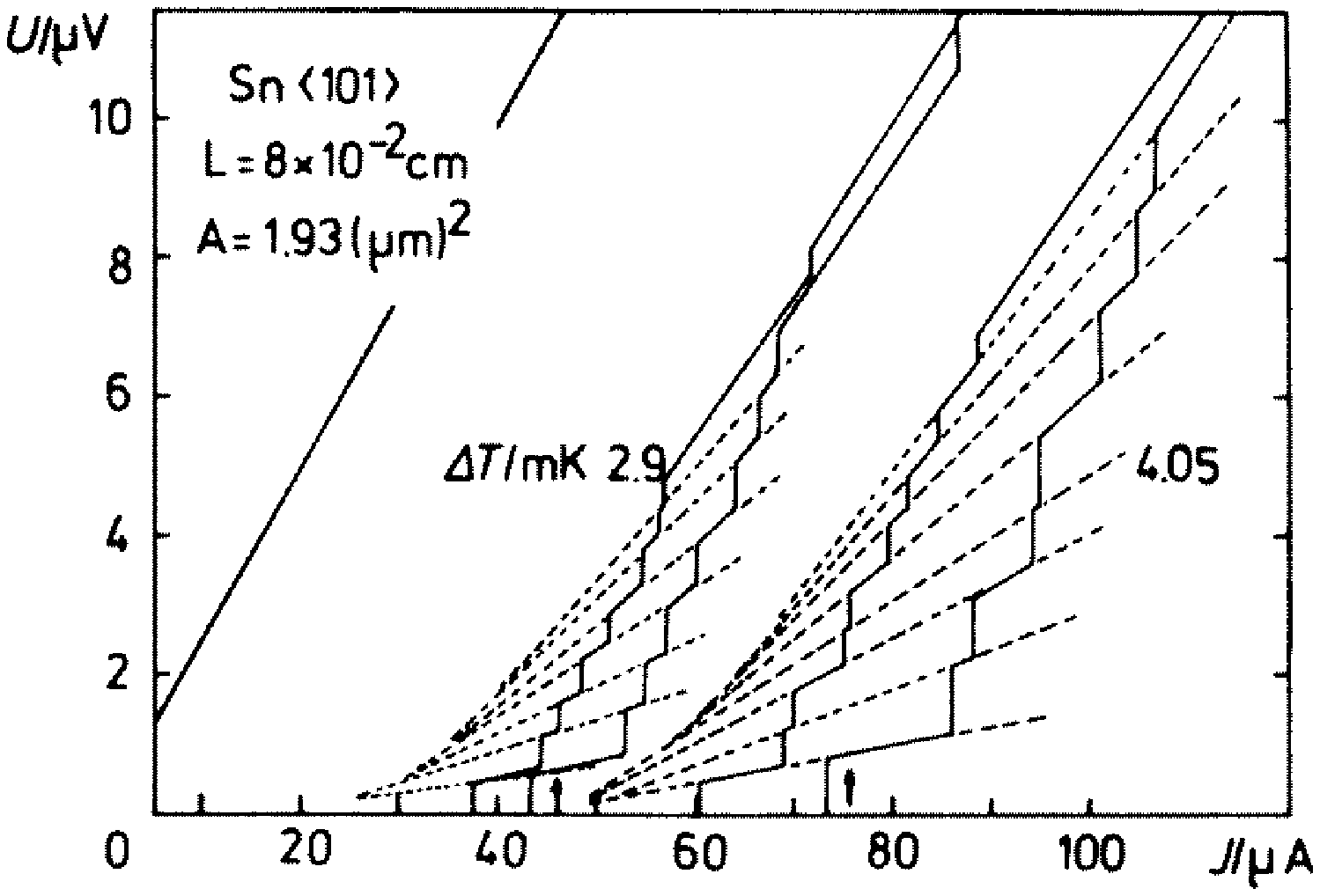}}
    \ncap[Forked Ascension in Wire CVC]{These are the dc current-voltage characteristics of a YBCO
    bridge from Jelila et al.~\cite{jelila98}.  The dotted lines show
    hysteresis for rising and falling currents.  This I--V is typical
    of pure superconductors not too far from transition temperatures.\label{cvcfig}}
    \end{figure}

When the current rises above the critical current, a single PSC
will enter.  More PSC enter at successively higher applied
currents.  The result is a series of steps in the I--V curve
called a ``forked ascension," shown in Fig.~\ref{cvcfig}.
Experiments on tin~\cite{skocpol74,skocpol74a} and
YBCO~\cite{jelila98} demonstrate similar voltage characteristics.
Each step represents the addition of a PSC.  The first solid
analysis of PSC by Skocpol, Beasley and Tinkham~\cite{skocpol74}
attributed the stability of the centers to thermal effects.
Because the middle of the phase slip is normal, it heats, possibly
above the transition temperature.  Averaging the additional
resistance of each PSC, they used the normal state resistivity to
find that the length scale of a PSC is the quasiparticle diffusion
length.  They mention, however, that in very clean systems near
$T_c$, steps are still observed ``due to some intrinsic inability
of the uniform state to nucleate in the presence of the phase-slip
process."  There is debate over when the normal or superconducting
systems are unstable with respect to the formation of phase slip
centers, but they appear to be intrinsic to the isothermal
behavior of the order parameter and a necessary state in the
adiabatic transition from normal to superconducting.

This does not mean, however, that a transition from normal
directly to superconducting is not possible.  There are two ways
to see the transition directly. As is shown in in Pals and
Wolter~\cite{pals79} and Jelila et al.~\cite{jelila94}, the order
parameter can take up to $200\:\text{ns}$ to respond to a
$3\:\text{ns}$ change in the applied current.  It is possible to
drop the current below the PSC nucleation current before the
system can respond. Secondly, applied currents in a system near
$T_c$ can be so small a PSC cannot form.  Each phase slip center
has a kind of activation energy, the amount of excess current
required to depress the order parameter enough that it reaches
zero so a phase slip can occur.  If the critical current is less
than the amount of current required to form a PSC, then the
transition will be direct from normal to superconducting.

Also, if a sample has superconducting contacts, it will be more
likely to become superconducting at higher temperatures and higher
applied currents~\cite{ivlev84}.  While most experiments etch
bridges into uniform substrates, some have applied normal contacts
to thin strips~\cite{dolan77}.  These would be more amenable to
measuring direct phase transitions.\footnote{Superconducting
contacts measure only the pair chemical potential while normal
contacts measure the full potential of the electric
current~\cite{watts-tobin81}. If the edges are in steady state,
however, the difference should be negligible.}

The transition from normal to the PSC state is not so different
from the transition from normal to superconducting.  While, in
decreasing currents, the PSC state forms much the same way the
Abrikosov lattice forms in a type-ii superconductor, the system's
reaction to a sudden decrease in applied current would be
nucleation.   The supercurrent at the contacts nucleates a local
phase change which spreads across the superconductor switching it
from normal to the resistive state.  The initial wavefront is
probably just a switching wave and is exactly what we study here.
The subsequent relaxation to the resistive state is not examined
here, however.
    \begin{figure}
    \centerline{\includegraphics{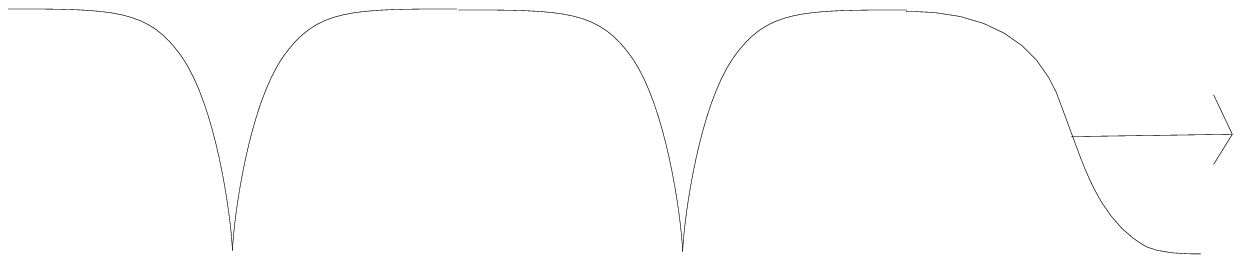}}
    \ncap[Switching Wave in a Resistive System]{Even when the
    system switches from the normal to the
    resistive state, the process should occur by means of a moving
    phase front.\label{resistiveswitchfig}}
    \end{figure}
A simple picture is shown in Figure~\ref{resistiveswitchfig}. In
any case, there seems to be little experimental data on this
region of the phase diagram.

Even if one can avoid forming PSC, heating of the normal metal
masks some of the intrinsic superconducting properties of thin
superconductors. Critical currents of cold superconductors can be
large enough that resistive heating in normal regions can control
the stability of stationary structures~\cite{skocpol74} and the
velocity of NS boundaries~\cite{broom60}.  Because we are
interested in the coexistence of superconducting and dissipative
normal regions, heat flux can be a significant factor in thermally
isolated systems.

There is a large literature devoted to heat flux in current-driven
superconductors.  Hot spots can form stable autosolitons or drive
phase transitions for the entire system.  We are more interested
in the behavior of isothermal systems because they describe
fundamental behavior of superconductivity.  The velocity of a
normal-superconducting interface, for instance, will be limited by
some combination of heat flux radiation, diffusion of the order
parameter, and relaxation times of superconducting pairs into
quasiparticles.  The experiments and analysis presented here will
be related to systems shown to be highly isothermal so that the
rate of heat flux will be of minimal importance.

There two ways commonly mentioned to combat heating effects. In
the second of their set of articles on nonuniform states in thin
strips in 1974~\cite{skocpol74a}, Skocpol et al.\ point out that
near $T_c$ critical currents are small so that applied voltages
will heat the sample less. More often, experimenters hope that a
superconducting strip on a substrate with good thermal capacitance
will dissipate heat efficiently.

More enlightening are time-resolved studies of voltage as an
applied current pulse is modulated from above the critical current
to currents below $J^*$ and $J_{\text{min}}$. One experiment which
seems a clear attempt to measure phase changes at applied currents
below the critical current is Jelila et al.~\cite{jelila94}.  The
measurements are so appropriate they must be discussed despite
inconclusive results. They applied a varying current pulse to a
thin strip of YBa$_2$C$_3$O$_x$ in order first to drive it normal
then to reduce the current and watch it return to the
superconducting or PSC states.  While their main focus was an
initial time delay of the material in responding to applied
currents, this delay was much better explained in a later PRL by
the same group~\cite{jelila98}. The earlier paper shows two graphs
of voltage versus time with resolution in nanoseconds.

The film itself was grown on a MgO substrate which was verified to
have very efficient heat conductivity~\cite{maneval94}.  The film
was a $30\:\text{nm}$ thick, $200\:\text{$\mu$m}$ long, and
$29\:\text{$\mu$m}$ wide microbridge.  The normal state resistance
of the bridge seems to be around $5\:\Omega$.  Experiments were
conducted at $4.2\:\text{K}$.  We discussed earlier that
temperatures close to $T_c$ are better suited to isothermal
measurements, but the signal to noise ratio was limited by shunt
resistors.

The paper shows two experiments on slightly different samples. The
first applies a supercritical $67\:\text{mA}$ pulse which drops to
$20\:\text{mA}$.  Here, the voltage falls within the fall time of
the pulse generator, $3\:\text{ns}$.  If we try to explain the
phase change as the progress of wave fronts, we can conclude the
wave fronts traveled faster than $3.3\times 10^4\:\text{m/s}$.

\begin{figure}
    \centerline{\includegraphics{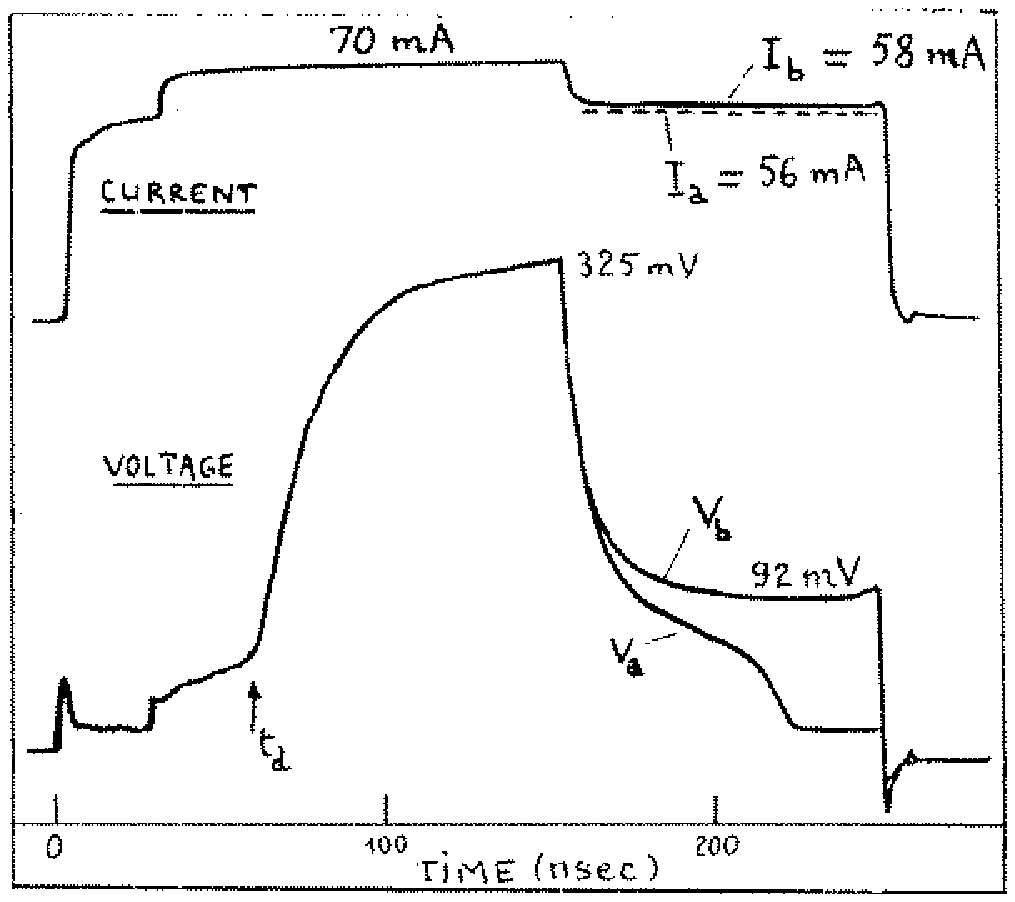}}
    \ncap[Response of a Microbridge to a Current Pulse]{This
    shows the voltage response as a function of time
    for a YBCO microbridge from Jelila et al.~\cite{jelila94}.
    The higher voltage drives the system normal.  The system then
    returns either to the superconducting state, $V_a$, or the PSC
    state, $V_b$.\label{pulsefig}}
    \end{figure}

The second experiment is shown in Figure~\ref{pulsefig}.
While the initial current of $70\:\text{mA}$ drives the system
normal we see a broad inductive rise in the voltage.  Then the
applied current is lowered to less than the critical current, and
the system shows one of two behaviors.  If the applied current is
greater than $57\:\text{mA}$, the system relaxes into a PSC state
with a voltage of $92\:\text{mV}$.  If the applied current is less
than $57\:\text{mA}$, the system returns to the superconducting
state.  The authors say only that ``the resistance drops to zero
in about $65\:\text{ns}$."  The shape of the curve is interesting.
At early times, we expect the shape to be dominated by rapid
phonon removal.  Later times, however, show a curious shape of the
curve which is not inductive.

It is unclear whether the system has had enough time to form the
PSC state before undergoing a transition into the pure
superconducting state.  Suppose that the last downturn of the
voltage, from $207\:\text{ns}$ to $224\:\text{ns}$, is related to
the movement of normal-superconducting phase fronts in the
material.  The shape of that part is roughly linear.  The change
of resistance with time of $0.0742\:\Omega/\text{ns}$ suggests
phase front velocities of about $1600\:\text{m/s}$ which is in the
general range of the calculations in this chapter.

What exactly is occurring during the phase change is not at all
clear.  There doesn't seem to be much data on phase changes at
nonzero currents.  Rather, most experiments examine the rate of
phase transition at zero bias or currents greater than the
critical current.

\section{Models for Phase Transition in a Wire}

\subsection{The TDGL Equations} \label{TDGLsection}

The starting point for our study is the set of TDGL equations for
the order parameter $\psi$, the scalar potential $\Phi$, and the
vector potential ${\bf A}$:
\begin{eqnarray}
\hbar \gamma \left( \partial_t + {i e^* \Phi \over \hbar} \right)
\psi &=& \frac{\hbar^2}{2m}\left(\del -\frac{ie^* {\bf A}}{\hbar
c} \right)^2\psi \nonumber \\ &&+ |a|\psi  -b |\psi|^2 \psi,
\label{TDGL1}
\end{eqnarray}
\begin{equation}
\del\times \del \times {\bf A} = \frac{4\pi}{c}( {\bf J}_n + {\bf
J}_s), \label{TDGL2}
\end{equation}
where the normal current ${\bf J}_n$ and the supercurrent ${\bf J}_s$
are given by
%\begin{mathletters}
\begin{eqnarray}
\label{TDGL3}
&&{\bf J}_n =   \sigma^{(n)}\left( - \partial_t {\bf A}/c
 - \del \Phi\right), \\
\label{TDGL4} &&{\bf J}_s = {\hbar e^* \over 2 m i} \left( \psi^*
\del \psi - \psi \del \psi^* \right) -\frac{e^{*2}}{mc}|\psi|^2
{\bf A},
\end{eqnarray}
%\end{mathletters}
and where $\gamma$ (which is assumed to be real) is a dimensionless
quantity characterizing the relaxation time of the order parameter,
$\sigma^{(n)}$ is the normal conductivity, and $a = a_0 (1- T/T_{c0})$.
 From these parameters we can form two important length scales,
the coherence length $\xi = \hbar/(2 m |a|)^{1/2}$ and the penetration
depth $\lambda = [mbc^2/4\pi (e^*)^2 |a|]^{1/2}$.

These equations assume relaxational dynamics for the order
parameter as well as a two-fluid description for the current. With
somewhat restrictive assumptions, they can be derived from the
microscopic BCS theory \cite{schmid66,gorkov68}. Further
simplification is possible in the limit of a thin, narrow film,
that is, when the thickness is less than the coherence length,
$d<\xi$, and the width is less than the effective penetration
depth \cite{pearl}, $w\ll \lambda^2/d$. In this case the current
carried by the film or wire is small, and we needn't worry about
the fields it produces. Therefore, Eq.~(\ref{TDGL2}) may be
dropped, and we need only specify the total current ${\bf J}={\bf
J}_n + {\bf J}_s$ (subject to $\del\cdot{\bf J}=0$), along with
the order-parameter dynamics, Eq.~(\ref{TDGL1}). This
approximation is commonly used for superconducting wires
\cite{ivlev84} and can be justified mathematically for
superconducting films \cite{chapman97}. In addition, we will be
considering processes in the absence of an applied magnetic field,
so that we may set ${\bf A}={\bf 0}$. With these simplifications,
we can now rewrite the equations in terms of dimensionless
(primed) quantities,
\begin{eqnarray}
&\psi = \sqrt{{|a| \over b}}  \psi^{\prime},
\ \ \ \ \ \ &\Phi = {\hbar e^* |a| \over m b \sigma^{(n)}}
 \mu^{\prime},
\nonumber \\
&x = \xi  x^{\prime},
\ \ \ \ \ \ &t = { m b \sigma^{(n)} \over e^{*2} |a|}
 t^{\prime},  \nonumber \\
&J = \sqrt{{2 \over m}} {e^* |a|^{3/2} \over b }
 J^{\prime},  \ \ \ \ &
\end{eqnarray}
which leads to
%\begin{mathletters}
\label{scaled}
\begin{eqnarray}
\label{TDGL-scaled}
&&        u(\partial_{t^{\prime}} + i\mu^{\prime})\psi^{\prime}
= (\nabla^{\prime 2} +1-|\psi^{\prime}|^2)\psi^{\prime},
 \\
\label{current-scaled} && {\bf J^{\prime}} = {\rm Im}
(\psi^{\prime *} \del^{\prime} {\psi^{\prime}}) - \del^{\prime}
{\mu}^{\prime}, \quad \del^{\prime}\cdot {\bf J^{\prime}}=0 .
\end{eqnarray}
%\end{mathletters}
Note that length is measured in units of coherence
length\footnote{We warn the reader that this choice is different
than in many applications of the TDGL equations, where the
penetration depth is chosen as the length scale.  In the thin film
limit the penetration depth drops out of the calculation, leaving
$\xi$ as the length scale.}. We will drop the primes hereafter.
The only parameters remaining in the problem are the scaled
current $J$ and a dimensionless material parameter
$u=\tau_{\psi}/\tau_J$, where $\tau_{\psi}= \hbar \gamma/|a|$ is
the order-parameter relaxation time and $\tau_J=\sigma^{(n)} m b /
e^{*2} |a|$ is the current relaxation time.
 We will treat $u$ as a phenomenological parameter and study the
nucleation and growth process as a function of $u$.
 The microscopic derivations of the TDGL equations  predict that
$u=5.79$ (nonmagnetic impurities) \cite{schmid66}, and $u=12$
(paramagnetic impurities) \cite{gorkov68}, but small $u$ is also
useful for modeling gapped superconductors \cite{ivlev80b}.

\subsection{Generalized TDGL Equations}
There have been fruitful generalizations of the TDGL applied to
superconducting wires. The simplest recognized that the change in
the magnitude of the order parameter is more closely related to
the relaxation time of the superconducting pairs while the change
in the phase of the order parameter is more closely related to
relaxations of the quasiparticle excitations.  These authors
substituted two relaxation constants, $u_\psi$ and $u_\phi$.

More fruitful were several competing derivations of TDGL which
account for superconductivity with a finite gap.  The first was
Kramer and Watts-Tobin~\cite{kramer78}, but alternative
derivations and discussions are available in Ivlev and
Kopnin~\cite{ivlev84} and Tidecks~\cite{tidecks90}. All derived
from microscopic theory, each accounting for somewhat different
pair-breaking mechanisms which lead to different contributions to
a gap parameter, $\Gamma$.  The general form of the equations is
    \begin{gather}
    -u\left(\frac{|\psi|^2}{\Gamma^2}+1\right)^{-1/2}
    \left(\pad{\psi}{t}+i\phi\psi+2\Gamma\pad{|\psi|^2}{t}\right)
    +\del^2\psi+\psi-|\psi|^2\psi = 0 \\
    \mathbf{j}=-\del\phi+\frac{1}{2i}\left(\psi^*\del\psi-\psi\del\psi^*\right).
    \end{gather}
When the gap parameter is zero, these equations reduce to the
standard TDGL.

Having more parameters clearly promises greater specificity to
particular metals, but the gap parameter explains significant
shortcomings in the how the TDGL describe the PSC state.  Whereas
we show below that the basic TDGL do not allow a normal to
superconducting transition above $J^*$, the equations with a gap
do demonstrate such a transition. They also better explain the
experimentally observed stability ranges of the resistive state.

The most creative and appropriate generalization of the TDGL for
this system is in a paper by Eckern, Schmid, Schmutz, and
Sch\"on~\cite{esss79}.  They derive a nonlinear Langevin equation
similar to the TDGL which is appropriate to dissipative
superconducting systems.  More interesting is that they are able
to develop a measure on this system that allows them to work with
the Langevin equation much like one treats a typical conservative
free energy.  They then proceed to describe the basic system
stability in terms of metastable and stable configurations and
calculate threshold solutions from critical nuclei, much as is
done here.  It is a fascinating attempt to coax sensible physics
out of a nonlinear dissipative system.

\subsection{Heat Equations}

As mentioned earlier, there is a large literature devoted to the
analysis of normal-superconducting boundaries driven by heat flux.
This analysis is entirely appropriate to a large class of
thermally isolated systems.  It seems most appropriate, for
instance, to current flow through a wire.  The analysis of these
systems also leads to coupled nonlinear equations, usually
parabolic diffusion equations.  They also display traveling
autosolitons as well as a wealth of other solitons common to
reaction-diffusion equations.

The first work on thermal effects in domain boundaries was done by
Skocpol, Beasley, and Tinkham~\cite{skocpol74,skocpol74a}.  The
did steady-state calculations for heat flow in thin bridges where
they assumed the normal-superconducting boundary was sharp.  They
predicted stable solitons where the center of the bridge remained
normal while the edges were superconducting.

Recently, Rudyi has both re-examined the stationary thermal
solitons~\cite{rudyi96} and switching waves~\cite{rudyi96a}, which
we study here for isothermal systems.  When looking at thermal
variations, one uses the temperature as an independent variable
rather than the order parameter.  Using the notation,
$\Theta(\theta)=(T(\theta)-T_0)/T_c$, where $T(\theta)$ is the
film temperature, $\theta=x-vt$ a self-similar variable, $T_0$ the
coolant temperature, and $T_c$ the critical temperature, Rudyi
models the system as
    \begin{gather}
    \Theta_s''(\theta)+\frac{v}{a_s}\Theta_s'(\theta)-b_s\Theta_s(\theta)=0
    \\
    \Theta_n''(\theta)+\frac{v}{a_n}\Theta_n'(\theta)-b_n\Theta_n(\theta)+W=0.
    \end{gather}
The constants concern heat transfer and thermal conductivity.
These equations predict a bistable, hysteretic system which
collapses through critical nuclei which become phase fronts. While
thermal propagation is a different mechanism from isothermal phase
propagation, the behavior of the system is almost identical. Most
thermal models are well understood as classic reaction-diffusion
systems~\cite{gurevich89} and, as such, they can be analyzed with
the standard autosoliton theory~\cite{vasiliev79,vasiliev79a}.  As
shown in Appendix~\ref{sec:diffusion}, the TDGL do not quite
conform.

While thermal flux can constrain the velocity of a
normal-superconducting interface or create nuclei in a filament,
it's effects can be minimized by good thermal coupling of the
superconductor to a substrate.  The order parameter relaxation and
current relaxation are then the dominant factors in the
stabilization of the normal-superconducting interface or the
growth of nuclei.

\section{Nucleation of the Superconducting Phase from the Normal
Phase} \label{nuclei}

In the presence of an applied current the normal phase in a
wire is linearly stable with respect to superconducting perturbations
for {\it any} value of the current \cite{gorkov70,kulik70}.
 The reason for this stability is that any quiescent superconducting
fluctuation will be accelerated by the electric field, its velocity
eventually exceeding the critical depairing velocity, resulting in
the decay of the fluctuation.
 The growth of the superconducting phase therefore requires a nucleus
of sufficient size that will locally screen the electric field and
allow the superconducting phase to continue growing; smaller nuclei
will simply decay back to the normal phase.
 The amplitude of the ``critical'' nucleus should decrease as the
current approaches zero, reaching zero only at $J=0$.
 We expect the {\it critical} nuclei to be unstable, stationary
(but nonequilibrium) solutions of the TDGL equations, which
asymptotically approach the normal solution as $x \rightarrow
\pm \infty$.
 These ``bump'' solutions of the TDGL equations are the subject of
this section.
 We include here an extensive numerical study of the amplitudes and
widths of the critical nuclei, as well as some analytical estimates
for these quantities.

\subsection{Numerical Results}
\label{nuclei-num}

Let us start by discussing the numerical work on the critical
nuclei.
 For the analytic work, we often find it convenient to use the
amplitude and phase variables, i.e. $\psi=f {\rm e}^{i \theta}$; but
they are ill-suited for the numerical work, since the calculation of
the phase becomes difficult when the amplitude is small.
 Following Likharev \cite{likharev74} we use instead $\psi=R+iI$, with
$R$ and $I$ real, and in one dimension Eqs.~(\ref{scaled}) become
%\begin{mathletters}
\label{numerics}
\begin{eqnarray}
u R_t &=& R_{xx} + u\mu I + R - (R^2 + I^2)R,  \\
u I_t &=& I_{xx} - u\mu R + I - (R^2 + I^2)I,  \\
J &=& R I_x - I R_x - \mu_x.
\end{eqnarray}
%\end{mathletters}

\begin{figure}
    \centerline{
    \includegraphics{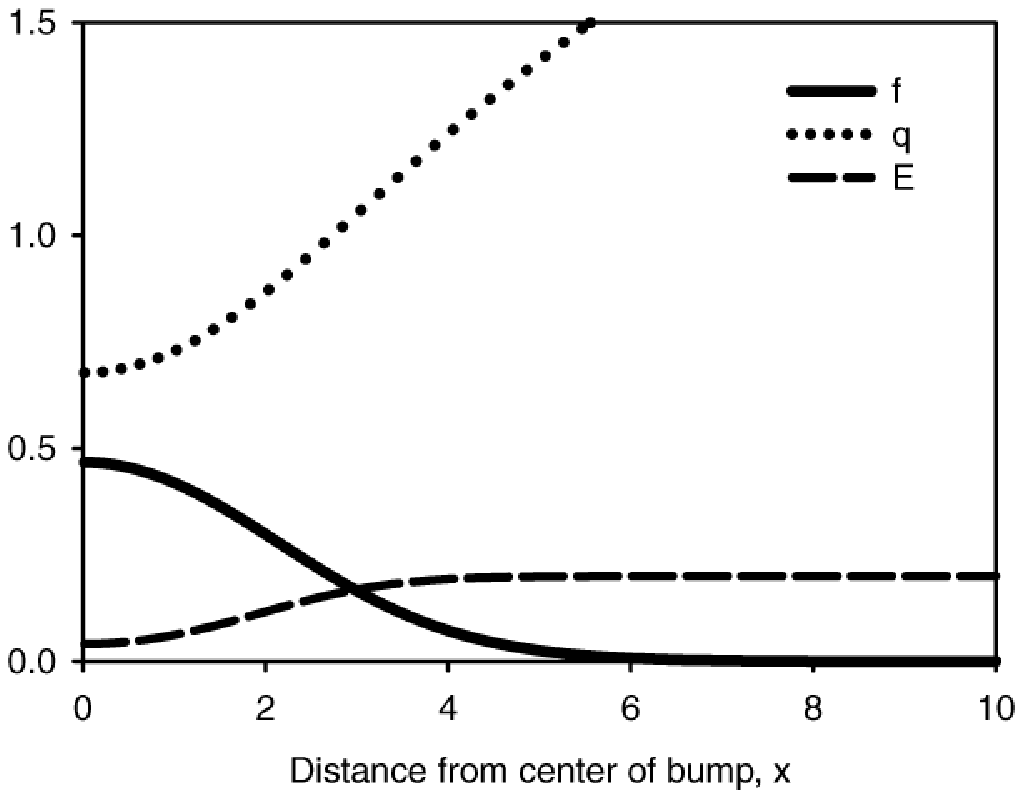}}
    \ncap[Bump Solution for $u=5.79$, $J=0.2$]{The
    bump's amplitude, $f(x)$, its superfluid velocity, $q(x)$, and the
    electric field, $E(x)$, for $u=5.79$ and $J=0.2$.\label{nuc-1}}
    \end{figure}

\begin{figure}
    \centerline{
    \includegraphics{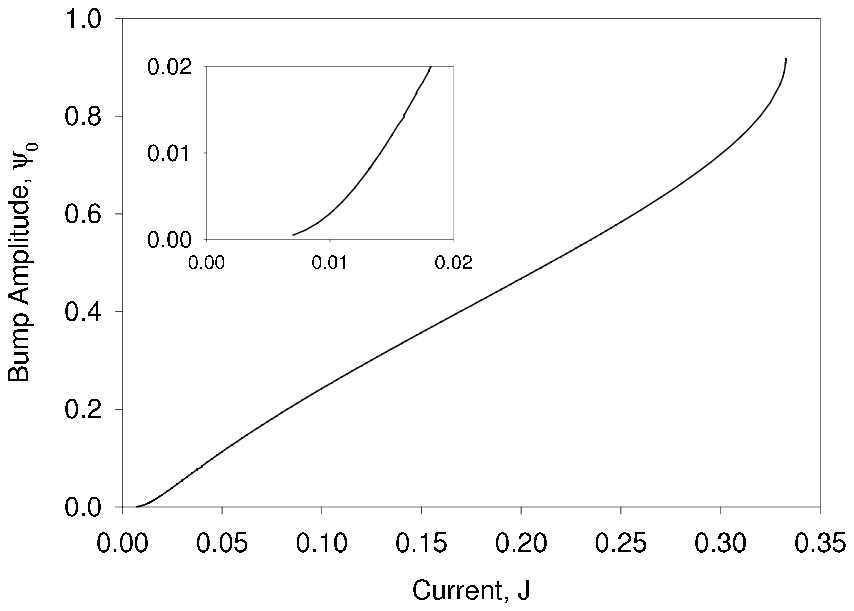}}
    \ncap[Maximum Amplitude of Bumps at $u=5.79$]{The
    maximum amplitude of the bumps $\psi_0$ as a function of
    $J$ for $u=5.79$. The inset shows the exponential dependence of
    the small-$J$ data, see Eq.~(\ref{expon-depend}).\label{nuc-2}}
    \end{figure}

Since the nuclei are unstable stationary states, they are
investigated only by time-independent means.
 Such solutions require a particular gauge choice---in this case
$\mu(x)=0$ where $\psi(x)$ has its maximum amplitude; they are
then sought using a relaxation algorithm \cite{num-rec}.
 Figure~\ref{nuc-1} shows a typical bump's amplitude, $f=
\sqrt{R^2+I^2}$, the associated superfluid velocity $q=
(RI_x-IR_x)/f^2$ and the electric field $E(x)=-\mu_x(x)$.
 The figure shows only half of the solution; $f(x)$, $q(x)$
and $E(x)$ are even about $x=0$.
 In Fig.~\ref{nuc-2} we plot the bump's maximum amplitude,
$\psi_0$, as a function of $J$; it grows as the current
rises, indicating the increasing stability of the normal phase.
 In the data presented by Watts-Tobin {\it et al.}\,\cite{watts-tobin81}
$\psi_0$ appears to vary linearly with $J$ for small $J$.
 However, in our numerical calculations at very small currents the
dependence deviates from linearity (see the inset of Fig.~\ref{nuc-2}),
and $\psi_0$ drops rapidly to zero as $J\rightarrow 0$, consistent
with the exponential behavior suggested in
Refs.~\cite{ivlev84,ivlev80}.
More precisely our small-$J$ data ($0.008\leq J\leq 0.015$) at $u=5.79$ is fit by
\begin{equation}
\label{expon-depend}
\psi_0(J) = B \exp ( - A/uJ),
\end{equation}
with $A=0.042$ and $B=0.19$.
A somewhat similar dependence (with $A=2/3$) was suggested by Ivlev
{\it et al.} \cite{ivlev84,ivlev80}; they were considering a distinct
quantity but one also related to critical fluctuations about the
normal phase (see the Appendix for more details).

\begin{figure}
    \centerline{
    \includegraphics{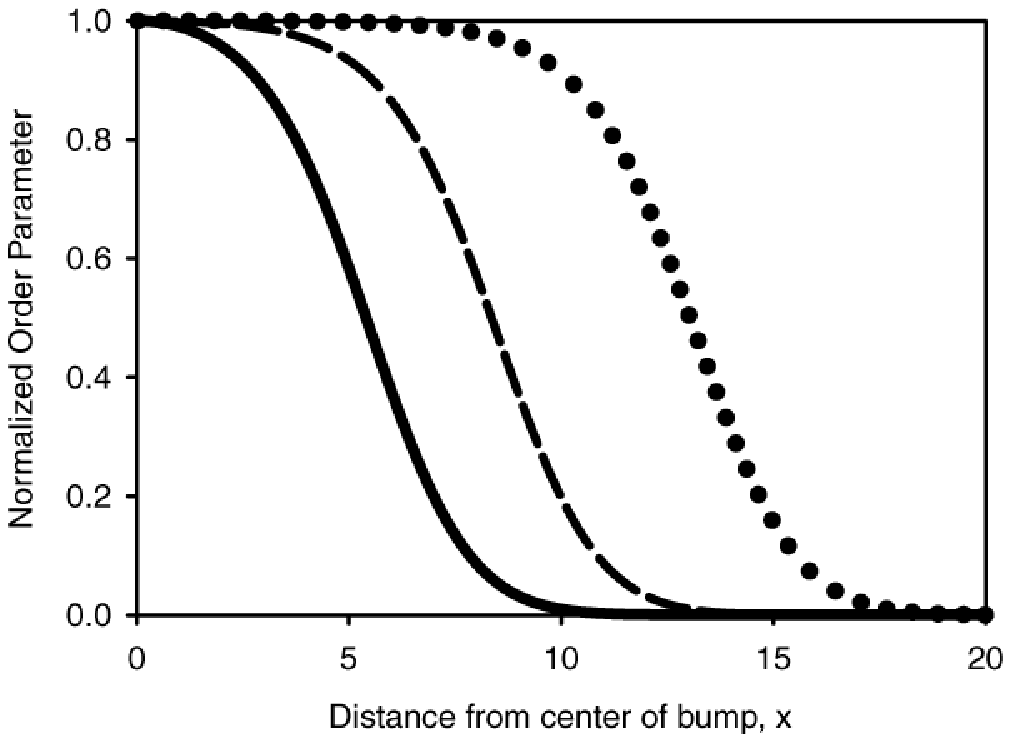}}
    \ncap[Bump Profiles as $J$ approaches $J^*$]{The
    bump profiles for $J^*-J=10^{-3}$ (solid),
    $J^*-J=10^{-5}$ (dashed) and $J^*-J=10^{-7}$ (dotted) at
    $u=5.79$.\label{nuc-3}}
    \end{figure}

The width of the bump diverges in the small-$J$ limit like
$(uJ)^{-1/2}$, as can be seen from the analysis below.
 So as $J$ is increased from zero, the width initially shrinks, but
eventually the width begins to grow again, diverging as the current
approaches the stall current $J^*$.
 In this limit the bump transforms into two well separated
interfaces (see Fig.~\ref{nuc-3}).

\subsection{Analysis in the $J\rightarrow 0$ Limit}
\label{nuc-smallj}

The equations for nuclei centered at the origin are
%\begin{mathletters}
\begin{eqnarray}
\label{lump-eq}
&& \psi_{xx} -iu \mu \psi +\psi -|\psi|^2\psi =0,  \\
&& \mu = -Jx + \int_0^x {\rm Im} \left(\psi^* \psi_{x'} \right)
dx' ,
\end{eqnarray}
%\end{mathletters}
where we have dropped the term $\psi_t$ and selected the gauge
$\mu(0)=0$.
 We saw in Fig.~\ref{nuc-2} that $\psi_0$ becomes very tiny
in the small-$J$ limit, thus the nonlinear terms can be neglected,
leading to
\begin{equation}
\psi_{xx} +  iuJx\psi + \psi = 0,
\label{1}
\end{equation}
a complex version of the Airy equation.
 Applying the WKB method results in the approximate solution
\begin{eqnarray}
\psi \sim &&[1+(uJx)^2]^{-1/8} \nonumber \\
\times &&\exp\left\{ \frac{2}{ 3\,uJ} \left[
\left[1 + (uJx)^2 \right]^{3/4} \cos \frac{3\alpha}{2}  -
1 \right] \right\} \nonumber \\
\times &&\exp \left\{i\left[ \frac{2}{3\, uJ}
\left[1 + (uJx)^2\right]^{3/4}\sin \frac{3\alpha}{2}
-\frac{\alpha}{4}\right] \right\},
\end{eqnarray}
where $\alpha = \tan^{-1}(uJx)$.
 The numerical data agrees quite well with this predicted shape in
the small-$J$ limit.
 For small $x$ the expression can be approximated by
\begin{equation}
\psi\sim \exp \left[i(1-uJ/4)x - uJx^2/4 \right].
\label{4}
\end{equation}
We see here that the width of the bump varies like $(uJ)^{-1/2}$
in this limit and that the superfluid velocity
$q \approx (1-uJ/4)$.
 For large $x$, on the other hand, where $\alpha \approx \pi/2$,
the expression becomes\typeout{multiple airy label.}
\begin{equation}
\label{airy} \psi\sim (uJx)^{-1/4} \exp \left[ - {\sqrt{2uJ}\over
3} |x|^{3/2}(1-i)\right], %\label{5}
\end{equation}
as one expects for the Airy function.
 Note that deep in the tail of the solution, we see a different
length scale $\lambda_{Airy} \sim (uJ)^{-1/3}$ arising.

Since the above analysis is of a linear equation, it can
not determine the amplitude of the nucleus;  for this purpose the
nonlinearities must be considered.
 In the appendix we outline an {\it ad hoc} calculation of the
small-$J$ limit of the bump amplitude.
 We take a $\psi$ of an unknown amplitude but of a fixed shape
inspired by the above analysis and assume that it is a stationary
solution of the full TDGL.
 We then determine its amplitude self-consistently.
 The resulting amplitude is
\begin{equation}
\label{self}
|\psi| \approx
\left(\frac{2J}{\pi u}\right)^{1/4}
\left(\frac{9}{8}  - \frac{1}{u} \right)^{-1/2}
\exp \left(- \frac{16}{81~uJ} \right).
\end{equation}
The factor $A=16/81$ is within a few percent of that
extracted from the numerical data.

\section{Stationary Interfaces}
\label{stationary}

As the current is raised, the width of the critical nucleus
grows and ultimately diverges as the stall current is reached,
resulting in well separated, stationary interfaces.
These interface solutions will be the subject of the rest of this
work.

\subsection{Numerical Methods and Results}
\label{stat-num}

Let us first discuss the numerical work on the interface
solutions.
 For given values of $u$ and $J$ we evolved the TDGL equations from an
initial guess which is purely superconducting on the left, $\psi(x)=
f_{\infty}~{\rm e}^{iq_{\infty}x}$ and  $\mu_x(x)=0$, and purely
normal on the right, $\psi(x)=0$ and $\mu_x(x)=-J$.
 The values $f_{\infty}$ and $q_{\infty}$ are related to the
applied current through
%\begin{mathletters}
\label{boundary-relations}
\begin{eqnarray}
J &=& f_{\infty}^2 \sqrt{1-f_{\infty}^2}, \\
q_{\infty} &=& \sqrt{1-f_{\infty}^2}.
\end{eqnarray}
%\end{mathletters}
Stability requires taking the larger positive root of the former
equation \cite{langer67} which places the following bounds on $J$,
$f_{\infty}$ and $q_{\infty}$:
%\begin{mathletters}
\label{bounds}
\begin{eqnarray}
&&0 \leq J \leq J_c = \sqrt{4/27} \approx 0.3849,
\label{j-bound} \\
&&1 \geq f_{\infty} \geq \sqrt{2/3} \approx 0.8165,
\label{f-bound} \\
&&0 \leq q_{\infty} \leq \sqrt{1/3} \approx 0.5774.
\label{q-bound}
\end{eqnarray}
%\end{mathletters}
We employed several schemes to integrate the equations in time
including both explicit (Euler) and implicit (Crank-Nicholson)
\cite{num-rec}.

Initially the front moves and changes shape but eventually it reaches
a steady state in which the interface moves at a constant velocity
without further deformation.
 By the time-dependent means we found locally stable,
constant-velocity solutions for currents less than $J_c$.
 To examine these solutions more accurately, we adopted a
time-independent method.
 First, we transformed coordinates to a moving frame, $x^{\prime}=
x-ct$; next, we chose $\mu=cq_{\infty}$ as $x\rightarrow -\infty$
which allows for a truly time-independent solution (i.e.\ one with
both amplitude and phase time-independent).
 Then we searched for stationary solutions using a relaxation
algorithm \cite{num-rec} where ($u,J$) are input parameters and $c$ is
treated as an eigenvalue.
 This approach requires an additional boundary condition to fix
translational invariance; we elected to fix $\mu$ on the rightmost
site.
 To find the stall currents $J^*$ we can set $c=0$ and take $J$ or $u$
as the eigenvalue.

\begin{figure}
    \centerline{
    \includegraphics{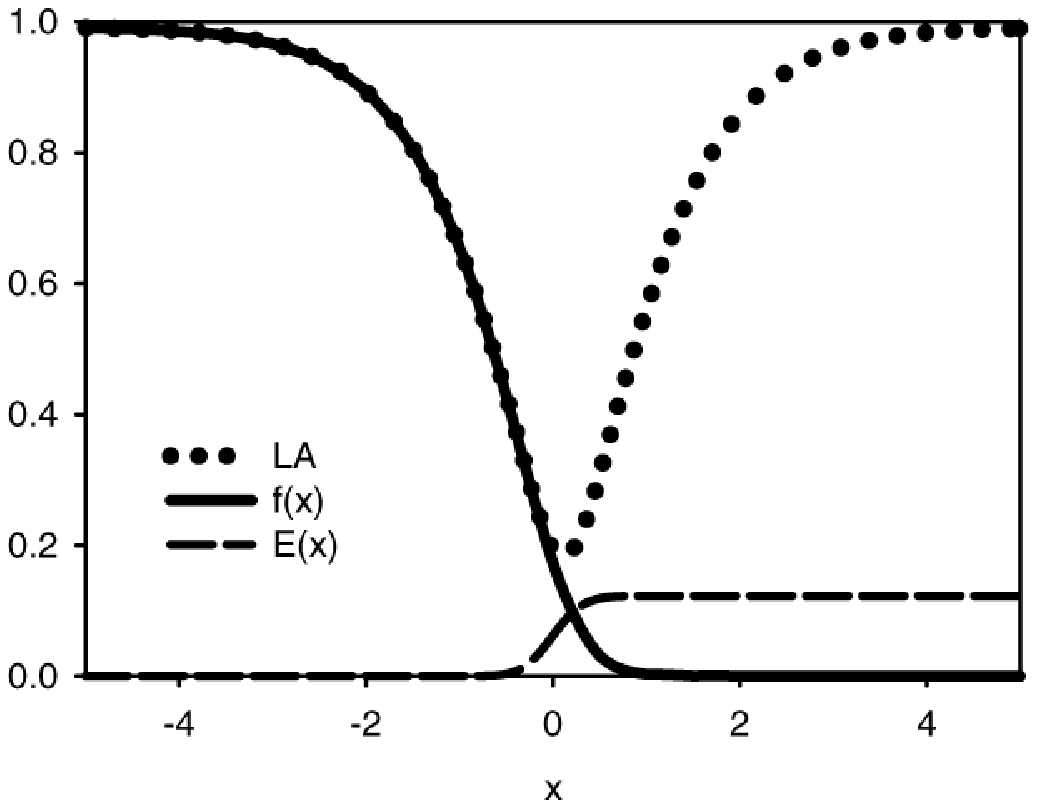}}
    \ncap[Stationary Normal-superconducting Interface for Large $u$]{The
    stationary NS interface solution when $u=500$ for which
    the stall current $J^*=0.12252$. Shown here are the numerically
    determined $f(x)$ and $E(x)$, as well as the Langer-Ambegaokar
    (LA) solution (Eq.~(\ref{langer}), the solution with no electric
    field) corresponding to the same current.\label{fig1}}
    \end{figure}

\begin{figure}
    \centerline{\includegraphics{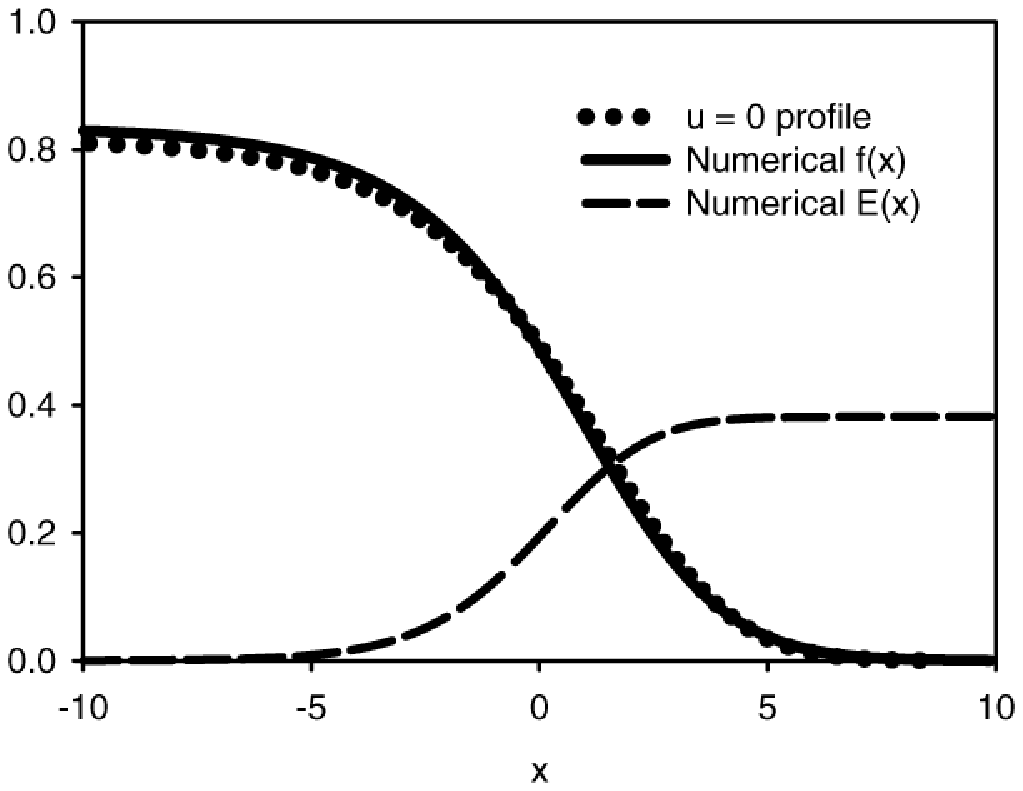}}
    \ncap[Stationary Normal-superconducting Interface for $u=1.04$]{The
    stationary NS interface solution when $u=1.04$ for which
    the stall current $J^*=0.3836$. Shown here are the numerically
    determined $f(x)$ and $E(x)$, as well as the function
    ${\hat f}_0(\hat x)$, the $u \rightarrow 0$ profile, derived
    from Eq.~(\ref{profile-inverse}) where $\hat x=
    u^{1/2}x$.\label{fig2}}
    \end{figure}

Figures \ref{fig1} and \ref{fig2} show the order-parameter amplitude
$f$ and the electric-field distribution $E=-\mu_x$ of the stalled
interface determined for $u=500$ and $u=1.04$ respectively.
 Note that while $f$ is very flat in the superconducting region, the
real and imaginary parts, $R$ and $I$, oscillate with a wavelength
$2 \pi /q_{\infty}$.
 Because of this additional length scale inherent in $R$ and $I$,
there is little to be gained from varying the mesh size.
 In fact, this length scale is compressed as we move to the right,
and we are only saved from the difficulties of handling rapidly
oscillating functions by the fact that the amplitudes decay so
quickly.

In the large-$u$ case (see Fig.~\ref{fig1}), $E(x)$ remains flat
throughout most of the space; it changes abruptly from one constant
to another only after $f(x)$ has become small.
 The variations of $f(x)$ are more gradual; however, the
greatest changes in $f_x$ occur in that same small area.
 This region of rapid change is known as a {\it boundary layer};
it marks where the current suddenly changes from superconducting to
normal, i.e., the position of the NS interface.
 As $u$ increases, the longer length scale over which $f$ varies
on the superconducting side remains essentially fixed, while
the boundary-layer thickness shrinks to zero.
 In the opposite limit, the small-$u$ case (see Fig.~\ref{fig2}),
$f(x)$ and $E(x)$ appear to vary together even in the superconducting
region; moreover, the length scale over which they vary grows as $u$
is decreased.
 We will postpone providing more of the numerical results on the
interfaces until some of the analytic arguments are available
for comparison.

\subsection{Asymptotic Analysis of the Interface Solutions:
preliminaries}
\label{stat-prelim}

Before addressing the large-$u$ and small-$u$ limits separately,
let us put the TDGL equations into a form convenient for analysis
and derive expressions for the length scales deep in the
superconducting and normal regions.
 The disparity of these length scales in the large-$u$ limit will
motivate the boundary-layer analysis in that regime; while an
inequality they satisfy will lead to the conclusion that
$J^* \rightarrow J_c$ in the small-$u$ limit.

We make the substitution $\psi = f e^{i\theta}$, which
yields
%\begin{mathletters}
\label{analytic}
\begin{eqnarray}
&& uf_t = f_{xx} - f (\theta_x)^2   +f - f^3 ,
\label{analytic-1} \\
&&u(\theta_t+\mu )f= 2f_x\theta_x + f \theta_{xx},
\label{analytic-2} \\
&& J =  f^2 \theta_x - \mu_x.
\label{analtyic-3}
\end{eqnarray}
%\end{mathletters}
Next we restrict our attention to stationary solutions.
 Note that only spatial derivatives of $\theta$ appear now, allowing
us to work with the superfluid velocity $q=\theta_x$ instead of
$\theta$.
 The equations become
%\begin{mathletters}
\label{fandq}
\begin{eqnarray}
&&  f_{xx} - q^2 f  +f - f^3=0 ,
\label{fandq-1} \\
&& u \mu f= 2f_x q + f q_{x},
\label{fandq-2} \\
&& J^* =  f^2 q - \mu_x,
\label{fandq-3}
\end{eqnarray}
%\end{mathletters}
where $J^*$ replaces $J$ as these equations apply to the stall
situation.
 Next multiply Eq.~(\ref{fandq-2}) by $f$ and note that the right
hand side is now $(f^2q)_x$ which we can express in terms of $\mu$
by differentiating Eq.~(\ref{fandq-3}); these steps lead to
\begin{equation}
\mu_{xx}=u f^2 \mu .
\label{fandmu}
\end{equation}

Now let us assume the following asymptotic forms as
$x\rightarrow -\infty$:
%\begin{mathletters}
\label{series}
\begin{eqnarray}
\label{series-f}
\lim_{x \to -\infty} f(x) &=& f_{\infty} -
f_1 ~{\rm e}^{x/\lambda_{f}} + \ldots , \\
\label{series-q}
\lim_{x \to -\infty} q(x) &=& q_{\infty} +
q_1 ~{\rm e}^{x/\lambda_{q}} + \ldots , \\
\label{series-mu}
\lim_{x \to -\infty} \mu(x) &=& \mu_{\infty} -
\mu_1 ~{\rm e}^{x/\lambda_{\mu}} + \ldots.
\end{eqnarray}
%\end{mathletters}
Substituting these expressions into Eqs. (\ref{fandq-1}),
(\ref{fandq-3}) and (\ref{fandmu}) and recalling the boundary
conditions yields (is label ``approach'' correct?)
%\begin{mathletters}
\begin{eqnarray}
\label{approach-1} &&\left(2f_{\infty}^2-\lambda_f^{-2} \right)
f_1 {\rm e}^{x/\lambda_f} -2 f_{\infty} q_{\infty}q_1 {\rm
e}^{x/\lambda_q} =0,
\\
\label{approach-2}
&&-2f_{\infty} q_{\infty} f_1 {\rm e}^{x/\lambda_f}+
f_{\infty}^2q_1 {\rm e}^{x/\lambda_q} +
\lambda_{\mu}^{-1} \mu_1 {\rm e}^{x/\lambda_{\mu}}=0, \\
\label{approach-3}
&&\lambda_{\mu}^{-2} - u f_{\infty}^2=0.
\end{eqnarray}
%\end{mathletters}
Eq.~(\ref{approach-3}) provides an expression for $\lambda_{\mu}$,
the electric-field screening length.
 Since $f_{\infty}$ is always of $O(1)$, we see that
$\lambda_{\mu}$ shrinks as $u \rightarrow \infty$ and diverges as
$u \rightarrow 0$, which is consistent with the behavior seen in
Figs.~\ref{fig1} and \ref{fig2}.

%{\it  A growing $\lambda_{\mu}$ implies that the normal component
%of the current, $\mu_x$, is more readily engaged; that $\lambda_{\mu}$
%grows when $u$ gets small is consistent since $u$ varies inversely as
%the normal conductivity. }

More than one decay length appears in  Eqs.~(\ref{approach-1}) and
(\ref{approach-2}).
 If they are not equal, the term with the shorter length is
exponentially small compared to the other(s) and will not
contribute to the $x \rightarrow -\infty$ limit.
 Since none of the terms in Eq.~(\ref{approach-2}) can equal
zero individually, we conclude that the longer two of $\lambda_f$,
$\lambda_q$ and $\lambda_{\mu}$ must be equal.
 Next, because the term multiplying ${\rm e}^{x/\lambda_q}$ in
Eq.~(\ref{approach-1}) cannot equal zero on its own, we determine
that $\lambda_q \leq \lambda_f$, making $\lambda_f$ one of the longer
lengths.
 Finally, if we assume that $\lambda_f=\lambda_{\mu} > \lambda_q$ we
find that $\lambda_f=2^{-1/2}f_{\infty}^{-1}$ and $\lambda_{\mu}=
u^{-1/2}f_{\infty}^{-1}$ and reach a contradiction (except at $u=2$).
 Thus provided the original assumption of an exponential approach is
valid, we conclude that
\begin{equation}
\label{inequality}
\lambda_f=\lambda_q \geq \lambda_{\mu}.
\end{equation}
This equality of $\lambda_f$ and $\lambda_q$ is reasonable given that
both $f$ and $q$ are related to the complex order parameter $\psi$.
 Also, having $\lambda_f > \lambda_{\mu}$ is consistent with the
large-$u$ data seen in Fig.~\ref{fig1}.
 If $\lambda_{\mu} \ne \lambda_f$ then
\begin{equation}
\label{LA-length}
\lambda_f^{-2} = 6 f_{\infty}^2 -4 = \lambda_{LA}^{-2}.
\end{equation}
We identify this length scale as $\lambda_{LA}$ since it
coincides with that occurring in the solution of Eqs. (\ref{fandq})
without any electric field ($\mu(x)=0$),
\begin{equation}
\label{langer}
f^2(x)=   f_{\infty}^2 - (3f_{\infty}^2 -2)
~{\rm sech}^2 \left(  \sqrt{\frac{3f_{\infty}^2-2}{2}}
x \right)  ,
\end{equation}
which was found by Langer and Ambegaokar \cite{langer67} in their
study of phase slippage.
 The asymptotic form of Eq.~(\ref{langer}) looks like
Eq.~(\ref{series-f}) with $\lambda_f$ given by Eq.~(\ref{LA-length}).
 As a matter of fact because $\lambda_f \gg \lambda_{\mu}$ in the
large-$u$ limit, the profile of $f(x)$ is only imperceptibly different
from the Langer-Ambegaokar (LA) solution in the superconducting
region and deviates from it only in the boundary layer, as is shown
in Fig.~\ref{fig1}.

Recall that $\lambda_{\mu}$ diverges as $u \rightarrow 0$; the
inequality $\lambda_f \geq \lambda_{\mu}$ implies that $\lambda_f$
must diverge as fast or faster in this limit.
 This scenario is consistent with the small-$u$ data shown in
Fig.~\ref{fig2} in which $f(x)$ and $E(x)$ vary on long length
scales.
 Eq.~(\ref{LA-length}) suggests that a diverging $\lambda_f$ implies
that $f_{\infty} \rightarrow \sqrt{2/3}$ and in turn that
$J \rightarrow J_c$ as $u \rightarrow 0$, which is also consistent
with what is found numerically.

In the other asymptotic limit, deep in the normal regime, $\psi$ is
very small and hence the nonlinear terms in Eqs.~(\ref{scaled}) can
be dropped as was done for the bumps in the small-$J$ limit.
 The result is a complex Airy equation, the asymptotic analysis of
which was supplied in Eq.~(\ref{airy}), where we saw the length
scale $\lambda_{Airy} \sim (u J^*)^{-1/3}$.
 Somewhat like $\lambda_{\mu}$, $\lambda_{Airy}$ shrinks as
$u \rightarrow \infty$ and expands as $u \rightarrow 0$ but with
different powers of $u$.
 The presence of the disparate length scales, $\lambda_f$,
$\lambda_{\mu}$ and $\lambda_{Airy}$, in the large-$u$ limit
motivates the use of the boundary-layer analysis that comes next.
 We will see that $\lambda_{Airy}$ scales in the same way as
the boundary-layer thickness.

\subsection{Asymptotic Behavior of the Stall Current as
$u\rightarrow\infty$}
\label{stat-largeu}

We have already seen in Fig.~\ref{fig1} that the large-$u$
profile can be divided into two regions---one slowly varying,
one rapidly varying, also known as the {\it outer} and {\it inner}
regions, respectively.
 Furthermore, it has been suggested that the ratio of the length
scales characterizing these regions decreases as $u \rightarrow
\infty$.
 These features make the problem ideally suited for
boundary-layer analysis, in which one identifies the terms that
dominate the differential equation in each region, analyzes the
reduced equations consisting of dominant terms and then matches
the behavior in some intermediate region.

We start by eliminating the superfluid velocity $q$  from
Eqs.~(\ref{fandq}), resulting in
%\begin{mathletters}
\begin{eqnarray}
\label{feqn-1}
&& f_{xx} - (J^* + \mu_x)^2  f^{-3}   +f - f^3 = 0, \\
\label{mueqn-1}
&& \mu_{xx} - u f^2\mu = 0.
\end{eqnarray}
%\end{mathletters}
Let us consider first the slowly varying, superconducting region.
 We saw in the preliminary analysis that for large $u$, $\mu(x)$
is exponentially small, so we drop it.
 Next, let us assume that $J^*$ is small and drop it; we can verify
in the end that this is self-consistent.
 The reduced equation is
\begin{equation}
\label{outer}
f_{xx}+f-f^3\approx 0,
\end{equation}
with solution $f(x)=-{\rm tanh}(x/\sqrt{2})$.

Moving in from the left toward the interface (into the
boundary-layer region), $f$ becomes small, and the second
term in Eq.~(\ref{feqn-1}) which was subdominant becomes dominant.
 In this inner region $f$ is small but rapidly varying, thus the
dominant terms are
\begin{equation}
\label{reduced-f}
f_{xx} \approx \frac{(J^*+\mu_x)^2}{f^3},
\end{equation}
along with Eq.~(\ref{mueqn-1}).
 Having identified the dominant terms, now we must make certain
they balance.
 We assume that in the boundary layer, all the quantities scale
as powers of $u$:
\begin{equation}
f \sim u^{-\alpha},  \quad \mu \sim u^{-\beta}, \quad
J^* \sim u^{-\gamma},  \quad x \sim u^{-\delta}.
\end{equation}
Balancing terms in Eq.~(\ref{reduced-f}), we find
$2\alpha = \gamma + \delta$, while balancing terms in
Eq.~(\ref{mueqn-1}) yields $2(\alpha + \delta) = 1$.
 Next, we need to ensure that the solutions in the boundary layer
match onto the solutions in the superconducting and normal regions.
 By expanding the superconducting solution near the interface, we see
that $f(x)\sim - x/\sqrt{2}$ as the boundary layer is approached;
matching to the boundary layer requires $f_x\sim 1$, so that
$\alpha = \gamma$.
 In the normal region, $\mu \approx - J^* x$, so that matching to the
boundary layer requires $\mu_x \sim J^*$, and $\beta = \gamma +
\delta$.
 Solving this set of equations, we conclude that $\alpha=\gamma=
\delta=1/4$ and $\beta=1/2$, i.e., the stall current $J^* \sim
u^{-1/4}$ for large $u$.
 Note that $J^* \rightarrow 0$ as $u \rightarrow \infty$, so
that we were justified in dropping $J^{*2}/f^3$ from
Eq.~(\ref{outer}).
 Substituting $J^* \sim u^{-1/4}$ into $\lambda_{Airy}$ gives
$\lambda_{Airy} \sim u^{-1/4}$, indicating that it may be
identified as the boundary-layer thickness.

    \begin{figure}[p]
    \centerline{
    \includegraphics{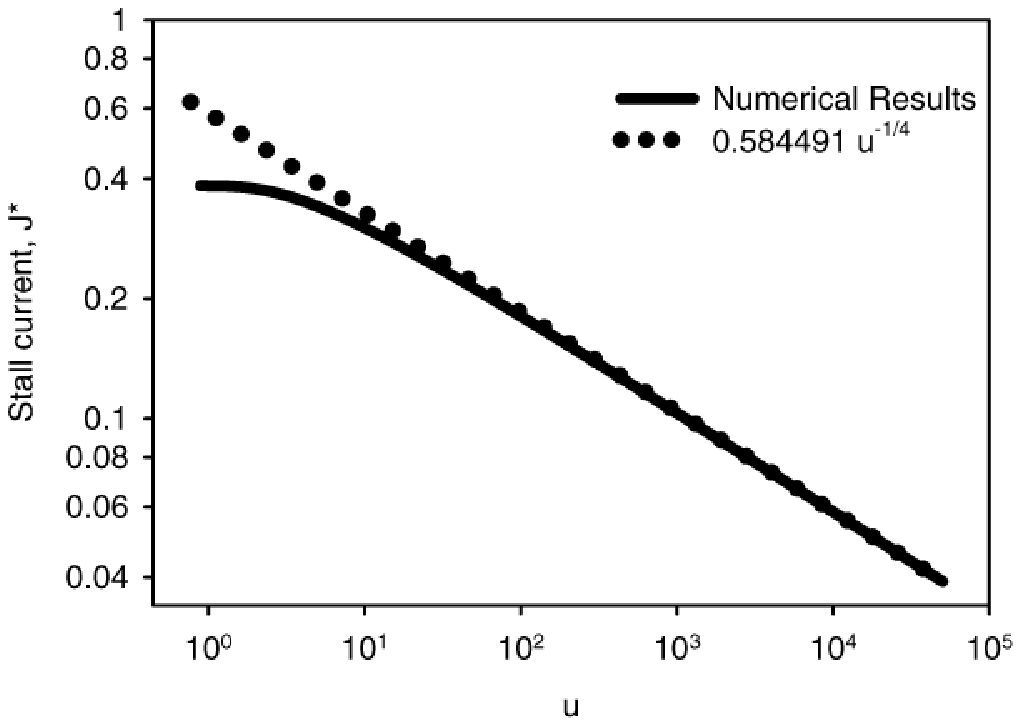}}
    \ncap[Asymptotic Dependence of Stall Current]{A log-log plot
    of the stall current $J^*$ vs. $u$.  The
    solid line shows the numerically determined $J^*$'s as a function
    of $u$ and the dotted line is $0.584491 ~u^{-1/4}$ (the large-$u$
    behavior predicted by matched asymptotic analysis).\label{drew3}}
    \end{figure}

    \begin{table}[p]
    \centerline{\begin{tabular}{lllll}
     $u $ & $J^*$   & $ J^* u^{1/4} $ & $ \eta $ & $\eta u^{-3/4}$ \\
    \hline 1     & 0.3838  & 0.3838   & 0.01871  & 0.01871 \\ 5     &
    0.3407  & 0.5094   & 0.6400  & 0.1914 \\ 10    & 0.3013  & 0.5359
    & 1.573  & 0.2797 \\ 50    & 0.2127  & 0.5655   & 8.258 & 0.4315
    \\ 100   & 0.1807  & 0.5715   & 15.59  & 0.4931 \\ 500   & 0.1224
    & 0.5788   & 62.51  & 0.5875 \\ 1000  & 0.1033  & 0.5807   & 111.3
    & 0.6259 \\ 5000  & 0.0693  & 0.5828   & 407.9  & 0.6847 \\ 10000
    & 0.0583  & 0.5833   & 708.4  & 0.7084 \\ 50000 & 0.0391  & 0.5840
    & 2487  &  0.7440
    \end{tabular}}
    \ncap[Numerical Results for Stall Current and Kinetic Coefficient]{Representative
    numerical results for the stall current $J^*$
    and kinetic coefficient $\eta$.\label{table2}}
    \end{table}
In order to determine the coefficient of the $u^{-1/4}$ in the
stall current we need to solve the boundary layer problem. Let us
rescale in the way suggested above:
\begin{eqnarray}
&&f = u^{-1/4}F,\quad \mu = u^{-1/2} M(X), \nonumber \\ &&J =
u^{-1/4} \tilde J, \quad x=u^{-1/4} X. \label{linner1}
\end{eqnarray}
Substituting these rescaled variables into Eqs.~(\ref{feqn-1}) and
(\ref{mueqn-1}), and then expanding $F$, $M$ and $\tilde J$ in powers
of $u^{-1/2}$, we obtain at the lowest order
%\begin{mathletters}
\begin{eqnarray}
\label{linner2} &&F_{0,XX} - \frac{({\tilde J}_0 + M_{0,X})^2}{
F_0^3}  =0,  \\ \label{linner3} &&M_{0,XX} -  F_0^2 M_0 = 0,
\end{eqnarray}
%\end{mathletters}
with the boundary conditions (from the outer regions)
\begin{eqnarray}
F_{0,X}(-\infty) & = -1/\sqrt{2},
\qquad M_0(-\infty) & = 0, \\
F_0(+\infty) & = 0,
\qquad M_{0,X}(+\infty) & = -{\tilde J}_0.
\label{inner_boundary}
\end{eqnarray}
(As before we need an extra boundary condition to fix the
translational invariance.) For an arbitrary ${\tilde J}_0$ the
solutions of Eqs.~(\ref{linner2}) and (\ref{linner3}) cannot
satisfy the boundary conditions; ${\tilde J}_0$ must be tuned to a
particular value before all of the boundary conditions are
satisfied, leading to a {\it nonlinear eigenvalue problem} for
${\tilde J}_0$. We have solved this eigenvalue problem numerically
and find that ${\tilde J}_0= 0.584491$.
 Therefore, to leading order we have for the stall current
\begin{equation}
J^* \approx 0.584491\, u^{-1/4}.
\label{asympt}
\end{equation}
This prediction agrees well with the numerical results and
disagrees with Likharev's conjecture of a $u^{-1/3}$ dependence
\cite{likharev74}, as can be seen in Fig.~\ref{drew3} and in Table
\ref{table2}. It is in principle possible to carry out this
procedure to successively higher orders, but the equations become
cumbersome. Instead we have simply opted to fit our numerical data
to a form inspired by the asymptotic analysis,
\begin{eqnarray}
J^* &=& 0.584491\, u^{-1/4} -  0.117461\, u^{-3/4}
- 0.12498\, u^{-5/4} \nonumber \\
&&+ 0.163043\, u^{-7/4}
+O\left(u^{-9/4}\right).
\end{eqnarray}
Even with the higher order terms, the asymptotics are appropriate
only for physically large values of $u$.

\subsection{Asymptotic Behavior of the Stall Current as
$u\rightarrow 0$}
\label{stat-smallu}

Now let us examine the opposite limit of $u \rightarrow 0$.
 In this case the electric-field screening length becomes long, and
Ivlev {\it et al.} \cite{ivlev80b} have proposed that this makes
the small-$u$ limit useful for modeling gapped superconductors.
 As already suggested the inequality of length scales, $\lambda_f
\geq \lambda_{\mu}$ implies that $J^* \rightarrow J_c$.
 We will begin our small-$u$ analysis by putting this result
on firmer ground and extracting as a byproduct the $u \rightarrow
0$ limit of the interface profile.

{\it The rescaled equations.}
Recall that deep in the superconducting region $\lambda_{\mu} \sim
u^{-1/2}$.
 This observation suggests that we rescale distance:
$x = u^{-1/2}\hat x$; furthermore, to ensure that the normal current
($-\mu_x$) scales in the same way as the total current we rescale
$\mu = u^{-1/2}\hat \mu$ as well.
 These rescalings yield
%\begin{mathletters}
\begin{eqnarray}
\label{louter1} && u \hat f_{\hat x \hat x} - {\hat q}^2 \hat f +
\hat f -{\hat f}^3 = 0, \\ \label{louter2} && \hat \mu \hat f = 2
\hat q {\hat f}_{\hat x} + \hat f {\hat q}_{\hat x}, \\
\label{louter3} && J^* = {\hat f}^2 \hat q - {\hat \mu}_{\hat x},
\end{eqnarray}
%\end{mathletters}
placing the small parameter $u$ in front of ${\hat f}_{\hat x
\hat x}$.
 If we expand these functions as series in powers of $u$
%\begin{mathletters}
\begin{eqnarray}
\label{lexpand1} \hat f &=& {\hat f}_0 + u{\hat f}_1  + \ldots, \\
\label{lexpand2} \hat q &=& {\hat q}_0 + u {\hat q}_1  + \ldots,
\\ \label{lexpand3} \hat \mu &=& {\hat \mu}_0 + u{\hat \mu}_1  +
\ldots, \\ \label{lexpand4} J^* &=& J_0^* + uJ_1^*  + \ldots,
\end{eqnarray}
%\end{mathletters}
then we find at the lowest order
%\begin{mathletters}
\begin{eqnarray}
\label{order11}
&&-{\hat q}_0^2 {\hat f}_0 + {\hat f}_0 - {\hat f}_0^3 = 0, \\
\label{order12}
&&{\hat \mu}_0 {\hat f}_0 = 2 {\hat q}_0 {\hat f}_{0,x} +
{\hat f}_0 {\hat q}_{0,x}, \\
\label{order13}
&&J_0^* = {\hat f}_0^2 {\hat q}_0 - {\hat \mu}_{0,\hat x}.
\end{eqnarray}
%\end{mathletters}

The solution of Eq.~(\ref{order11}) is either ${\hat f}_0 = 0$ (the
normal phase) or ${\hat f}_0 = (1-{\hat q}_0^2)^{1/2}$ (the
superconducting phase).
 Let us focus on the superconducting solutions.
 By eliminating ${\hat q}_0$, we obtain the first order equations
%\begin{mathletters}
\begin{eqnarray}
\label{order14}
{\hat f}_{0,\hat x} &=& \frac{{\hat f}_0 \sqrt{1-{\hat f}_0^2}
~{\hat \mu}_0}{2-3 {\hat f}_0^2}, \\
\label{order15}
{\hat \mu}_{0,\hat x} &=& {\hat f}_0^2 \sqrt{ 1 - {\hat f}_0^2} -
J_0^*.
\end{eqnarray}
%\end{mathletters}
Because ${\hat f}_0$ ranges from $f_{\infty}$ to $0$ and $f_{\infty}
\geq \sqrt{2/3}$, we know that ${\hat f}_0$ either starts at or passes
through $\sqrt{2/3}$.
 (Strictly speaking we should be writing here $f_{\infty,0}$, the
lowest order term in the expansion of $f_{\infty}$.)
 Thus, the effect of the pole in Eq.~(\ref{order14}) must be
considered.
 If it is not canceled by a zero in ${\hat \mu}_0$,
${\hat f}_{0,\hat x}$ diverges at ${\hat f}_0=\sqrt{2/3}$.

We can obtain an expression for ${\hat \mu}_0({\hat f}_0)$ by dividing
Eq.~(\ref{order15}) by Eq.~(\ref{order14}), which leads to
\begin{equation}
{\hat \mu}_0\, d{\hat \mu}_0 = \frac{\left[{\hat f}_0^2
\sqrt{1-{\hat f}_0^2 } - J^*_0\right]\left(2-3{\hat f}_0^2\right)}
{ {\hat f}_0 \sqrt{1-{\hat f}_0^2}}
d{\hat f}_0.
\label{combine}
\end{equation}
Integrating both sides and recalling the boundary condition
$\mu_{\infty}=0$, we find
\begin{eqnarray}
\frac{{\hat \mu}_0^2}{2} &=&
{\hat f}_0^2-\frac{3}{4}\,{\hat f}_0^4- f_{\infty}^2+
\frac{3}{4}\,f_{\infty}^4
\nonumber \\
&&+2\,J^*_0 \ln \left[ \frac {1+\sqrt {1-{\hat f}_0^2}}{{\hat f}_0}
\right]-3\,J^*_0\sqrt {1-{\hat f}_0^2}
\nonumber \\
&&-2\,J^*_0\ln \left[
\frac {1+\sqrt {1-f_{\infty}^2}}{ f_{\infty}}\right]
+3\,J^*_0\sqrt {1-f_{\infty}^2},
\label{mess}
\end{eqnarray}
where $J^*_0 = f_\infty^2 \sqrt{1-f_\infty^2}$.
 To keep ${\hat f}_{0,x}$ from diverging, we insist that
${\hat \mu}_0({\hat f}_0=\sqrt{2/3})=0$ which can be shown from
Eq.~(\ref{mess}) to imply $f_{\infty}=\sqrt{2/3}$, i.e.\
the small-$u$ limit of the stall current is the critical depairing
current.
 Note that the pole in Eq.~(\ref{order14}) and the compensating
zero in ${\hat \mu}_0({\hat f}_0)$  occur at the boundary
($x \rightarrow -\infty$).

We can rearrange Eq.~(\ref{order14}) as follows
\begin{equation}
\label{profile-inverse}
\int_{{\hat f}_0(0)}^{{\hat f}_0(\hat x)} \frac{(2-3f^2)~df}
{f \sqrt{1-f^2} ~{\hat \mu}_0(f)}= {\hat x}.
\end{equation}
Then we can substitute in Eq.~(\ref{mess}) for ${\hat \mu}_0(f)$,
numerically integrate the resulting expression and finally invert
it in order to calculate ${\hat f}_0(\hat x)$, the $u \rightarrow 0$
profile.
 Figure \ref{fig2} includes a comparison of ${\hat f}_0(\hat x)$ and
the profile of a small-$u$ numerical solution.

To find the asymptotic behavior of ${\hat f}_0$ and ${\hat \mu}_0$ in
the superconducting region, Taylor expand ${\hat \mu}_0({\hat f}_0)$
around $f_{\infty}$
\begin{equation}
{\hat \mu}_0({\hat f}_0) = -3 \sqrt{2} \left({\hat f}_0 -\sqrt{2/3}
\right)^2 +\ldots.
\end{equation}
Notice that ${\hat \mu}_0({\hat f}_0)$ is a second order zero, so
that ${\hat f}_{0, \hat x}=0$, as it should at the boundary.
 As a consequence, the integral supplying the inverse profile,
Eq.~(\ref{profile-inverse}), has a pole; integrating the
expression in its neighborhood yields $\sqrt{6} \ln (\sqrt{2/3}-
{\hat f}_0)$, leading to
\begin{equation}
\label{sol1}
{\hat f}_0(\hat x)\sim \sqrt{2/3} - A_0~{\rm exp}
\left( \hat x/\sqrt{6} \right),
\end{equation}
where $A_0$ is an integration constant undetermined because of the
translational invariance.
 Note ${\hat f}_0(\hat x)$ has the form assumed in the preliminary
analysis with $\lambda_{f,0}=\sqrt{u/6}$.
 Putting this result into Eq.~(\ref{order14}) leads to
\begin{equation}
\label{sol2}
{\hat \mu}_0(\hat x)\sim  - 3\sqrt{2} A_0^2~{\rm exp}
\left( 2 \hat x/ \sqrt{6}\right),
\end{equation}
where $\lambda_{\mu,0}=u^{1/2} f_{\infty}$ in agreement with the
expression found previously.

Let us examine Eqs.~(\ref{order14}) and (\ref{order15}), which
are strictly speaking superconducting solutions, in the normal
(small-${\hat f}_0$) limit.
 Eq.~(\ref{order15}) leads to ${\hat \mu}_0(\hat x)\approx -J_c \hat x$,
and inserting this into Eq.~(\ref{order14}) reveals that ${\hat f}_0
\rightarrow 0$ in the following way
\begin{equation}
\label{sol3}
{\hat f}_0(\hat x)\sim {\rm exp} \left( - J_c {\hat x}^2/4 \right).
\end{equation}
This same dependence was seen earlier in the analysis of the bump
shapes in the small-$J$ limit, Eq. (\ref{4}).

What is surprising here is that what are ostensibly the ``outer''
equations for the superconducting region also satisfy the boundary
conditions in the normal region and interpolate in between.
 This is consistent with the numerical observation that there does
not seem to be a boundary layer at small $u$, that the $u f_{xx}$
term is apparently {\it not} a singular perturbation.
 With this in mind, we pursue the perturbative expansion to higher
orders.

{\it The $O(u)$ equations.}
The eigenvalue $J^*_0$ was determined by examining the behavior deep
in the superconducting region and did not require imposing the
boundary conditions on the normal side.
 Furthermore, the spatial dependence of the solution in this
region is of the form assumed in Eqs.~(\ref{series}).
 We exploit these features to obtain higher order terms.
 The $O(u)$ equations are
%\begin{mathletters}
\label{order-u}
\begin{eqnarray}
&& {\hat f}_{0, \hat x \hat x} - 2 {\hat q}_0 {\hat f}_0
{\hat q}_1 - {\hat q}_0^2 {\hat f_1} - 3 {\hat f}_0^2 {\hat f}_1 = 0,
\\
&& {\hat \mu}_0 {\hat f}_1 + {\hat f}_0 {\hat \mu}_1 =
2 {\hat q}_0 {\hat f}_{1, \hat x} + 2 {\hat f}_{0, \hat x}
{\hat q}_1 + {\hat f}_0 {\hat q}_{1, \hat x} + {\hat q}_{0, \hat x}
{\hat f}_1 \\
&& J^*_1 = 2 {\hat f}_0 {\hat q}_0 {\hat f}_1 + {\hat f}_0^2
{\hat q}_1 - {\hat \mu}_{1, \hat x}.
\end{eqnarray}
%\end{mathletters}
The asymptotic form of ${\hat f}_0(x)$ is
\begin{equation}
{\hat f}_0(x) = {\hat f}_0^{(0)}+{\hat f}_0^{(1)}~{\rm e}^{x/\sqrt{6}}
+ {\hat f}_0^{(2)}~{\rm e}^{2x/\sqrt{6}} + \ldots
\end{equation}
and similarly for ${\hat q}_0(x)$ and ${\hat \mu}_0(x)$.
 Eqs.~(\ref{order-u}) can be satisfied if the asymptotic form
of ${\hat f}_1(x)$ is
\begin{eqnarray}
{\hat f}_1(x) = {\hat f}_1^{(0)}+&&\left({\hat f}_1^{(1)}+
{\hat g}_1^{(1)} \hat x \right)~{\rm e}^{x/\sqrt{6}} \nonumber \\
&&+ \left({\hat f}_1^{(2)}+{\hat g}_1^{(2)} \hat x \right)
~{\rm e}^{2x/\sqrt{6}} + \ldots,
\end{eqnarray}
and similarly for ${\hat q}_1(x)$ and ${\hat \mu}_1(x)$.
 At $O(u^2)$, ${\hat f}_2(x)$ would have second-order polynomials
multiplying the exponentials, and so on.
 Substituting these expressions into the differential equations
allows us to determine the unknown constants (except for those
associated with the translational invariance).
 For $f_{\infty}$ it yields the series
\begin{equation}
f_{\infty}= \sqrt{\frac{2}{3}} + \frac{u}{24 \sqrt{6}} +
+ \frac{u^2}{768 \sqrt{6}} + \ldots,
\end{equation}
which corresponds to
\begin{equation}
J^*=\frac{2}{3 \sqrt{3}} -\frac{u^2}{ 576 \sqrt{3}}
-\frac{u^3}{5184 \sqrt{3}} + \ldots.
\label{small_u_series}
\end{equation}
Note that the first correction to the $u \rightarrow 0$ limit
of $J^*$ is of $O(u^2)$, since the lowest term $J_c$
is at the maximum of $J^*(f_{\infty})=f_{\infty}^2
\sqrt{1-f_{\infty}^2}$.

\begin{figure}
     \centerline{
     \includegraphics{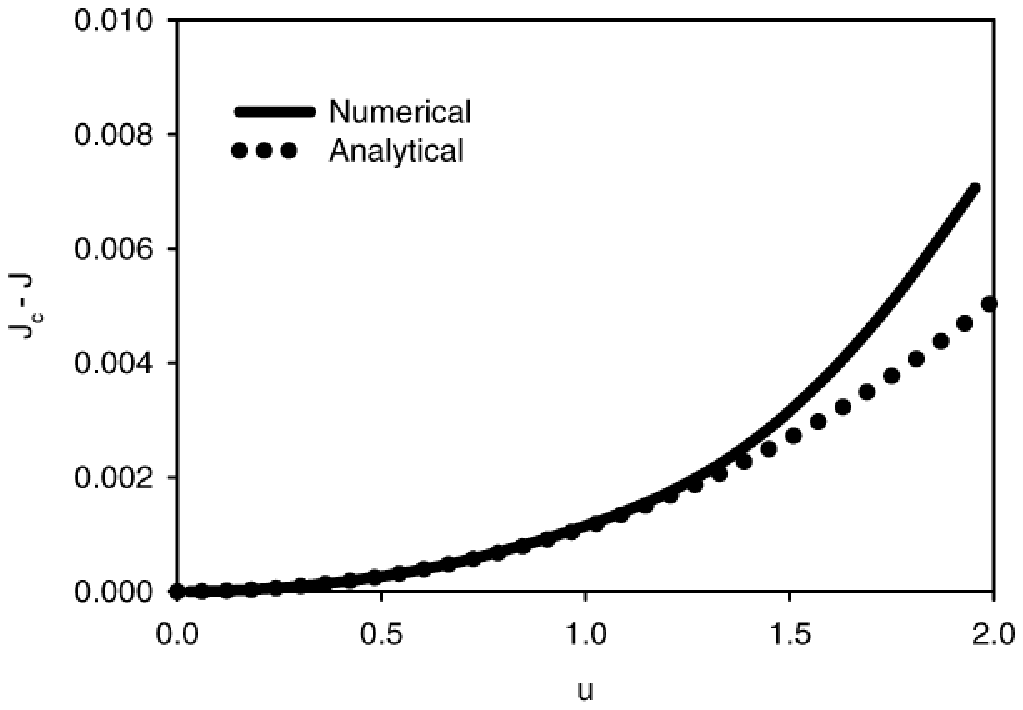}}
     \ncap[Asymptotics for Small-$u$ Analysis]{$(J_c-J)$ for the numerical data (solid line) and for the
     result of the small-$u$ analysis, Eq.~(\ref{small_u_stall})
     (dashed line).\label{small-u-fig}}
     \end{figure}

The series found through the asymptotic perturbative expansion
above can be obtained by another method.
 Looking back at Eqs. (\ref{sol1}) and (\ref{sol2}), we note that
the ratio of decay lengths $\lambda_f/\lambda_{\mu}=2$.
 If we insert the expressions we have for these length scales, Eqs.
(\ref{approach-3}) and (\ref{LA-length}), we find as $u \rightarrow
0$
\begin{equation}
\frac{ \lambda_{f}}{\lambda_{\mu}}=
\left[ \frac{uf_{\infty}^2}{6f_{\infty}^2-4}\right]^{1/2}=2.
\end{equation}
Solving for $f_{\infty}$, and then calculating $J^*$, we find
\begin{equation}
J^*= J_c (1-u/8)^{1/2} (1-u/24)^{-3/2},
\label{small_u_stall}
\end{equation}
with $J_c= \sqrt{4/27}$, which when expanded for small-$u$ agrees
with the series (\ref{small_u_series}) found above.
 We plot the small-$u$ numerical data and this expression together
in Fig.~\ref{small-u-fig}.
 The fit is surprisingly good at small $u$, suggesting to us that
the corrections to Eq.~(\ref{small_u_stall}) are exponentially small
as $u\rightarrow 0$.

\section{Moving Interfaces}
\label{moving}

At currents other than $J^*$, the NS interfaces move with a constant
velocity.
 For such solutions the operator $\partial_t$ can be replaced by
$-c\partial_x$, so that Eqs. (\ref{analytic}) become
%\begin{mathletters}
\label{speed}
\begin{eqnarray}
&& -cuf_x = f_{xx} - f q^2  +f - f^3 ,
\label{speed-1} \\
&&u(-cq+\mu )f= 2f_x q + f q_x,
\label{speed-2} \\
&& J =  f^2 q - \mu_x.
\label{speed-3}
\end{eqnarray}
%\end{mathletters}
While the boundary conditions on $f$ and $q$ remain the same,
that on the scalar potential becomes $\mu_{\infty}=cq_{\infty}$.
 Actually, it is more convenient to use instead $\tilde \mu =
\mu -c q$, which is the gauge-invariant potential in the
constant-velocity case.

\begin{figure}
     \centerline{
     \includegraphics{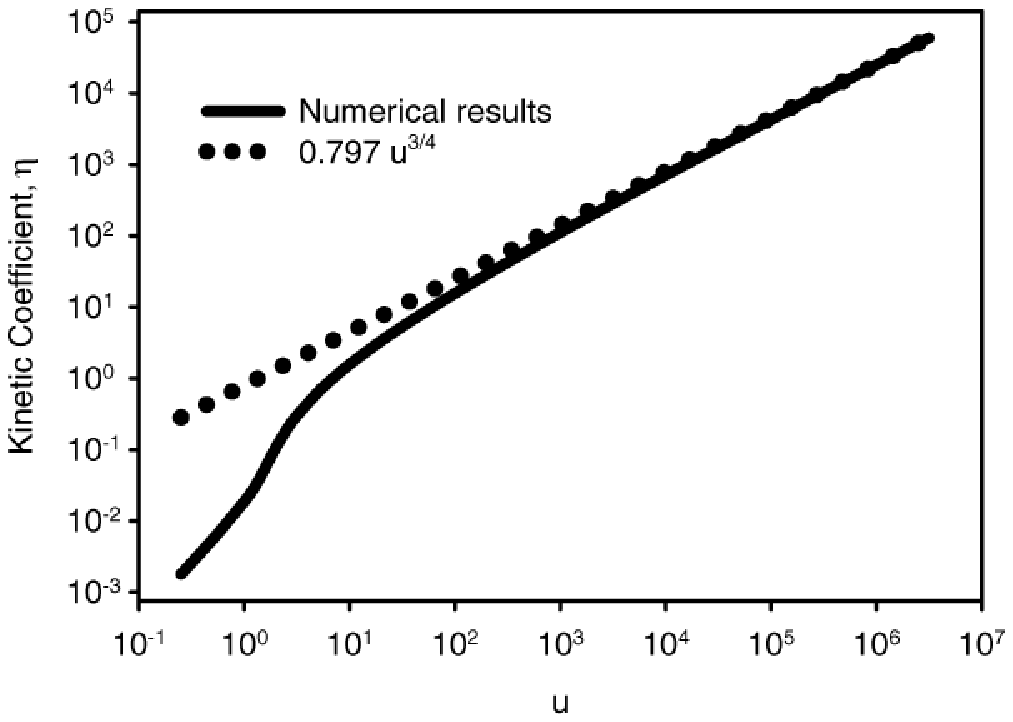}}
     \ncap[Asymptotics for Large-$u$ Kinetic Coefficient]{A
     log-log plot of the numerically determined kinetic
     coefficient as a function of $u$ (the solid line) along with an
     asymptotic fit of $0.797 u^{3/4}$ (the dotted line).\label{fig5}}
     \end{figure}

The superconducting phase invades the normal phase if $J < J^*$
and vice versa if $J > J^*$. For currents near $J^*$, the
interface speed is proportional to $(J-J^*)$. In this linear
response regime, one can define a kinetic coefficient (which
Likharev \cite{likharev74} refers to as a ``viscosity")
    \begin{equation}
    \eta = \left( \fud{c}{J} \right)^{-1}_{J=J^*}.
    \end{equation}
Figure~\ref{fig5} shows the numerically determined kinetic
coefficient as a function of $u$. For large-$u$, we find $\eta
\sim u^{3/4}$ for which we provide an argument below.

The velocity of the interface does not depend on the direction of
the applied current.  This can be seen from the symmetry of the
current in the equations we use, but it seems not to make physical
sense.  In fact an article by Gurevich and Mints~\cite{gurevich81}
addresses effects of the collision of quasiparticles with the NS
boundary, and they do cause a small asymmetry in the normal zone
propagation.  The TDGL do not, however, account for kinetics
across a phase boundary, so the effect is not included in
Likharev's equations.

\begin{figure}
     \centerline{
     \includegraphics{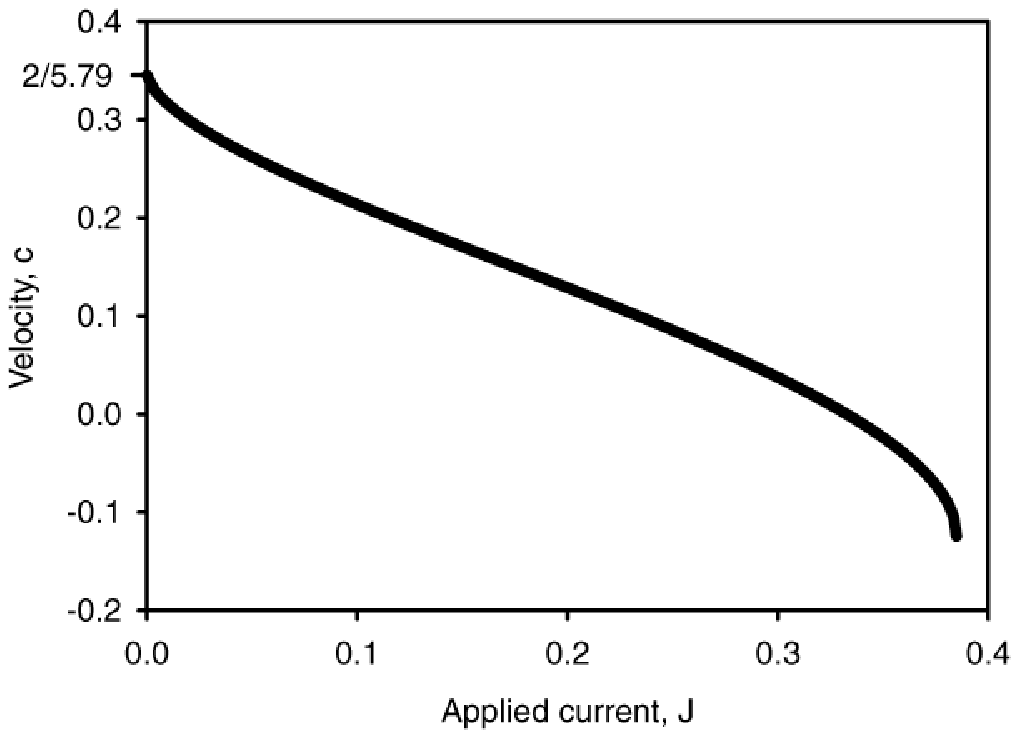}}
     \ncap[Velocity of the Front Versus Current for $u=5.79$]{The
     velocity of the front versus the current $J$ for
     $u=5.79$.\label{fig6}}
     \end{figure}

Farther from  the stall current, the velocities deviate from this
linear behavior, as seen in Fig.~\ref{fig6}.
 The greatest departure occurs in the limits $J \rightarrow 0$
and $ J \rightarrow J_c$.
 In fact, Likharev \cite{likharev74} conjectured that the interface
speed diverges in both of these limits; we find that it is bounded.

{\it The $J \rightarrow 0$ limit}.
The moving interface equations, Eqs. (\ref{speed}), simplify in
the $J \rightarrow 0$ limit, since that limit implies that
both $q \rightarrow 0$ and $\mu \rightarrow 0$, leaving only
\begin{equation}
f_{xx} + uc f_x +f -f^3 = 0.
\label{fisher-kpp}
\end{equation}
If we replace $uc$ in the above equation by a speed $v$, then
we have the steady-state version of Fisher's equation
\cite{fisher-kpp}, which is known to have propagating front
solutions with $v=2$ \cite{aronson78}.
 In our case this implies that as $J \rightarrow 0$, $c=2/u$,
which is in good agreement with the numerical data shown in
Fig.~\ref{fig6}.

We can combine the above result with an earlier one to suggest
that $\eta\sim u^{3/4}$ as $u \rightarrow \infty$.
 In the large-$u$ limit, we have information on the following two
points: (1)  the stalled interface ($J=J^*\sim u^{-1/4},c=0$); and
(2) the interface in the absence of current ($J=0,c=2/u$).
 In going from (1) to (2), the changes in current and velocity are
$\Delta J \sim u^{-1/4}$ and $\Delta c  \sim u^{-1}$.
 As $u \rightarrow \infty$, both of these changes are small so that
$\eta$ might be approximated by
\begin{equation}
\eta \approx {\Delta J \over \Delta c } \sim  u^{3/4},
\end{equation}
yielding the behavior seen in the numerical data
(see Fig.~\ref{fig5} and Table~\ref{table2}).

\begin{figure}
     \centerline{
     \includegraphics{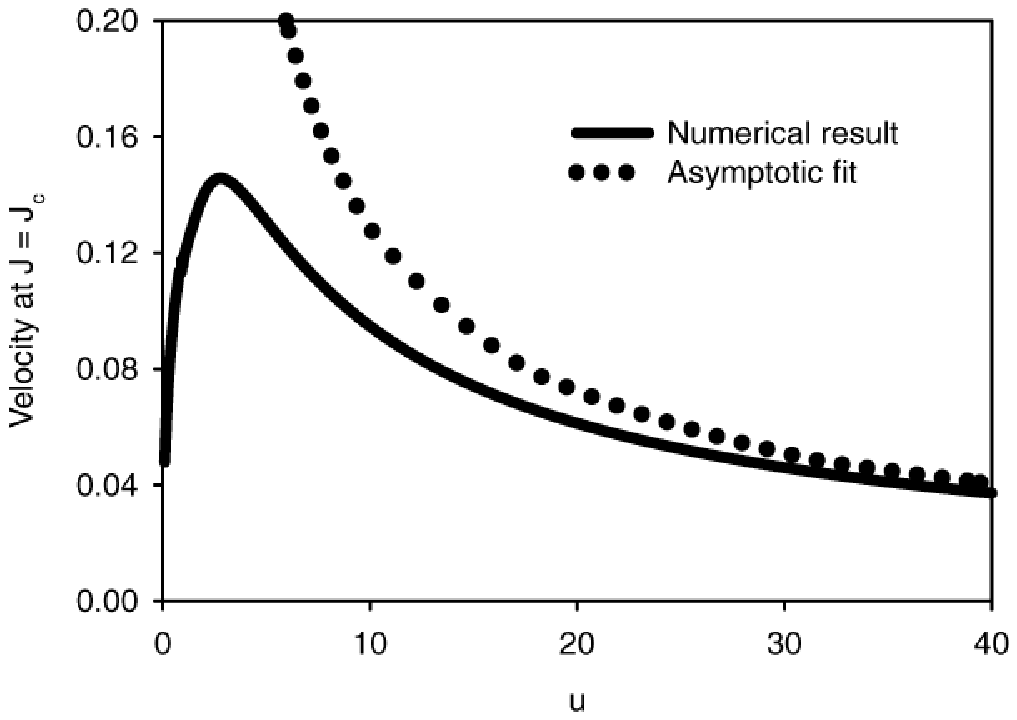}}
     \ncap[Velocity as $J\rightarrow J_c$ as a Function of $u$]{The
     velocity as $J \rightarrow J_c$ as a function of $u$.
     For large $u$, the velocity asymptotically approaches $0.92
     u^{-0.85}$.\label{fig7}}
     \end{figure}

{\it The $J \rightarrow J_c$ limit.}
 The numerical work indicates that the velocity is finite as
$J \rightarrow J_c$; the limiting velocity is shown in
Fig.~\ref{fig7} as a function $u$.
 We can find an analytic bound on this velocity as follows.
 First, take Eqs.~(\ref{speed}), use the gauge-invariant potential
$\tilde \mu$, and find the constant-velocity analog of
Eq.~(\ref{fandmu}).
 Then substitute the asymptotic forms, Eqs.~(\ref{series}), into
the resulting equations, leading to
%\begin{mathletters}
\begin{eqnarray}
&&\left(cu\lambda_f^{-1} + \lambda_f^{-2}-2f_{\infty}^2 \right)
f_1 {\rm e}^{x/\lambda_f} = 2 f_{\infty}q_{\infty}q_1
{\rm e}^{x/\lambda_q}, \\
&&\left( uf_{\infty}^2 -\lambda_{\mu}^{-2}\right)\tilde \mu_1
{\rm e}^{x/\lambda_{\mu}} = c q_1 \lambda_q^{-2}
{\rm e}^{x/\lambda_q}, \\
&& 2f_{\infty}q_{\infty}f_1 {\rm e}^{x/\lambda_f} +
\left( f_{\infty}^2 -c \lambda_q^{-1}\right)q_1
{\rm e}^{x/\lambda_q} \nonumber \\
&& \ \ \ \ \ \ \qquad \qquad \qquad
- \tilde \mu_1 \lambda_{\mu}^{-1}
{\rm e}^{x/\lambda_{\mu}}=0.
\end{eqnarray}
%\end{mathletters}
Arguments similar to those following
Eqs.~(\ref{approach-1}--\ref{approach-3}) lead one to the
conclusion that in this case $\lambda_f=\lambda_q=\lambda_{\mu}$.
 The above equations can then be shown to yield the following
relation
\begin{eqnarray}
\label{c-bound}
&&u^2 c^2 +\left(2u\lambda^{-1}
-2uf_{\infty}^2\lambda -u^2 f_{\infty}^2 \lambda \right)c
\nonumber \\
&&+\left[2 \left(uf_{\infty}^2 \lambda^2-1 \right)
\left(3 f_{\infty}^2-2 \right)  -
uf_{\infty}^2 +\lambda^{-2}\right]=0,
\end{eqnarray}
where we have used $q_{\infty}^2=1-f_{\infty}^2$.
 We find the bound by:  (1) solving Eq.~(\ref{c-bound}) for $c$;
(2) substituting in $f_{\infty}=\sqrt{2/3}$ (which corresponds to
$J=J_c$); and (3) extremizing that result with respect to the
decay length $\lambda$.
 The small-$u$ limit of the resulting bound is $-\sqrt{2u}/9$,
and the large-$u$ limit is $-1/2\sqrt{3}$.
 The square-root dependence of the velocity in the small-$u$
limit agrees with the data.
 Now we can consider going from the stall current $(J^*,c=0)$
to the critical depairing current $(J_c,c\sim u^{1/2})$ which
results in changes $\Delta J \sim u^2$ and $\Delta c \sim u^{1/2}$,
suggesting that the small-$u$ kinetic coefficient $\eta \sim u^{3/2}$,
which is in rough agreement with the numerical data.
 We have also observed that as a function of $J$ the speed appears
to approach its bound via a square root dependence
\begin{equation}
c(J)= A + B (J_c-J)^{1/2}
\end{equation}
for all $u$.

%%%%%%%%%%%%%%%%%%%%%%

\section{Summary and Remarks}
\label{conclusions}

In this chapter we have studied in detail the nucleation and
growth of the superconducting phase in the presence of a current.
The finite amplitude critical nuclei grow as the current is
increased, with the amplitude eventually saturating as the stall
current $J^*$  is approached, leading to the formation of
interfaces separating the normal and superconducting phases. The
stall current can be calculated in the limit of large $u$ using
matched asymptotic expansions, demonstrating once again the
utility of this technique for problems in inhomogeneous
superconductivity. We have also derived an analytic expression for
the stall current for small $u$, which we believe to be correct up
to exponentially small corrections. Deviations from the stall
current cause the interfaces to move, and we have calculated the
mobility of these moving interfaces for a wide range of $u$.
Finally we have shown that the interface velocity $c= 2/u$ as
$J\rightarrow 0$ and that $c$ is bounded as $J\rightarrow J_c$, in
contrast to some conjectures in the literature.

As in the magnetic-field analogy, the issue of stability and
dynamics of the current-induced NS interfaces will be more
complicated and interesting in the two-dimensional case. Some
preliminary work in this direction has been reported by Aranson
{\it et al.}\cite{aranson96}, who find that the current has a
stabilizing effect on the NS interface. This can be interpreted as
a positive surface tension for the interface, due entirely to {\it
nonequilibrium} effects. They provide a heuristic derivation of an
interesting free-boundary problem for the interfacial dynamics (a
variant of the Laplacian growth problem);  however, this
free-boundary problem is sufficiently complicated that they were
unable to solve it to compare with their numerical results.
Clearly, further work in this direction would be helpful in
understanding the nucleation and growth of the superconducting
phase in two-dimensional superconducting films.

\appendix
\newcommand{\ba}{\mathbf{A}}

%\begin{document}
\chapter{Detailed Derivation of the TDGL}
\thispagestyle{empty}
\section{Derive TDGL from \GL Free Energy}
\label{sec:derivetdgl}
This note shows the details of deriving the time-dependent GL equations from
the GL free energy.
\begin{gather}
    \pad{\psi}{t} = -\Gamma\frac{\delta G}{\delta \psi^*} =
    -\Gamma\left(\pad{}{\psi^*}-\pad{}{x_i}\pad{}{\pad{\psi^*}{x_i}}\right)G
\end{gather}
Start with the GL free energy
\begin{equation}
 f = f_n + \alpha|\psi|^2+\frac{\beta}{2}|\psi|^4+\frac{|\mathbf{h}|^2}{8\pi}+
 \frac{1}{2m_s}\left|\left(-i\hbar\del-\frac{e_s\mathbf{A}}{c}\right)\psi\right|^2.
\end{equation}
Now work out the functional derivative with respect to $\psi^*$
\begin{align}
    & \pad{}{\psi^*}(\psi^*\psi)  = \psi \\
    & \pad{}{\psi^*}(\psi^{*2}\psi^2) = |\psi|^2\psi \\
    & \pad{}{\psi^*}\left[(i\hbar\del+\frac{e}{c}\mathbf{A})\psi
        \cdot(-i\hbar\del+\frac{e}{c}\mathbf{A})\psi^*\right] =
        \frac{e}{c}\mathbf{A}\cdot(i\hbar\del+\frac{e}{c}\mathbf{A})\psi \\
    & \pad{}{x_i}\pad{}{\pad{\psi^*}{x_i}}\left[(i\hbar\del+\frac{e}{c}\mathbf{A})\psi
        \cdot(-i\hbar\del+\frac{e}{c}\mathbf{A})\psi^*\right] =
        -i\hbar\del\cdot(i\hbar\del+\frac{e}{c}\mathbf{A})\psi
\end{align}
Put it together to find
\begin{align}
    \frac{\delta G}{\delta\psi^*} & = \alpha\psi+\beta|\psi|^2\psi
    +\frac{1}{2m}\left(\frac{2ie\hbar}{c}\mathbf{A}\cdot\del\psi+\left(\frac{e}{c}\mathbf{A}\right)^2
    \psi-\hbar^2\del^2\psi\right) \\
    & = \alpha\psi+
        \beta|\psi|^2\psi+\frac{1}{2m}\left(i\hbar\del+\frac{e}{c}\mathbf{A}\right)^2\psi
\end{align}
Now do the next independent variable, $\mathbf{A}$.  First we write out some
simple definitions for reference.
\begin{align}
    \del\times\ba  = &\left(\pad{A_z}{y}-\pad{A_y}{z}\right)\hat{i}
    -\left(\pad{A_z}{x}-\pad{A_x}{z}\right)\hat{j}
    +\left(\pad{A_y}{x}-\pad{A_x}{y}\right)\hat{k} \\
    \del\times\del\times\ba = &
    \left[\pad{}{y}\left(\pad{A_y}{x}-\pad{A_x}{y}\right)
    +\pad{}{z}\left(\pad{A_z}{x}-\pad{A_x}{z}\right)\right]\hat{i} \\
    & -\left[\pad{}{x}\left(\pad{A_y}{x}-\pad{A_x}{y}\right)
    -\pad{}{z}\left(\pad{A_z}{y}-\pad{A_y}{z}\right)\right]\hat{j} \\
    & +\left[-\pad{}{x}\left(\pad{A_z}{x}-\pad{A_x}{z}\right)
    -\pad{}{y}\left(\pad{A_z}{y}-\pad{A_y}{z}\right)\right]\hat{k}
\end{align}
Now let's look at the functional derivative of $(\del\times\ba)^2$.
When we look at a single component, it becomes a little clearer.
\begin{align}
    \pad{}{x}\pad{(\del\times\ba)^2}{\pad{A_x}{x}} & = 0 \\
    \pad{}{y}\pad{(\del\times\ba)^2}{\pad{A_x}{y}} & =
        \pad{}{y}2(\del\times\ba)\cdot(-\hat{k}) \\
    \pad{}{z}\pad{(\del\times\ba)^2}{\pad{A_x}{z}} & =
        \pad{}{z}2(\del\times\ba)\cdot\hat{j}
\end{align}
You can see we are reconstructing a cross product
\begin{gather}
    \frac{\delta(\del\times\ba)^2}{\delta A_x} =
    2\left[\pad{(\del\times\ba)_z}{y}-
    \pad{(\del\times\ba)_y}{z}\right]
\end{gather}
to get
\begin{gather}
    \frac{\delta}{\delta\ba}(\del\times\ba)^2 = 2\del\times\del\times\ba
\end{gather}
Similarly,
\begin{gather}
    \frac{\delta}{\delta\ba}(\del\times\ba)\cdot\mathbf{H}
 = \del\times\mathbf{H}
\end{gather}
Now figure out the rest of the derivative for $\ba$.
\begin{gather}
\pad{}{\ba}\left[(i\hbar\del+\frac{e}{c}\mathbf{A})\psi
        \cdot(-i\hbar\del+\frac{e}{c}\mathbf{A})\psi^*\right] =
        \frac{e}{c}\psi(-i\hbar\del+\frac{e}{c}\mathbf{A})\psi^*
        + \text{c.c.}.
\end{gather}
We can add the two pieces together to find the current
\begin{gather}
-\frac{ie\hbar}{2mc}(\psi\del\psi^*-\psi^*\del\psi)+\frac{e^2}{mc^2}\ba|\psi|^2
\end{gather}
Putting the second equation together, we find
\begin{gather}
    \frac{\delta G}{\delta\ba} = \frac{1}{4\pi}\del\times(\del\times\ba-\mathbf{H})
    -\frac{ie\hbar}{2mc}(\psi\del\psi^*-\psi^*\del\psi)+\frac{e^2}{mc^2}\ba|\psi|^2
\end{gather}

\section{Find Dimensionless Variables}
\label{sec:dimensionless}
We start now with the known form of the TDGL,
\begin{gather}
    \frac{1}{\Gamma}\left(\pad{\psi}{t}+\frac{ie}{\hbar}\phi\psi\right)
        +\alpha\psi+\beta|\psi|^2\psi
        +\frac{1}{2m}(i\hbar\del+\frac{e}{c}\ba)^2\psi = 0 \\
    \frac{\sigma}{c}\pad{\ba}{t}
        +\sigma\del\phi
        +\frac{c}{4\pi}\del\times\del\times\ba
        -\frac{ie\hbar}{2m}(\psi\del\psi^*-\psi^*\del\psi)+
        \frac{e^2}{mc}\ba|\psi|^2 = 0
\end{gather}
We begin by requiring $\psi$ vary between zero and one so that $\psi=\psi_0\psi'$ where
$\psi_0 = \sqrt{-\smash[b]{\alpha/\beta}}$.  That determines the overall factor on the
first equation.
\begin{gather}
    \frac{\alpha}{\alpha\Gamma}\left(\pad{\psi'}{t}+\frac{ie}{\hbar}\phi\psi'\right)
        +\alpha\psi'-\alpha|\psi'|^2\psi'
        -\alpha(\frac{i\hbar}{\sqrt{-2m\alpha}}\del+\frac{e}{\sqrt{-2m\alpha c^2}}\ba)^2\psi' = 0
\end{gather}
We can immediately determine $\ba=A_0\ba'$ to be
\begin{gather}
    A_0 =\sqrt{\frac{-2m\alpha c^2}{e^2}}.
\end{gather}
We also see that the gradient's factor is of the form
\begin{gather}
    \frac{\hbar}{\sqrt{-2m\alpha}}\frac{1}{\lambda} = \frac{1}{\kappa} =
    \frac{\xi}{\lambda}
\end{gather}
where $x=\lambda x'$.
\begin{gather}
    -\frac{1}{-\alpha\Gamma}\left(\pad{\psi'}{t}+\frac{ie}{\hbar}\phi\psi'\right)
        +\psi'-|\psi'|^2\psi'
        -(\frac{i}{\kappa}\del'+\ba')^2\psi' = 0
\end{gather}

Looking now to the second equation, we can divide through by $A_0$ to see
\begin{gather}
    \frac{\sigma}{ct_0}\pad{\ba'}{t'}+\sqrt{\frac{-e^2}{2m\alpha c^2}}\frac{\sigma\phi_0}{\lambda}\del'\phi'+
        \frac{c}{4\pi \lambda^2}\del'\times\del'\times\ba'
        +\frac{ie\hbar\alpha}{2m\lambda\beta}\sqrt{\frac{-e^2}{2m\alpha c^2}}
            (\psi'\del'\psi^{'*}-\psi^{'*}\del'\psi')
        -\frac{e^2\alpha}{mc\beta}\ba'|\psi'|^2 = 0
\end{gather}
The last term is the only one we already know.  Multiplying all terms by the
inverse of its prefactor gives
\begin{gather}
    -\frac{\sigma}{ct_0}\frac{mc\beta}{e^2\alpha}\pad{\ba'}{t'}
    +\frac{m\sigma\phi_0}{\kappa\hbar e|\psi_0|^2}\del'\phi'
    -\frac{mc^2\beta}{4\pi e^2\alpha\lambda^2}\del'\times\del'\times\ba'
        -\frac{i}{2\lambda}\sqrt{\frac{-\hbar^2}{2m\alpha}}
        (\psi'\del'\psi^{'*}-\psi^{'*}\del'\psi')
        +\ba'|\psi'|^2 = 0
\end{gather}
Now we know from the third term that we must define
\begin{gather}
    x_0 =\lambda =  \sqrt{\frac{-mc^2\beta}{4\pi e^2\alpha}}.
\end{gather}
With that, the third term simplifies dramatically to yield
\begin{gather}
    \frac{4\pi \sigma\lambda^2}{c^2t_0}\pad{\ba'}{t'}
    +\frac{m\sigma\phi_0}{\kappa\hbar e|\psi_0|^2}\del'\phi'
    +\del'\times\del'\times\ba'
        -\frac{i}{2\kappa}(\psi'\del'\psi^{'*}-\psi^{'*}\del'\psi')
        +\ba'|\psi'|^2 = 0
\end{gather}
The last definitions look clear.
\begin{equation}
    t_0 = \frac{4\pi\sigma\lambda^2}{c^2}\qquad
    \text{and}\qquad
    \phi_0 = \kappa|\psi_0|^2\frac{\hbar e}{\sigma m}
\end{equation}
If we return to the equation for $\psi$ to finish changing its variables, we
find
\begin{equation}
    \frac{\alpha}{\alpha\Gamma}\left(\pad{\psi'}{t}+\frac{ie}{\hbar}\phi\psi'\right)
    = \frac{1}{\alpha t_0\Gamma}\left(\pad{\psi'}{t'}
    +\frac{ie\phi_0t_0}{\hbar}\phi'\psi'\right).
\end{equation}
We are hoping to find that
\begin{equation}
    \frac{ie\phi_0t_0}{\hbar} = i\kappa = i\frac{\lambda}{\xi},
\end{equation}
which it is.  Our final equations, when this becomes
clearer, will be (dropping primes)
\begin{gather}
    \gamma\left(\pad{\psi}{t}+i\kappa\phi\psi\right)
        -\psi+|\psi|^2\psi
        +(\frac{i}{\kappa}\del+\ba)^2\psi = 0 \\
    \pad{\ba}{t}
    +\del\phi
    +\del\times\del\times\ba
        -\frac{i}{2\kappa}(\psi\del\psi^*-\psi^*\del\psi)
        +\ba|\psi|^2 = 0
\end{gather}

\chapter{Likharev's Equations as an Active Kinetic Equation}

\label{sec:diffusion}

Likharev's equations behave like nonlinear diffusion equations.
They don't appear to be the same at first glance, however.  We
would like to see whether they could resemble more traditional
diffusion equations, so we begin with Likharev's equations
\begin{eqnarray}
    u f_t & = & f_{xx} - f\theta_x^2+f-f^3 \label{lik1}\\
    u (\theta_t+\mu)f^2 & = & (f^2\theta_x)_x \label{lik2}\\
    J & = & f^2\theta_x - \mu_x.
\end{eqnarray}
and try to coax them into a form like that discussed in Gurevich
and Mints~\cite{gurevich89}
\begin{eqnarray}
    \tau_\psi\frac{\partial\psi}{\partial t} & = &
l_\psi^2\frac{\partial^2\psi}{\partial x^2}-F(\psi,\phi,\beta) \\
    \tau_\phi\frac{\partial\phi}{\partial t} & = &
l_\phi^2\frac{\partial^2\phi}{\partial x^2}-R(\psi,\phi,\beta)
\end{eqnarray}
where $\beta$ is a parameter representing our $(u,J)$.  We may be
able to put our equations in this form if we change variables from
$(f,\theta)$ to $(f,j_s)$ where $j_s = f^2\theta_x$ is the
supercurrent.  Equation~\ref{lik1} is already in the correct form
if we substitute $\theta_x = j/f^2$.
\begin{equation}
    u f_t  =  f_{xx} +f\left(1-f^2- \frac{j^2}{f^4}\right)
\end{equation}
If we take the derivative of \ref{lik2}, we get
\begin{equation}
    u(\theta_{xt}+\mu_x)f^2 + 2\frac{f_x}{f}(f^2\theta_x) = (f^2\theta_x)_{xx}
\end{equation}
We can get rid of $\theta$ by constructing the derivative of the supercurrent
\begin{equation}
    (f^2\theta_x)_t = 2ff_t\theta_x + f^2\theta_{xt}.
\end{equation}
We substitute $f_t$ from \ref{lik1}
\begin{equation}
    2uff_t\theta_x = 2f\theta_x(uf_t) = 2f\theta_x(f_{xx}-f\theta_x^2+f-f^3)
\end{equation}
into the derivative of the supercurrent to find
\begin{equation}
    u(f^2\theta_x)_t = (f^2\theta_x)_{xx} - 2\frac{f_x}{f}(f^2\theta_x)_x +
u(J-f^2\theta_x)f^2+2f\theta_xf_{xx}-2f^2\theta_x^3+2(1-f^2)f^2\theta_x.
\end{equation}
It is time to substitute $f^2\theta_x = j$ to find
\begin{equation}
    uj_t = j_{xx}-\frac{2}{f}(f_xj_x-f_{xx}j)-(2+u)f^2j+2j+uJf^2-\frac{2j^3}{f^4}.
\end{equation}
We could condense this a little to
\begin{equation}
    uj_t = j_{xx}-2\frac{j^2}{f}\left(\frac{f_x}{j}\right)_x+u(J-j)f^2+
    2j\left(1-f^2- \frac{j^2}{f^4}\right).
\end{equation}
What remains is of the form of an active kinetic system
\begin{equation}
    \pad{x_i}{t}=F_i(x_1,x_2,\dots,x_n)+
    \pad{}{r}\left(\sum_{j=1}^n D_{ij}\pad{x_j}{r}\right)\quad
    (i=1,2,\dots,n)
\end{equation}
as described in Vasil'ev, Romanovsk\u{\i\i}, and
Yakhno~\cite{vasiliev79}, but the numerous spatial derivatives
complicate the first integrals typically used to examine such
equations.

\chapter{Amplitude of the Critical Nuclei in the $J \rightarrow 0$
Limit} \label{appendix}

In this appendix we provide a self-consistent calculation of the
amplitude of the critical nuclei in the $J \rightarrow 0$ limit.
 Choosing the gauge appropriate for bumps centered at $x=0$
and combining Eqs.~(\ref{scaled}) into one equation yields
\begin{eqnarray}
\label{start} &&\left[-u \partial_t +iuJx + \partial_x^2  +1
\right] \psi(x,t) = |\psi(x,t)|^2~ \psi(x,t) \nonumber \\ && \ \ \
\ \ \ +iu \left[\int_0^{x} dy ~{\rm Im} \left(\psi^*(y,t)
\partial_y \psi (y,t) \right) \right] \psi(x,t).
\end{eqnarray}
The propagator for the linear operator appearing on the left hand
side of Eq.~(\ref{start}) satisfies the condition
\begin{eqnarray}
\left[-u \partial_t  +iuJx + \partial_x^2  +1 \right]&&
G(x,x';t-t') \Theta(t-t') \nonumber \\ = &&-u~ \delta(x-x')
\delta(t-t')
\end{eqnarray}
and is given by
\begin{eqnarray}
G(x,x';\tau) = &&\left( \frac{u}{4 \pi \tau} \right)^{1/2} \exp
\left[ \frac{\tau}{u} - \frac{J^2 \tau^3}{12u} \right] \nonumber
\\ && \times \exp \left[ \frac{iJ \tau(x+x')}{2} -
\frac{u(x-x')^2}{4 \tau} \right] .
\end{eqnarray}
Ivlev {\it et al.} \cite{ivlev80,ivlev84} used this linear
propagator to evolve perturbations having widths of $O(1)$ and
carrying no current.
 Without the nonlinear terms such perturbations initially grow
but ultimately reach a maximum size and then decay away.
 Ivlev {\it et al.} suggested that the amplitudes of the critical
nuclei are exponentially small in the $J \rightarrow 0$ limit by
asking what sized initial perturbations are of $O(1)$ at their
maxima.
 Their arguments motivated us to use the propagator in a more
careful estimate of the amplitude that includes the nonlinear
terms as an essential ingredient.

We can convert Eq.~(\ref{start}) into an integral equation by
multiplying both sides of Eq.~(\ref{start}) (with $x \rightarrow
x'$ and $t \rightarrow t'$) by $ G(x,x';t-t')$ and integrating
over all $x'$ and integrating $t'$ from $0$ to $t$.
 After some manipulation these steps lead to
\begin{eqnarray}
\label{final}
 \psi(x,t) && =
\int_0^t dt' \int_{-\infty}^{\infty} dx'  ~G(x,x',t-t') \nonumber
\\ \times&&\left\{ \psi(x',t')\delta(t-t')  -\frac{1}{u}
|\psi(x',t')|^2~ \psi(x',t') \right. \nonumber \\
 -i  && \left. \left[\int_0^{x'} dy ~{\rm Im} \left(
\psi^*(y,t') \partial_y \psi(y,t') \right) \right] \psi(x',t')
\right\} ,
\end{eqnarray}
where $t>0$.

In order to estimate the amplitude of the threshold solutions, we
will substitute into Eq.~(\ref{final}) the following form
\begin{equation}
\label{form} \psi(x,t) = \psi_0 \exp \left\{  -\frac{uJx^2}{4} +
ix \right\}.
\end{equation}
Note that this form is stationary and has a fixed Gaussian shape
(which is inspired by our WKB approximation, see Eq.~(\ref{4}))
but it has an arbitrary amplitude which we will determine
self-consistently.

Let us take the $t \rightarrow \infty$ limit and focus on $x=0$
since our interest is in the amplitude.
 After substituting Eq.~(\ref{form}) into Eq.~(\ref{final}), we
can do both integrals for the first term on the right hand side
exactly, and it can be seen to decay to zero in the $t \rightarrow
\infty$ limit.
 Next, we perform the $x'$ integration of the second term on the
right hand side ($II$), which yields
\begin{equation}
II=-\frac{\psi_0^3}{uJ} \int_0^{\infty}
\frac{d\tau}{\sqrt{1+3\tau}} \exp \left\{
\frac{24\tau^2-4\tau^3-3\tau^4}{12uJ(1+3\tau)} \right\},
\end{equation}
where $\tau=Jt$.
 We now apply the method of steepest descent to obtain
\begin{equation}
II \approx   -\frac{\sqrt{\pi}\psi_0^3}{\sqrt{2uJ}} \exp \left\{
\frac{32}{81~uJ} \right\} .
\end{equation}

In the third term on the right hand side ($III$) of
Eq.~(\ref{final}), we make the substitution $y=vx'$ and then
perform the $x'$ integration giving
\begin{eqnarray}
III=\frac{\psi_0^3}{uJ^2}&&\int_0^{\infty} d\tau  \int_0^1 dv
\frac{(2\tau+\tau^2)}{\sqrt{[1+\tau(1+2v^2)]^3}} \nonumber \\ &&
\times \exp \left\{ \frac{24v^2\tau^2-4\tau^3-(1+2v^2)\tau^4}
{12uJ[1+(1+2v^2)\tau]}\right\}.
\end{eqnarray}
The maximum of the term in the exponential of $III$ occurs at
$v=1$ (which is an endpoint).
 Linearizing about that maximum provides
\begin{eqnarray}
III \approx \frac{\psi_0^3}{uJ^2}&& \int_0^{\infty} \frac{ d \tau
\tau(2+\tau)} {\sqrt{(1+3 \tau)^3}} \exp \left\{
\frac{24\tau^2-4\tau^3-3\tau^4}{12uJ(1+3\tau)} \right\} \nonumber
\\ && \times \int_0^1 dw \exp \left\{ -\frac{\tau^2(2+\tau)^2w}
{uJ(1+3\tau)^2}\right\},
\end{eqnarray}
where $w=1-v$.
 After the $w$ integration, we apply the method of steepest
descent to the $\tau$ integration to obtain
\begin{equation}
III \approx \frac{\sqrt{\pi u}\psi_0^3}{\sqrt{2J}} \frac{9}{8}
\exp \left\{ \frac{32}{81~uJ} \right\}.
\end{equation}
Putting all of these results back into Eq.~(\ref{final}) gives
\begin{equation}
\psi_0 \approx \frac{\sqrt{\pi u}\psi_0^3}{\sqrt{2J}} \exp \left\{
\frac{32}{81~uJ} \right\} \left[\frac{9}{8}  - \frac{1}{u}
\right],
\end{equation}
which provides the expression given in the text, Eq.~(\ref{self}).
 This calculation clearly runs into trouble when $u<8/9$; however,
the numerical coefficients in front of these integrals are
sub-leading terms, and they can be varied by adding sub-leading
terms to the initial Gaussian guess.

\singlespacing

\bibliographystyle{plain}
\nocite{*}
\bibliography{gl,likharev}

\end{document}